\documentclass[]{scrartcl} 
  \pdfoutput=1
   \usepackage[]{graphicx}
 \usepackage[]{natbib}
\usepackage{url}
\usepackage{aas_macros}
\usepackage{multirow}
\usepackage{epstopdf}
\usepackage[affil-it]{authblk}

\begin{document}
\title{Molecular Detectability in Exoplanetary Emission Spectra}
\author{M. Tessenyi$^{a}$, G. Tinetti$^{a}$, G. Savini$^{a,b}$, E. Pascale$^{b}$}
\affil{$^a$University College London, Department of Physics and Astronomy, Gower Street, London WC1E 6BT, UK}
\affil{$^b$School of Physics and Astronomy, Cardiff University, Queens Buildings, The Parade, Cardiff CF24 3AA, UK}

\begin{abstract}
Of the many recently discovered worlds orbiting distant stars, very little is yet known of their chemical composition.
With the arrival of new transit spectroscopy and direct imaging facilities, the question of molecular detectability as a function of signal-to-noise (SNR), spectral resolving power and type of planets has become critical.
In this paper, we study the detectability of key molecules in the atmospheres of a range of planet types, and report on the minimum detectable abundances at fixed spectral resolving power and SNR.
The planet types considered --- hot Jupiters, hot super-Earths, warm Neptunes, temperate Jupiters and temperate super-Earths --- cover most of the exoplanets characterisable today or in the near future.
We focus on key atmospheric molecules, such as $CH_4$, $CO$, $CO_2$, $NH_3$, $H_{2}O$, $C_{2}H_2$, $C_{2}H_6$, $HCN$, $H_{2}S$ and $PH_3$.
We use two methods to assess the detectability of these molecules: a simple measurement of the deviation of the signal from the continuum, and an estimate of the level of confidence of a detection through the use of the likelihood ratio test over the whole spectrum (from $1$ to $16\mu m$).
We find that for most planetary cases, SNR=5 at resolution R=300 ($\lambda < 5\mu m$) and R=30 ($\lambda > 5\mu m$) is enough to detect the very strongest spectral features for the most abundant molecules, whereas an SNR comprised between 10 and 20 can reveal most molecules with abundances $10^{-6}$ or lower, often at multiple wavelengths.
We test the robustness of our results by exploring sensitivity to parameters such as vertical thermal profile, mean molecular weight of the atmosphere and relative water abundances. We find that our main conclusions remain valid except for the most extreme cases.
Our analysis shows that the detectability of key molecules in the atmospheres of a variety of exoplanet cases is within realistic reach, even with low SNR and spectral resolving power.

\end{abstract}
\maketitle

\label{intro}
\section{Introduction}

The exoplanet field has been evolving at an astonishing rate: nearly a thousand planets have been detected \citep{schneider2013} and twice as many are awaiting confirmation \citep{kepler,batalha2012,fressin2013}. 
Astronomers have begun classifying these planets by mass, radius and orbital parameters, but these numbers tell us only part of the story as we know very little about their chemical composition. Spectroscopic observations of exoplanet atmospheres can provide this missing information, critical for understanding the origin and evolution of these far away worlds. 
At present, transit spectroscopy and direct imaging are the most promising methods to achieve this goal.
Ground and space-based observations (VLT, Keck, IRTF, Spitzer, and the Hubble Space Telescope) of exoplanets have shown the potentials of the transit method: current observations of hot gaseous planets have revealed the presence of alkali metals, water vapour, carbon monoxide and dioxide and methane in these exotic environments \citep[e.g.][]{charbonneau_detection_2001,knutson_using_2007,tinetti_2007,beaulieu_primary_2007,redfield_sodium_2007,grillmair_strong_2008,snellen_ground-based_2008,swain2008b,swain2009a,swain2009b,bean2010,beaulieu_2009,crossfield2010,stevenson2010,snellen2010,tinetti_2010,berta2012,crouzet2012,dekok2013,deming2013,swain2013,waldmann2013}. However, the instruments used in the past ten years were not optimised for this task, so the available data are mostly photometric or low resolution spectra with low signal to noise.
Additionally, multiple observations are often required, during which many effects can alter the signal: from the weather on the planet to other sources of noise including instrument systematics and stellar variability. 
The interpretation of these --- often sparse ---  data is generally a challenge \citep{swain2009a,swain2009b,madhu_seager_09,lee12,line12}.
\\
With the arrival of new facilities such as Gemini/GPI, VLT/SPHERE, E-ELT and \emph{JWST}, and possibly dedicated space instruments such as \emph{EChO} \citep{Tinetti2012b}, many questions need to be tackled in a more systematic way. Among these stands out the question of molecular detectability: \emph{what are the objective criteria that need to be met to claim a molecular detection in an exoplanet?} 
In this paper we aim to address this question by focusing on the signatures of a selection of key molecules, with a range of abundances, over a broad wavelength range (1 to 16 $\mu m$).
To capture the extent of possible chemical compositions of exoplanet atmospheres, we have chosen five planetary cases: hot Jupiter, hot super-Earth, warm Neptune, temperate Jupiter and temperate super-Earth.
While our study has been inspired by transit spectroscopy with a hypothetical \emph{EChO}-like space-based instrument, the methodology and results of this paper are applicable to observations with 
other instruments and techniques, including direct imaging.
\section{Methods}
\label{sec:methods}
We select five planets out of a range of sizes (Jupiter, Neptune and super-Earth sizes) and temperatures (hot, warm and temperate), listed in Table \ref{tab:tempsize}, to describe comprehensively the chemical compositions that can be expected in exoplanet atmospheres.
The atmospheric components and their spectroscopic signals depend strongly on the planetary temperature and size, we thus focus on cases delimiting these parameters.
Other cases can be constrained by these five planet types.
\begin{table}[h!]
\small
\centering
\begin{tabular}{l c c c}
\hline
\hline
Temperature/Size &  Jupiter-like &  Neptune-like &  super-Earth \\
\hline
Hot ($\ge$800 K) &   \textbf{HJ} &   HN &  \textbf{HSE}\\
Warm (350-800 K) & WJ & \textbf{WN}  & WSE \\
Temperate (250-350K) & \textbf{TJ} & TN  & \textbf{TSE}\\
\hline
\end{tabular}
\caption{ \footnotesize Subdivision of planetary atmospheres according to temperature and planet size.
The difficulty in the observations increases from left to right and from top to bottom.
The categories highlighted in bold are the subject of our study. The observability of other planet types can be extrapolated from these cases.
Planets with temperatures below ``temperate''  have a signal too weak for both transit spectroscopy and direct detection, we consider warmer candidates for this study.
}
\label{tab:tempsize}
\end{table}
The planetary and stellar parameters assumed for these targets, listed in Table \ref{tab:target_params}, are obtained from observations when possible; calculated values are used otherwise.
We used HD 189733b \citep{bouchy189} as a template for the hot Jupiter case, GJ 436b \citep{butler436} for the warm Neptune case, and Cnc 55e \citep{winn2011} for the hot super-Earth case.
We also consider the case of a temperate super-Earth orbiting a late type star. Such a planet could be subjected to intense radiation and be tidally locked; however, an atmosphere on this type of planet is plausible, as has been discussed in the literature (e.g. \citet{joshi1997,wordsworth2010,segura2010}).
\begin{table}[h!]
\small
\hspace{-0.2in}
\begin{tabular}{l c c c c c}
\hline
\hline
 & \multicolumn{2}{c}{Hot} & Warm & \multicolumn{2}{c}{Temperate}\\
Star & Jupiter & super-Earth & Neptune &  Jupiter & super-Earth\\
\hline
Spectral Type 		& K1V  	& G8V	& M2.5V 	& K4V	& M4.5V\\
Radius (R$_\odot$)	& 0.79 	& 0.94	& 0.46 	 	& 0.75	& 0.22\\
Mass (M$_\odot$)	& 0.8	& 0.91	& 0.45		& 0.8	& 0.22\\
Temperature (K)		& 4980 	& 5196  	& 3684 		& 4780	& 3300\\
Distance (pc)		& 19.3	& 12.34	& 10.2		& 10	& 10\\
\hline
\multicolumn{6}{l}{Planet}\\
\hline
Radius (R$_{jup}$ / R$_\oplus$) & 1.138 / 12.77 	& 0.194 / 2.18	& 0.365 / 4.10	&	1.138 / 12.77	& 0.16 / 1.8\\
Temperature (K)		& 1350 	& 2100		& 750 		& 250	& 250\\
Semi-major axis (au)	& 0.031	& 0.016		& 0.029 		& 0.4 	& 0.046\\
Period (days)		& 2.2 	& 0.74		& 2.6		& 102	& 7.6 \\
Transit duration (hr)	& 1.83	& 1.76		& 1.03		& 7.9 	& 1.39\\
Bulk atm. composition & $H_2$ & $H_{2}O$ & $H_2$ & $H_2$ & $N_2$\\
$\mu$ ($u$) 		& 2.3 	& 18.02			& 2.3			& 2.3		& 28.01\\
\hline
Surfaces ratio  & 2.20$\times 10^{-2}$ & 4.48$\times 10^{-4}$ & 6.55$\times 10^{-3}$ & 2.43$\times 10^{-2}$ & 5.6$\times 10^{-3}$\\
\hline
\end{tabular}
\caption{\footnotesize Stellar and planetary parameters assumed for this study. The planetary radii are given both in units of Jupiter radii and Earth radii, and the temperatures listed are an average temperature from the adopted temperature-pressure profiles. The mean molecular weight of the atmosphere considered is indicated by $\mu$. The star/planet ratio $({R_{pl}}/{R_*})^2$ is also listed here to facilitate the comparison among the targets studied.} 
\label{tab:target_params}
\end{table}

In this study, we focus on emission spectroscopy in the infrared, obtainable through secondary eclipse observations or direct imaging.
For transiting planets, the emission spectra can be obtained by subtracting the stellar signal from the combined light of star+planet.
In practice, the measurements and simulations are given as the flux emitted by the planet in units of the stellar flux:
\begin{equation}
F_{II} (\lambda) = {\left( \frac{R_p}{R_\star} \right)}^2 \frac{F_p(\lambda) }{F_\star(\lambda)}
\label{eq:fluxcontrast}
\end{equation}
where $F_p$  and $F_\star$ are the planetary and stellar spectra.
This equation highlights the influence of both the surfaces ratio and the relative temperatures of the planet and star for secondary eclipse measurements. 
\subsection{Models}
\label{sec:models}
\subsubsection{Planetary and Stellar Spectra}
\begin{figure}[!h]
\hspace*{-0.6in}
\includegraphics[trim=0 7cm 0 1.8cm, clip=true,width=6.8in,angle=180]{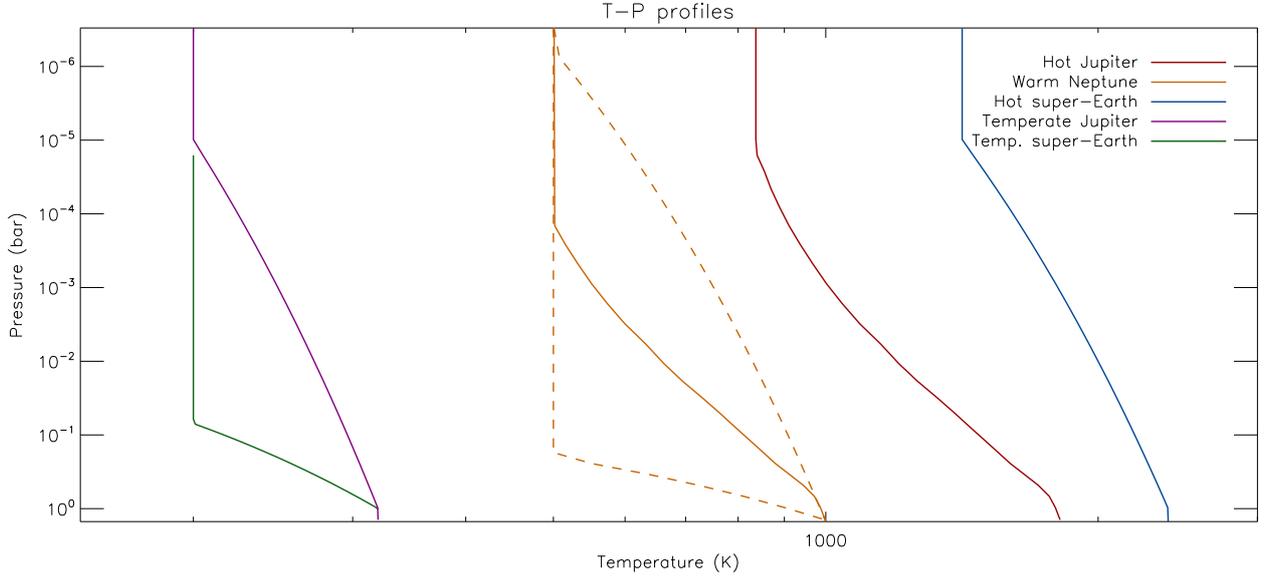}
\caption{\footnotesize Temperature-pressure (T-P) profiles of the five target types presented. \emph{From left to right:} temperate super-Earth and Jupiter, warm Neptune with three possible profiles: a steep dry adiabatic profile (dashed, left), a more isothermal profile (dashed, right) and a simulated one \citep{beaulieu2011} in between (solid), a hot Jupiter profile \citep{Burrows2008} and a hot super-Earth profile. }
\label{fig:tpprofiles}
\end{figure}

With the range of planetary temperatures and sizes considered, the temperature-pressure (T-P) profile will vary significantly for the five planet cases. The T-P profile describes the change in temperature as a function of pressure in a given atmosphere.
Figure \ref{fig:tpprofiles} shows the T-P profiles assumed for the planets. 
To investigate the effect that the thermal gradient has on the observed signal, two additional more extreme T-P profiles are presented for the Warm Neptune case: a dry adiabatic profile with a steep lapse rate reaching 500 K at $\sim$0.1 bar, and a profile with a lapse rate closer to isothermal, reaching 500K at $10^{-6}$ bar.
Results for these additional profiles are presented in section \ref{sec:alttp}.\\
In the case of super-Earths, the atmosphere --- if present --- could be dominated by a variety of molecules, such as hydrogen ($\mu=2.3u$), water vapour ($\mu=18.02u$), nitrogen ($28.01u$) or carbon dioxide ($44u$). 
A change in the main atmospheric component will impact both the atmospheric scale height ($H$) and the atmospheric lapse rate ($\gamma$).
For our tests we have assumed a dry adiabatic lapse rate: 
\begin{equation}
H=\frac{kT}{\mu g}  \,\,\,\,\,\,\,\,\,\,\,\,\,\,\,\,\,\,\,\,\,\,\,\,\,\,\,\,\,\,\,\,\,\,\,\,   \gamma=-\frac{dT}{dz} = \frac{g}{c_p}
\end{equation}
where $k$ is the Boltzmann constant, $g$ is the gravitational acceleration, $T$ the temperature in degrees Kelvin, $\mu$ the mean molecular weight of the atmosphere, $z$ the altitude and $c_p$ the specific heat of the gas.
We tested the impact on molecular detectability in an atmosphere composed of hydrogen, water vapour, nitrogen or carbon dioxide.
The parameters derived for each of the cases are shown in Table \ref{tab:mugamma}.
\begin{table}[h!]
\small
\centering
\begin{tabular}{l c c c}
\hline
\hline
Main constituent & \,\,\,\, $\mu$ ($u$)\,\,\,\,	& \,\,\,\,  $H$ (km) \,\,\,\, & \,\,\,\, $\gamma$ (K/km) \,\,\,\,\\
\hline
Hydrogen 		&  2.3 	&   76.6 &  1.1 \\
Water vapour 	& 18.02 	&    9.8  & 8.1 \\
Nitrogen 		& 28.01 	&     6.3 & 14.5\\
Carbon dioxide 	& 44 		&     4.0 & 17.8\\
\hline
\end{tabular}
\caption{ \footnotesize Temperate super-Earth atmospheric parameters considered, from a hydrogen dominated atmosphere to a carbon dioxide dominated atmosphere.
$\mu$ is the molecular weight, $H$ the atmospheric scale height and $\gamma$ the corresponding dry adiabatic lapse rate.
}
\label{tab:mugamma}
\end{table}\\

We calculated the infrared emission spectra using a line-by-line radiative transfer model (See e.g. \citet{goudy_yung}, Chapter 6) developed for disk-averaged, terrestrial planetary spectra \citep{tinetti_infrared_2006} and subsequently adapted to simulate hot, gaseous planets \citep{tinetti_faraday}.
The model covers a pressure range from 10 to $10^{-6}$ bars.
The molecular absorption is computed based on the mixing ratio, local density and temperature in accordance with the assumed T-P profile.
the wavelength dependent molecular opacity is estimated through the ExoMol \citep{tennyson_exomol} and HITRAN 2008 \citep{rothman} line-lists.\\
For every planetary case, an individual spectrum is generated for each molecule (Table \ref{tab:methodmols}) assuming five mixing ratios, ranging from $10^{-7}$ to $10^{-3}$.
Stellar spectra are obtained from observed and simulated models \citep{hauschildt99,kurucz_solar_1995}.
\begin{table}[h!]
\small
\centering
\begin{tabular}{l l}
\hline
\hline
Planet & Molecules considered \\
\hline
Hot Jupiter & $CH_4$, $CO$, $CO_2$, $NH_3$, $H_{2}O$, $C_{2}H_2$, $C_{2}H_6$, $HCN$, $H_{2}S$ and $PH_3$\\
Hot super-Earth & $H_{2}O$, $CO$ and $CO_2$\\
Warm Neptune & $CH_4$, $CO$, $CO_2$, $NH_3$, $H_{2}O$, $C_{2}H_2$, $C_{2}H_6$, $HCN$, $H_{2}S$ and $PH_3$\\
Temperate Jupiter & $H_{2}O$, $CH_4$, $CO_2$, $C_{2}H_2$ and $C_{2}H_6$\\
Temperate super-Earth & $H_{2}O$, $CO_2$, $NH_3$ and $O_3$\\
\hline
\end{tabular}
\caption{\footnotesize Molecules considered in the atmospheres of the planets studied. For all planets and molecules, a uniform mixing ratio is assumed across the temperature-pressure range. }
\label{tab:methodmols}
\end{table}
The planetary and stellar parameters and spectra are used to calculate the photon flux from the planet and star as a function of wavelength, and are presented as a planet/star contrast spectrum (equation \ref{eq:fluxcontrast}).\\
We consider the 1 to 16 $\mu m$ wavelength range to best capture the key molecular features present in a planetary atmosphere with a temperature between 250K and 3000K.
This spectral interval is also compatible with the currently available or foreseen instruments for transit spectroscopy and direct imaging.
The spectral resolution is set to R=300 and R=30 for the 1 to 5 and 5 to 16 $\mu m$ spectral intervals, respectively, and lowered to R=20 in the 5 to 16 $\mu m$ spectral interval for the temperate super-Earth.
These choices optimise the performances of potential instruments with the number of photons typically available.\\
The only source of noise assumed in this work is photon noise, and an overall optical efficiency of 0.25 has been considered (e.g. reflectivity of mirrors, throughput of optical system, detector quantum efficiency, etc.). 
For a given duration of observation and for every resolution bin, the signal to noise ratio (SNR) is calculated for the star and for the planet:
\begin{eqnarray}
\label{eq:snrplanet}
SNR_*&=&N_* / \sqrt{N_*}\\
SNR_p&=&F_{II}\times SNR_* = \frac{N_p}{\sqrt{N_*}} 
\end{eqnarray}
where $N_*$ is the number of photons received from the star, $N_p$ is the number of photons received from the planet, and $F_{II}$ is the planet/star contrast spectrum (see equation \ref{eq:fluxcontrast}).
One sigma error bars are computed for the planet/star contrast spectrum in every resolution bin: 
\begin{equation}
\label{eq:sigma}
\sigma= \frac{F_{II}}{SNR_p}
\end{equation}
To address the question of molecular detectability, the results in section \ref{sec:results} are presented as function of fixed SNR$_p$ (from hereon referred to as SNR) in the spectral intervals where the molecular features are located.
In this way, our results are completely independent from the duration of the observations and the instrument design.
However, to give an estimate of the observational requirements needed to achieve these SNR values, we show in appendix A the typical SNR values obtainable with a dedicated space-based instrument.
\subsection{Molecular Detectability}
In a planet/star contrast spectrum, the molecular features appear as departures from the continuum. At a fixed T-P profile, the absorption depth or emission feature will depend only on the abundance of the molecular species.
We use two approaches to determine the minimum detectable abundance for each molecule: individual bins and likelihood ratio test.
\subsubsection{Individual bins}
\label{sec:methodfixsnr}
This is the most intuitive and conservative approach: we measure in every bin the difference between the planetary signal with or without the absorption of a selected molecule. We claim a detection if a difference of at least 3-sigma (see equation \ref{eq:sigma}) is found between the continuum and the molecular signature in a given bin.
\begin{figure}[h!]
\hspace*{-0.45in}
\includegraphics[width=3.3in]{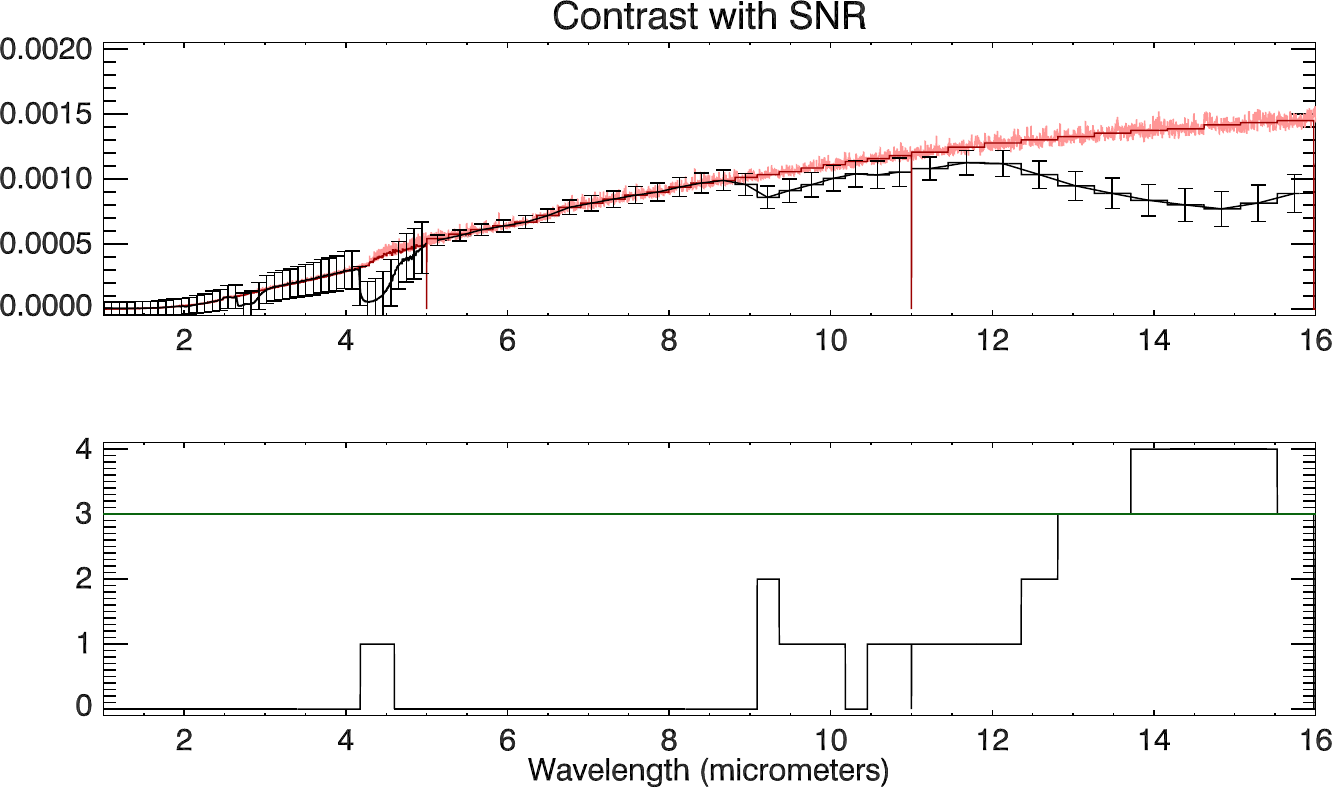}\includegraphics[width=3.3in]{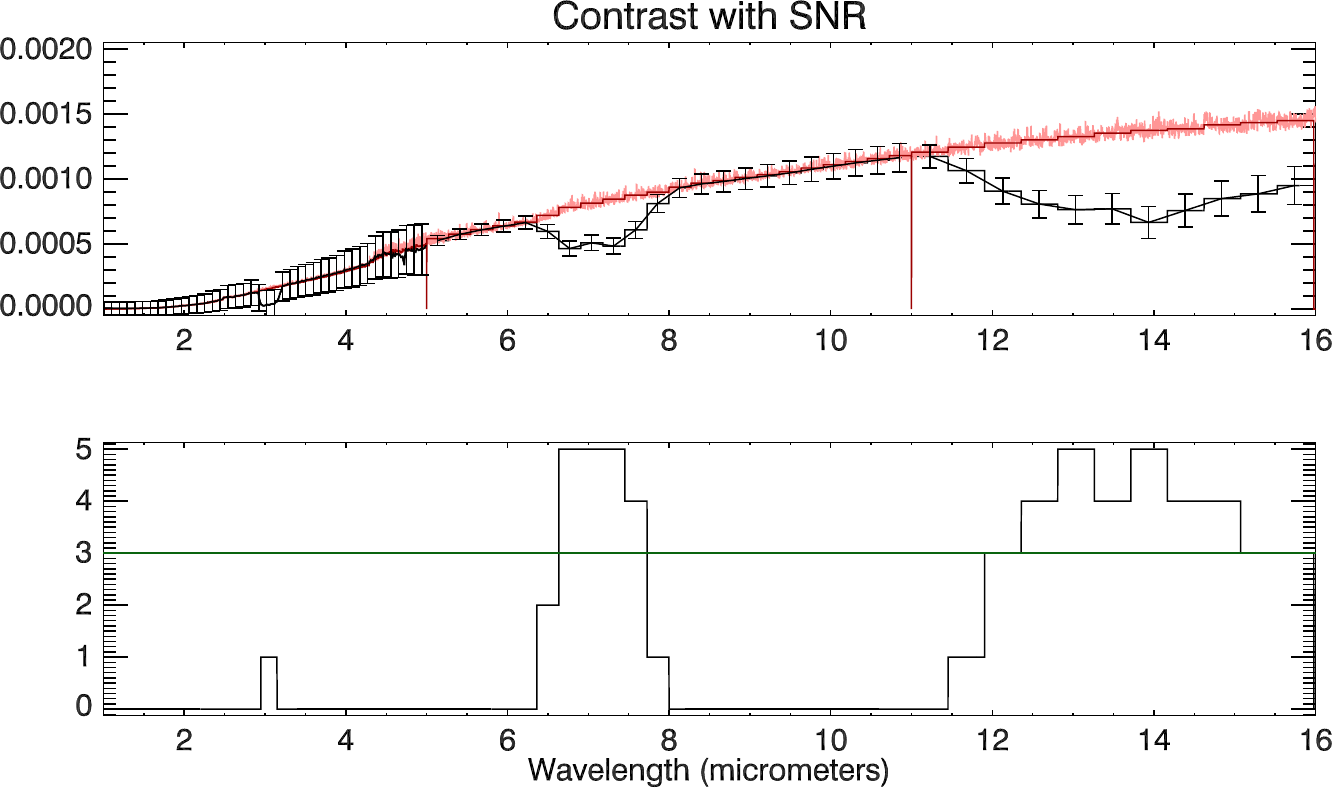}
\caption{\footnotesize
Individual bin method to detect the presence of a molecule in the atmosphere of a Warm Neptune.
The upper panels show contrast spectra where two different molecules absorb.
The error bars are computed with fixed SNR=10.
\emph{Left: } $CO_2$ with mixing ratio=$10^{-5}$, \emph{Right:} $HCN$ with mixing ratio=$10^{-4}$.
The planet continuum is shown in red.
The lower panels show the departure of the molecular signal from the continuum in units of sigma (see eq. \ref{eq:sigma}).
A 3-sigma departure is required to claim a detection. This threshold is shown here as the green horizontal line.
}
\label{fig:3sigma}
\vspace{-0.1in}
\end{figure}
While the depth of the feature will depend on the abundance of the molecule (at fixed thermal profile), the SNR in that bin will determine the value of sigma. 
We present in our results the minimum molecular abundance detectable as a function of fixed SNR=5, 10 or 20 and wavelength.
Figure \ref{fig:3sigma} shows an example of $CO_2$ and $HCN$ in the atmosphere of a Warm Neptune, with a fixed SNR=10.
If the departure from the continuum is less than 3-sigma, we cannot claim a detection.
However, given that most spectral features span over multiple bins, the likelihood ratio test can use this information in a more optimal manner. 
\subsubsection{Likelihood Ratio Test}
As in the individual bin method, the idea here is to test the hypothesis of a molecular detection in a noisy observation.
Also, for every molecule considered, the tests described here are repeated for the five abundance levels, to determine the minimum detectable abundances.
The likelihood ratio test \citep{np1928} provides the confidence with which we can reject the ``null hypothesis'', i.e.  no molecular features are present in our observation.
We consider a detection to be valid if we can reject the null hypothesis with a 3-sigma confidence.\\
In this paper, we simulate the null hypothesis by a blackbody curve at the planetary temperature. 
The ``alternative hypothesis'' is represented by a planetary spectrum containing features carved by a specific molecule at a particular abundance.
As we are not using observational data, the planetary and stellar spectra are simulated with the methods described in section \ref{sec:models}.\\
We perform a likelihood ratio test over the selected wavelength range under two assumptions:
first, we consider a signal that has been emitted by a planet with no molecular features present,
and second, we consider a signal of a planetary spectrum containing features of a molecule at a selected abundance.
These tests are repeated $\sim10^5$ times to build up an empirical understanding of the noise distribution.
To reproduce the observational setting, we combine the planetary signal with a stellar signal.
We generate poisson noise for both the star+planet signal and for the star only signal, with means equal to the respective signals.
The noisy planetary signal is the difference between these two noisy signals, on which we perform two calculations:\\
the likelihood of observing the null hypothesis ($H_0$), i.e. the noisy planet signal as a blackbody curve,
and the likelihood of observing the alternative hypothesis ($H_1$), i.e. the noisy planet signal as a spectrum containing molecular features.\\
The general form of the likelihood ratio test is given as:
\begin{equation}
D = -2 \ln \left( \frac{L_0}{L_1} \right) =-2 \ln (L_0) + 2 \ln (L_1)
\label{eq:lrtlog}
\end{equation}
where $L_0$ and $L_1$ are the likelihoods of observing the null hypothesis and the alternative hypothesis, respectively.
Both $L_0$ and $L_1$ are calculated using the Gaussian distribution, as it is a good approximation to the distribution of the difference of two poisson random variables with large means, over all the bins $i$:
\begin{equation}
L_{0} = \prod\limits_{i=1}^n \frac{1}{\sigma_i \sqrt{2 \pi}} \exp{\frac{-(x_i - \mu_{i,0})^2}{2 \sigma_i^2}}\\
\label{eq:l0}
\end{equation}
\begin{equation}
L_{1} = \prod\limits_{i=1}^n \frac{1}{\sigma_i \sqrt{2 \pi}} \exp{\frac{-(x_i - \mu_{i,1})^2}{2 \sigma_i^2}}
\label{eq:l1}
\end{equation}
where for both equations, $x_i$ is the observed (noisy) data in bin $i$, $\mu_i$ is the expected value of the signal in the bin, and $\sigma_i^2$ is the sum of variances of the star+planet and star variances ($ \sigma^2 = 2\sigma_{star}^2 + \sigma^2_{planet} = 2 \mu_{star} + \mu_{planet}$), which are both poisson distributions.
Both equations \ref{eq:l0} and \ref{eq:l1} can be expressed in the logarithm form:
\begin{equation}
\ln (L_{0,1})= \sum -\frac{(x_i-\mu_{i_{0,1}})^2}{2\sigma_i^2}-\ln \sigma_i -\frac{\ln 2\pi}{2}
\end{equation}

Using equation \ref{eq:lrtlog}, we thus obtain a value $D$. We repeat these steps $\sim10^5$ times, generating a new noisy signal at each iteration. We build up a distribution of the likelihood difference values $D$ for the planetary signal generated from a blackbody curve.
\\
Under the second assumption, the planetary signal is replaced with a planetary spectrum containing features of a molecule at a selected abundance.
Noise is added as described above, and we compute the likelihood of the null hypothesis ($H'_0$) and the likelihood of the alternative hypothesis ($H'_1$).
Using equation \ref{eq:lrtlog}, we obtain a likelihood ratio value that we call $D'$.
These steps are repeated $\sim10^5$ times, generating a new noisy signal for each iteration.
With these results we build a distribution of the likelihood difference values $D'$ for the planetary signal including molecular features.
The two distributions ($D$ and $D'$) are expected to be approximately symmetric as they are obtained by the same test, by switching the null hypothesis and the alternative hypothesis in the signal generation process.

\begin{figure}[!h]
\hspace{-0.4in}
\includegraphics[width=3.1in]{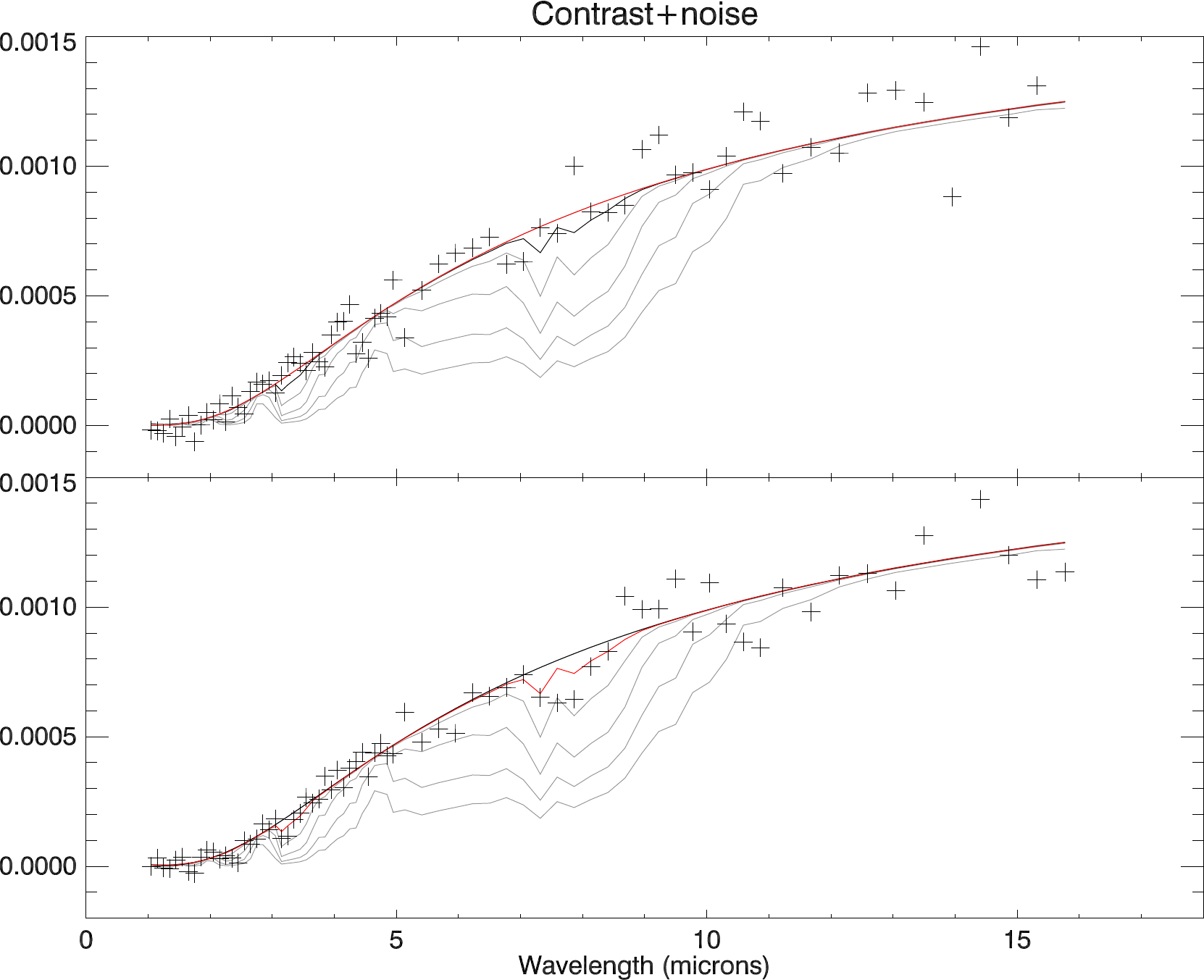}\includegraphics[width=3.4in]{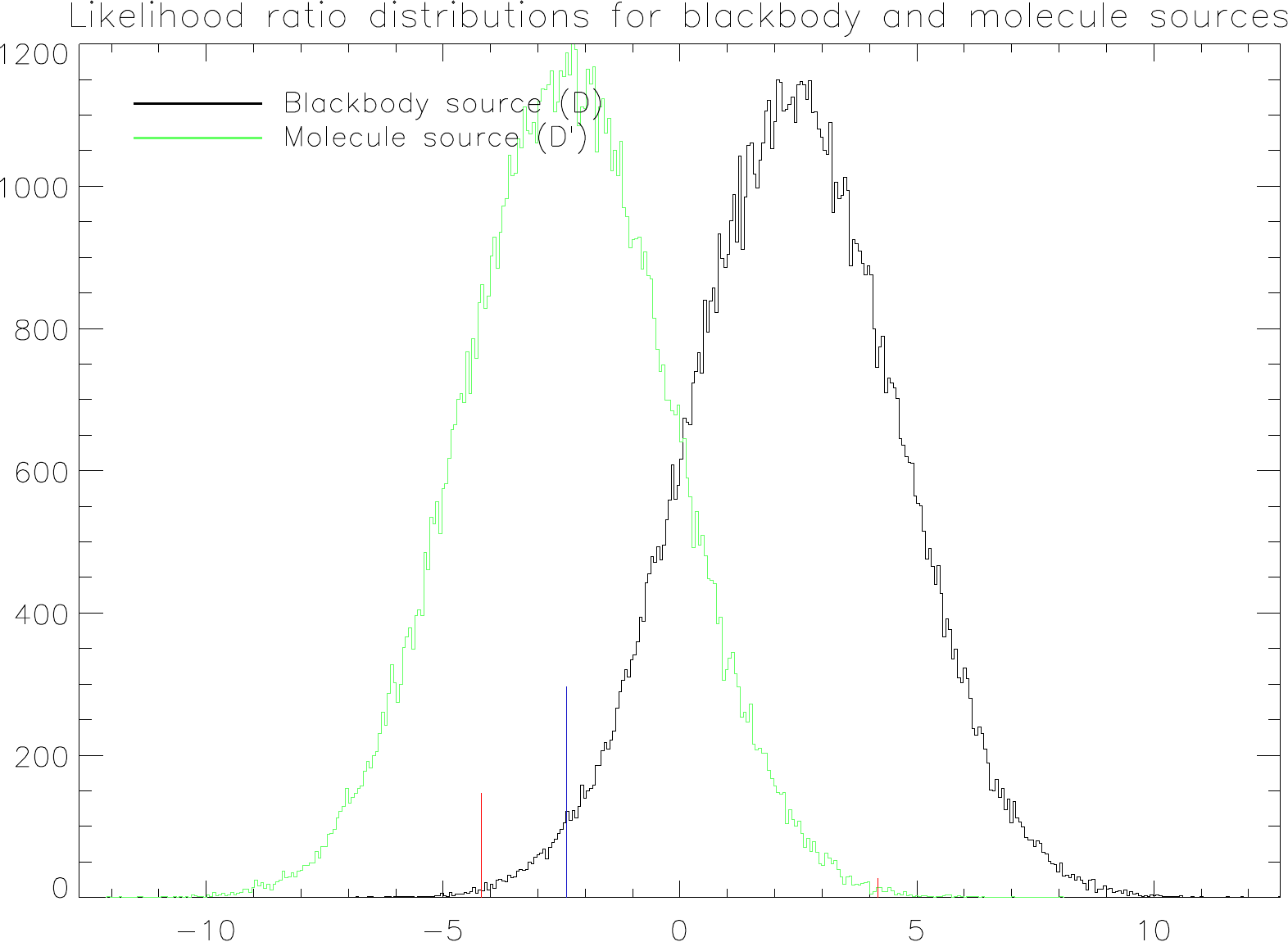}
\caption{\footnotesize Likelihood ratio test results for a Warm Neptune with $CH_4$ in the atmosphere.
\emph{Left:} One transit simulation of the planetary signal.
\emph{Top:} Planet/star contrast spectra generated with 5 abundances (in grey). The planetary signal is generated by a blackbody in red.
\emph{Bottom:} The planetary signal here is generated by a molecular spectrum with abundance $10^{-7}$ (red).
In both plots, the resolution in the 1 to 5 $\mu m$ channel has been lowered to R=30 for clarity purposes.
\emph{Right:}  The two LR distributions including the null hypothesis ($D$, black) and the alternative hypothesis ($D'$, green).
The red line on the null hypothesis distribution marks the 3-sigma limit, and the blue line on the alternative hypothesis distribution marks the median.
Here the two distributions overlap, and more than 50\% of the alternative hypothesis distribution has crossed the 3-sigma detection limit.
Given the result, the null hypothesis is not rejected, and we cannot claim a detection.
}
\label{fig:lr2}
\vspace{-0.1in}
\end{figure}

The level of distinction between the two considered signals will depend, as in the individual bin method, on the amount of noise and the strength of the molecular features.
If the noise is large on the simulated observations, the two distributions will overlap as the likelihood of the hypotheses $H_0$ and $H_1$ are similar. 
If the signal is strong compared to the noise, there will be little or no overlap between the distributions $D$ and $D'$:
the null hypothesis will typically be the most likely in the first test, and the alternative hypothesis will typically be the most likely in the second test.
As we investigate in this paper the smallest abundance at which a detection could be obtained, we only require the rejection of the null hypothesis with a 3-sigma confidence.
We do not require a 3-sigma confidence level on the alternative hypothesis; we place a maximum type-2 error (not rejecting the null hypothesis when the alternative hypothesis is true) on our alternative hypothesis of 50\%.
The $D$ distribution is used to delimit the critical value of the null hypothesis, and the $D'$ distribution is used to limit the type-2 error.
With this threshold, half of the observations will give an inconclusive result, and the other half will reject the null hypothesis with 3-sigma certainty.
\begin{figure}[!h]
\hspace{-0.4in}
\includegraphics[width=3.1in]{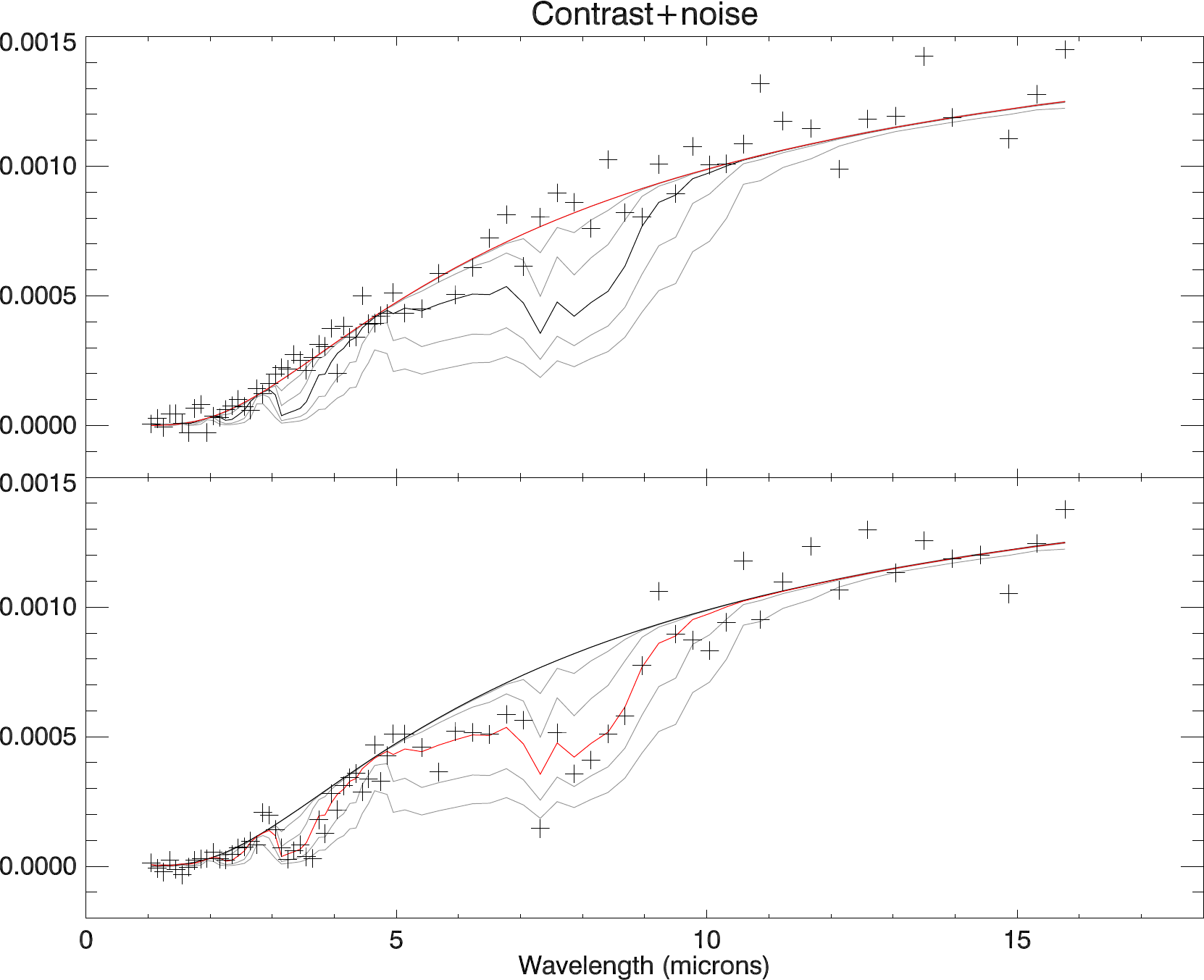}\includegraphics[width=3.4in]{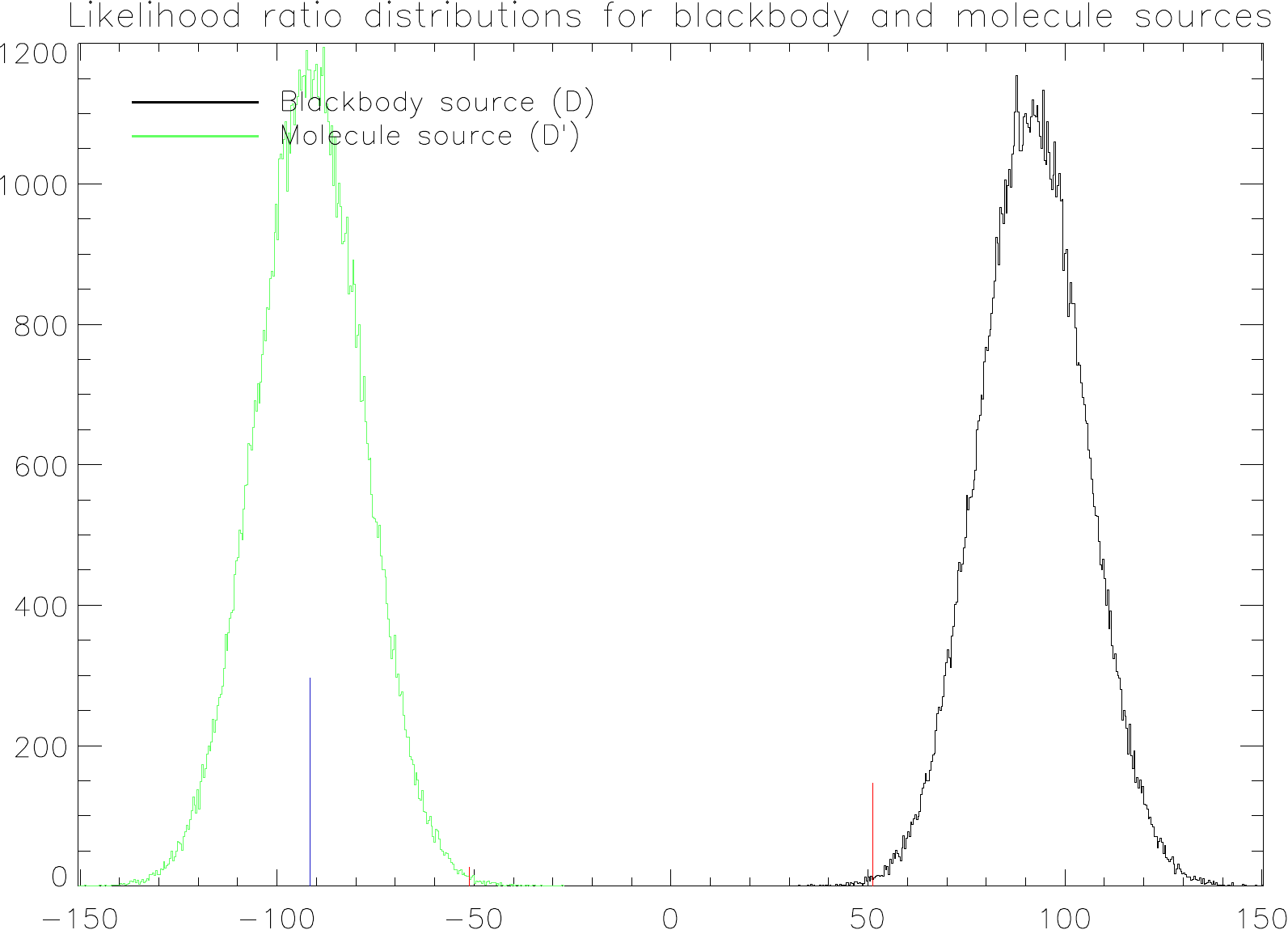}
\caption{\footnotesize Likelihood ratio test results for a Warm Neptune with $CH_4$ in the atmosphere.
\emph{Left:} One transit simulation of the planetary signal.
\emph{Top:} Planet/star contrast spectra generated with 5 abundances (in grey). The planetary signal is generated by a blackbody in red.
\emph{Bottom:} The planetary signal here is generated by a molecular spectrum with abundance $10^{-5}$ (red).
In both plots, the resolution in the 1 to 5 $\mu m$ channel has been lowered to R=30 for clarity purposes.
\emph{Right:} The two LR distributions including the null hypothesis ($D$, black) and the alternative hypothesis ($D'$, green).
The red line on the null hypothesis distribution marks the 3-sigma limit, and the blue line on the alternative hypothesis distribution marks the median.
Here the two distributions are clearly separated, and the null hypothesis of a blackbody planet signal is rejected.
Given the result, the detection of this molecule at this abundance is possible for this observation.}
\label{fig:lr1}
\vspace{-0.1in}
\end{figure}
Figure \ref{fig:lr1} shows an example of a Warm Neptune with $CH_4$ absorbing at abundance $10^{-5}$ (lower left panel).
The distribution indicated as ``blackbody source'' corresponds to the distribution of $D$ values (Figure \ref{fig:lr1}, right panel).
On the same plot, the distribution indicated as ``molecule source'', corresponds to the distribution of $D'$ values.
The two distributions are clearly separated, given that the the noise on the lower left-hand side plot doesn't appear to follow the blackbody signal, and the noise on the upper left-hand side plot doesn't appear to follow the molecular spectrum.
If a smaller abundance is considered, e.g. $10^{-7}$ rather than $10^{-5}$ (Figure \ref{fig:lr2}), the distinction between the two signals from the noisy observation is hard to make.
The two distributions here overlap quite significantly.
Both Figure \ref{fig:lr1} and \ref{fig:lr2} show a vertical red line marking the 3-sigma deviation from the mean on the ``blackbody source'' distribution, and a blue vertical line marking the median on the ``molecule source'' distribution.
\\We compare the performance of the likelihood ratio test to the individual bin method in Section \ref{sec:results2}.
\subsubsection{Detectability Limits in a Wet Atmosphere}
In the previous sections we describe the detectability limit tests of a single molecule at a time.
However, many molecules are usually present in an atmosphere and they may have overlapping spectral features.
In those cases, disentangling the various molecular signals in the spectrum may be a challenging task.
The presence of water vapour in particular may severely interfere with an accurate retrieval of other species, as water absorbs from the visible to the far infrared.
In comparison, other molecules show sparser spectral features, and we can usually separate their signatures by selecting spectral regions with no significant overlap. 
The choice of a broad spectral coverage and appropriate spectral resolving power are essential to enable an optimal retrieval process.
If these two requirements are not met, the retrieved solutions may not be unique and may present degeneracies. 
A full analysis on spectral retrieval capabilities and limits is outside the scope of this paper, we refer to \citet{terrile2008,swain2009a,swain2009b,madhu_seager_09,lee12,line12} for currently available methods in this domain.
\\
As a test case, we investigate the impact of a water vapour signal on the detectability of key molecules, such as $CO$, $CO_2$, $CH_4$ and $NH_3$, in the atmosphere of a warm Neptune.
We calculate the minimum detectable abundances of these molecules in a wet atmosphere (water vapour abundances ranging from $10^{-3}$ to $10^{-7}$) and compare those to the results presented in section \ref{sec:results} for a water free atmosphere.
In these tests, the combined ($H_{2}O$ + molecule) spectra are compared to a water only spectrum, and any deviations from this baseline are tested for $3\sigma$ detectability.
\\The results for these tests are presented in Section \ref{sec:results3}.\\

\section{Results I - Molecular detectability at fixed SNR}   
\label{sec:results}
In this section,we present the minimum mixing ratio detectable for a selected molecule, absorbing in a planetary atmosphere, as a function of wavelength and SNR (SNR of planet, SNR$_p$). The SNR here is fixed at 5, 10 and 20.
We repeat these calculations for the five planet cases: warm Neptune, hot Jupiter, hot and temperate super-Earth, and temperate Jupiter.
\subsection{Warm Neptune}
\label{sec:wnresults}
\label{sec:molecules}
\begin{figure}[h!]
\hspace*{-0.6in}
\includegraphics[width=3.5in]{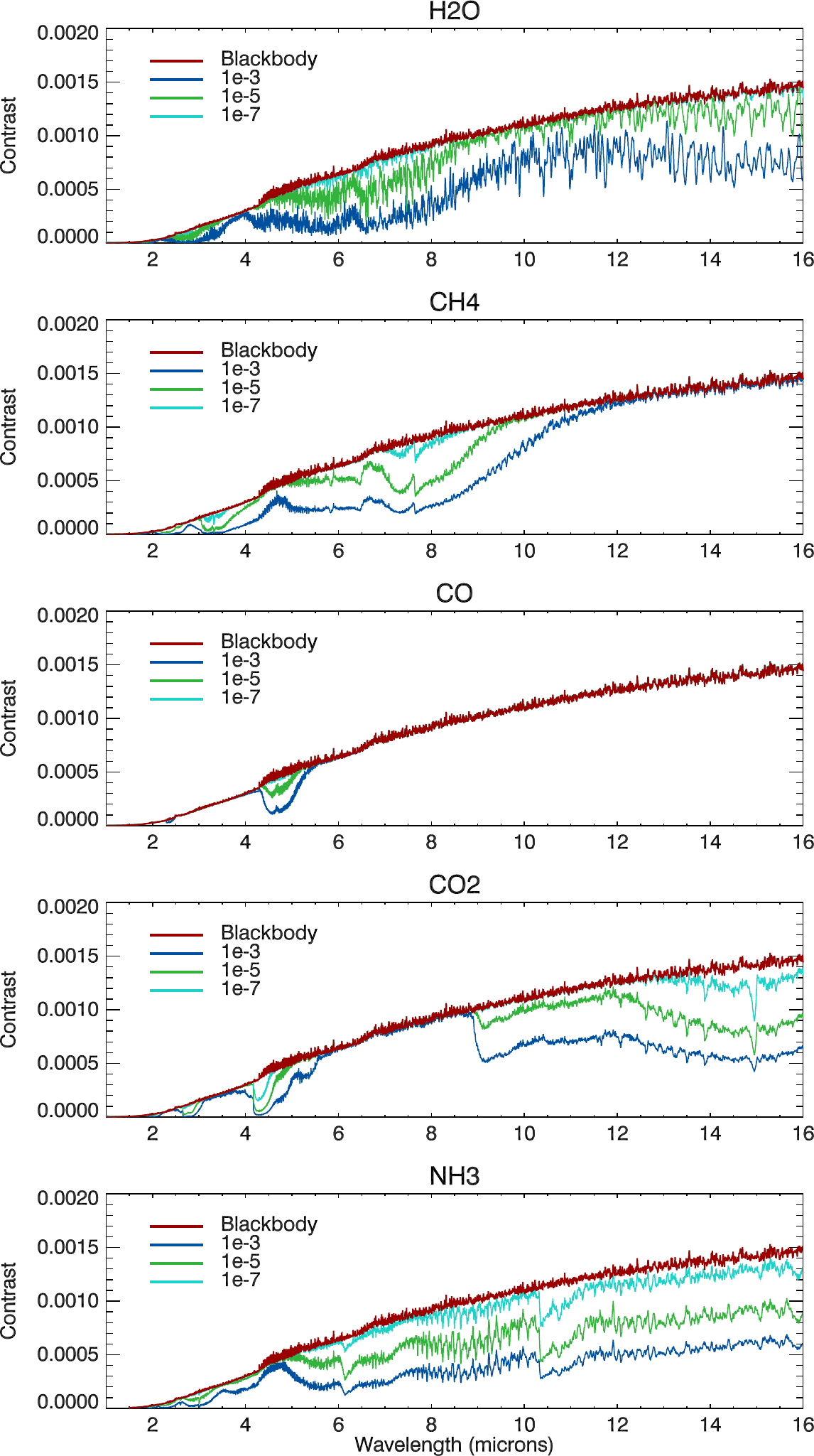}\includegraphics[width=3.5in]{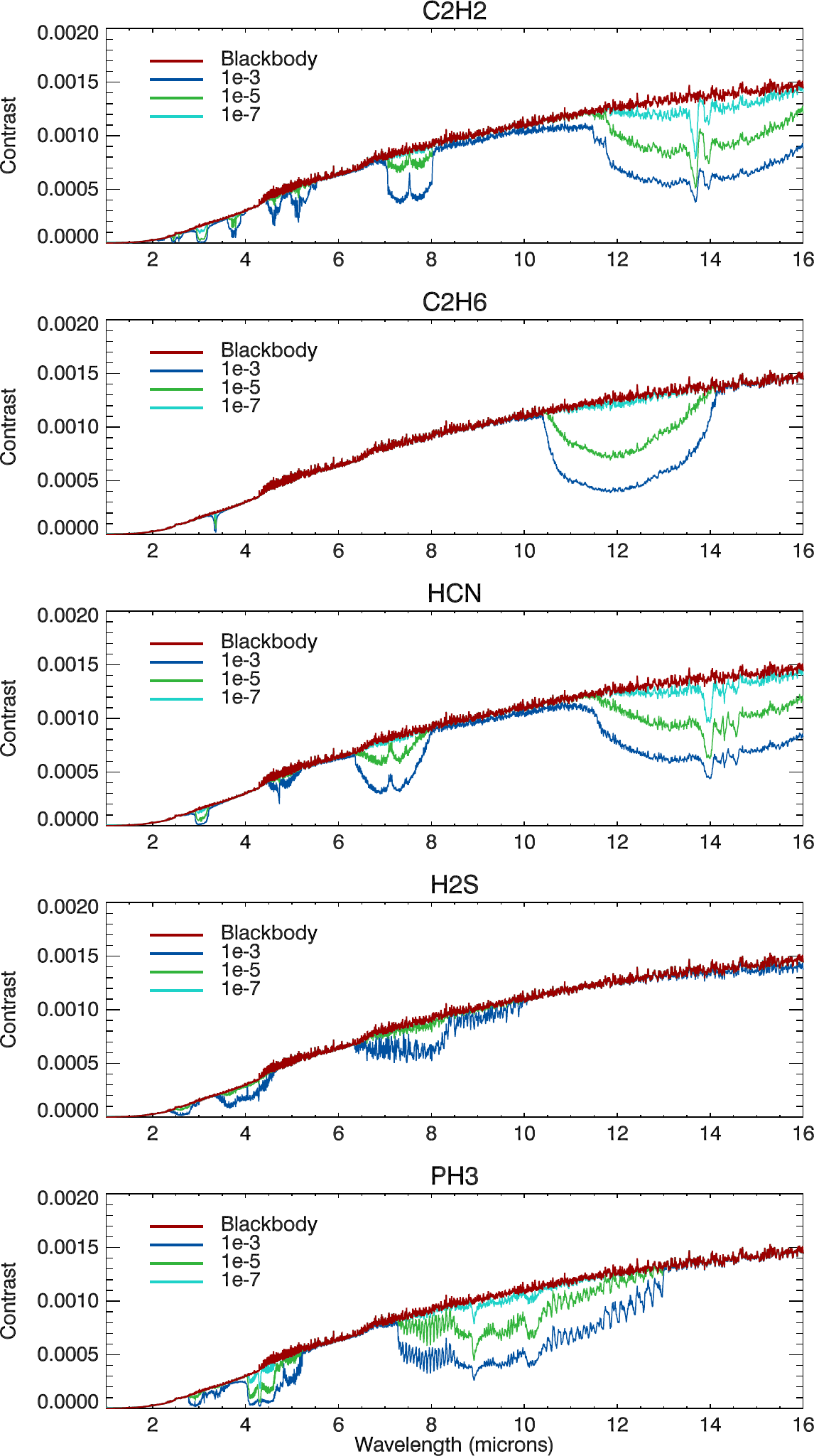}
\caption{\footnotesize Warm Neptune: planet/star contrast spectra simulating the effect of the 10 considered molecules: $CH_4$, $CO$, $CO_2$, $NH_3$, $H_{2}O$, $C_{2}H_2$, $C_{2}H_6$, $HCN$, $H_{2}S$ and $PH_3$.
The red line shows a planetary blackbody emission with no molecules present, divided by a stellar spectrum.
The green-blue colored lines depict the molecular features at different abundances.
For clarity purposes, only three abundances are plotted out of the five calculated.}
\label{fig:wn_molecules}
\end{figure}

We present in Figure \ref{fig:wn_molecules} the contrast spectra corresponding to a warm Neptune case with the following molecules: methane ($CH_4$), carbon monoxide ($CO$), carbon dioxide ($CO_2$), ammonia ($NH_3$), water ($H_{2}O$), hydrogen cyanide ($HCN$), acetylene ($C_{2}H_2$), ethane ($C_{2}H_6$), hydrogen sulfide ($H_{2}S$) and phosphine ($PH_3$).
For each molecule we present a continuum line corresponding to a blackbody emission from the planet with no molecular absorption, and three planet/star contrast spectra generated with different abundances: $10^{-7}$, $10^{-5}$, and $10^{-3}$.
While we study several abundances, for clarity we display only three values on the plots.
\label{sec:wn_table}
In Table \ref{tab:wn_fixed} we list the lowest abundances detectable as a function of SNR.
\begin{table}[!h]
\hspace*{-0.3in}
\footnotesize
\begin{tabular}{lcclcclccclcc}
\hline
\hline
 & \multicolumn{2}{c}{$CH_4$} &  & \multicolumn{2}{c}{$CO$} & & \multicolumn{3}{c}{$CO_2$} & & \multicolumn{2}{c}{$PH_3$}\\
SNR \,\,\,\,\, & 3.3 $\mu m$ & 8 $\mu m$ &  \,\,\,\,\, & 2.3 $\mu m$ & 4.6 $\mu m$ & \,\,\,\,\, & 2.8 $\mu m$ & 4.3 $\mu m$ & 15 $\mu m$ & \,\,\,\,\, & 4.3 $\mu m$ & 10 $\mu m$\\
\hline
20 & $10^{-7}$ & $10^{-6}$ 	& &	$10^{-4}$	& $10^{-6}$ 	& & $10^{-7}$ & $10^{-7}$ & $10^{-7}$ 	& & $10^{-7}$ & $10^{-6}$ \\
10 & $10^{-7}$ & $10^{-6}$ 	& &	$10^{-3}$	& $10^{-5}$ 	& & $10^{-6}$ & $10^{-7}$ & $10^{-6}$ 	& & $10^{-7}$ & $10^{-6}$ \\
5 & $10^{-7}$ & $10^{-5}$ 	& &	$10^{-3}$	& $10^{-4}$ 	& & $10^{-6}$ & $10^{-7}$ & $10^{-5}$ 	& & $10^{-7}$ & $10^{-5}$ \\
\hline
\\
\end{tabular}\\
\hspace*{-0.26in}
\begin{tabular}{lccclccclccc}
\hline
\hline
 & \multicolumn{3}{c}{$NH_3$} &  & \multicolumn{3}{c}{$HCN$} & & \multicolumn{3}{c}{$H_{2}O$}\\
SNR \,\,\,\,\, & 3 $\mu m$ & 6.1 $\mu m$ & 10.5 $\mu m$ &  \,\,\,\,\, &  3 $\mu m$ & 7 $\mu m$ & 14 $\mu m$ & \,\,\,\,\, & 2.8 $\mu m$ & 5 - 8 $\mu m$ & 11 - 16 $\mu m$  \\
\hline
20 & $10^{-7}$ & $10^{-6}$ & $10^{-7}$	& &	$10^{-7}$	& $10^{-5}$ & $10^{-7}$	& & $10^{-6}$ & 	$10^{-6}$	&  $10^{-5}$\\
10 & $10^{-6}$ & $10^{-6}$ &$10^{-6}$	& &	$10^{-6}$	& $10^{-5}$ & $10^{-6}$	& & $10^{-6}$ & 	$10^{-5}$	& $10^{-4}$\\
5 & $10^{-5}$ & $10^{-5}$ & $10^{-5}$	& &	$10^{-6}$	& $10^{-4}$ & $10^{-5}$	& & $10^{-5}$   & 	$10^{-5}$	& $10^{-4}$\\
\hline
\\
\end{tabular}\\
\hspace*{0.05in}
\begin{tabular}{lcclccclccc}
\hline
\hline
 & \multicolumn{2}{c}{$C_{2}H_6$} &  & \multicolumn{3}{c}{$H_{2}S$} & & \multicolumn{3}{c}{$C_{2}H_2$} \\
SNR \,\,\,\,\, & 3.3 $\mu m$ & 12.2 $\mu m$ &  \,\,\,\,\, & 2.6 $\mu m$ & 4.25 $\mu m$ & 8 $\mu m$ & \,\,\,\,\, & 3 $\mu m$ & 7.5 $\mu m$ & 13.7 $\mu m$ \\
\hline
20 & $10^{-6}$ & $10^{-6}$	& &	$10^{-5}$	& $10^{-4}$ & $10^{-4}$	& & $10^{-7}$ & $10^{-5}$ & $10^{-7}$\\
10 & $10^{-5}$ & $10^{-5}$	& &	$10^{-5}$	& $10^{-4}$ & $10^{-3}$	& & $10^{-7}$ & $10^{-4}$ & $10^{-6}$\\
5 & $10^{-5}$ & $10^{-5}$	& &	$10^{-4}$	& $10^{-3}$ & -	& & $10^{-7}$ & $10^{-3}$ & $10^{-5}$\\
\hline
\\
\end{tabular}
\caption{\footnotesize Warm Neptune: Minimum detectable abundance at fixed SNR=5, 10 and 20.}
\label{tab:wn_fixed}
\end{table}
\subsubsection{Alternative TP profiles}
\label{sec:alttp}
\begin{figure}[h!]
\hspace*{-0.6in}
\includegraphics[width=3.5in,trim=0 40 0 0]{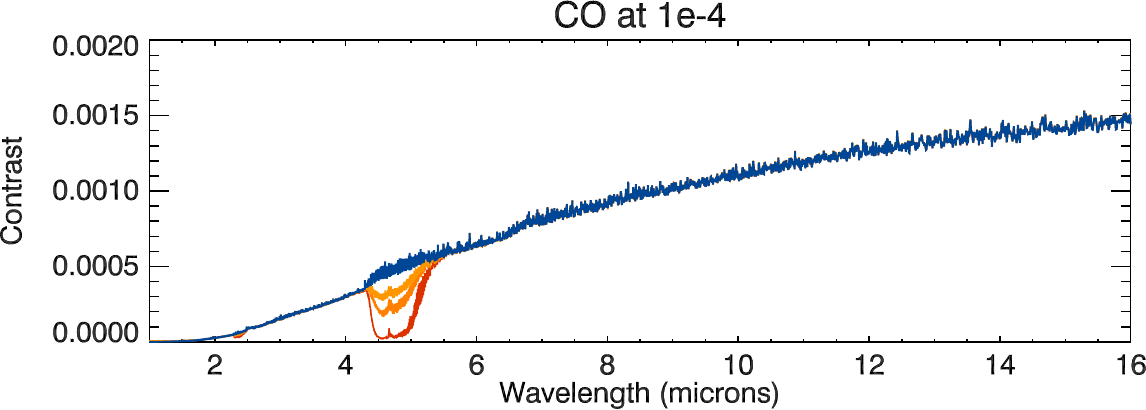}\includegraphics[width=3.5in,trim=0 40 0 0]{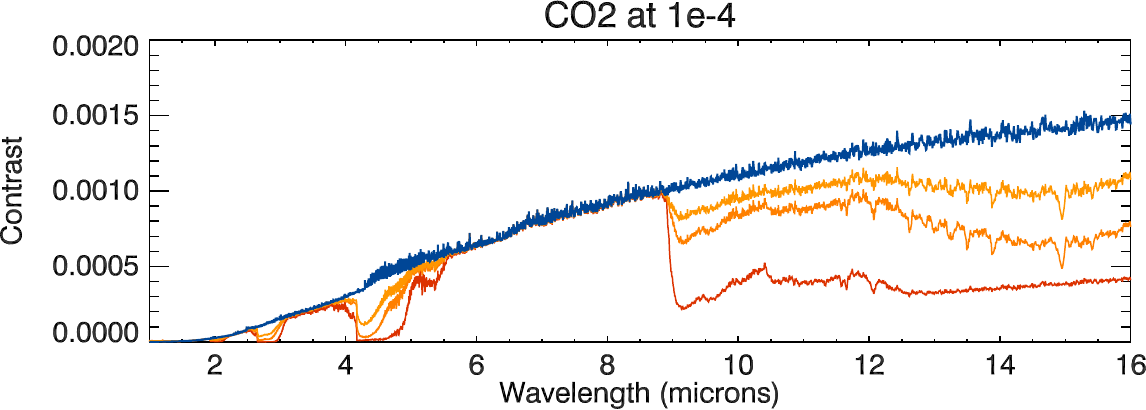}\\
\caption{\footnotesize Alternative TP profiles
(Warm Neptune): planet/star contrast spectra simulating the effect of carbon monoxide (\emph{left}) and carbon dioxide (\emph{right}). 
The blue line shows a planetary blackbody emission with no molecules present, divided by a stellar spectrum.
The three spectra show the strength of absorption with the furthest from the continuum corresponding to the dry adiabatic profile (in red), and the nearest to the more isothermal profile (yellow).}
\label{fig:alternative_tp}
\end{figure}

We repeat these calculations for two alternative TP profiles.
In Figure \ref{fig:alternative_tp} and Table \ref{tab:wn_alt_tp}, we show the outcome for $CO$ and $CO_2$, when a steep dry adiabatic profile and a more isothermal profile are used.
Not surprisingly, a steeper thermal gradient is equivalent to an increase in the molecular abundance. A more isothermal profile causes the opposite effect.
This shows that simultaneous temperature retrieval is very important for the analysis of secondary transit observations.
\begin{table}[!h]
\centering
\footnotesize
\begin{tabular}{lcclccc}
\hline
\hline
& \multicolumn{2}{c}{$CO$} & & \multicolumn{3}{c}{$CO_2$}\\
SNR \,\,\,\,\, & 2.3 $\mu m$ & 4.6 $\mu m$ & \,\,\,\,\, & 2.8 $\mu m$ & 4.3 $\mu m$ & 15 $\mu m$\\
\hline
20 & 	$10^{-(4/\textbf{4}/5)}$	& $10^{-(5/\textbf{6}/6)}$ 	& & $10^{-(7/\textbf{7}/7)}$ & $10^{-(7/\textbf{7}/7)}$ & $10^{-(6/\textbf{7}/7)}$\\
10 &	$10^{-(3/\textbf{3}/4)}$	& $10^{-(4/\textbf{5}/6)}$ 	& & $10^{-(6/\textbf{6}/7)}$ & $10^{-(7/\textbf{7}/7)}$ & $10^{-(5/\textbf{6}/7)}$\\
5 &	$10^{-(-/\textbf{3}/4)}$	& $10^{-(3/\textbf{4}/6)}$ 	& & $10^{-(5/\textbf{6}/7)}$ & $10^{-(6/\textbf{7}/7)}$ & $10^{-(3/\textbf{5}/7)}$\\
\hline
\\
\end{tabular}
\caption{\footnotesize Alternative TP profiles: Warm Neptune minimum detectable abundances at fixed SNR=5, 10 and 20, for $CO$ and $CO_2$, with three TP profiles, at the wavelengths of specific features. The minimum abundance for the three profiles are presented as $10^{-(x,{\bf{y}},z)}$, where $x$ is the result for the more isothermal profile, ${\bf y}$ the intermediate profile presented in Table \ref{tab:wn_fixed}, and $z$ the result for the dry adiabatic profile. }
\label{tab:wn_alt_tp}
\end{table}
\\

\newpage
\subsection{Hot Jupiter}
\begin{figure}[h!]
\hspace*{-0.6in}
\includegraphics[width=3.5in]{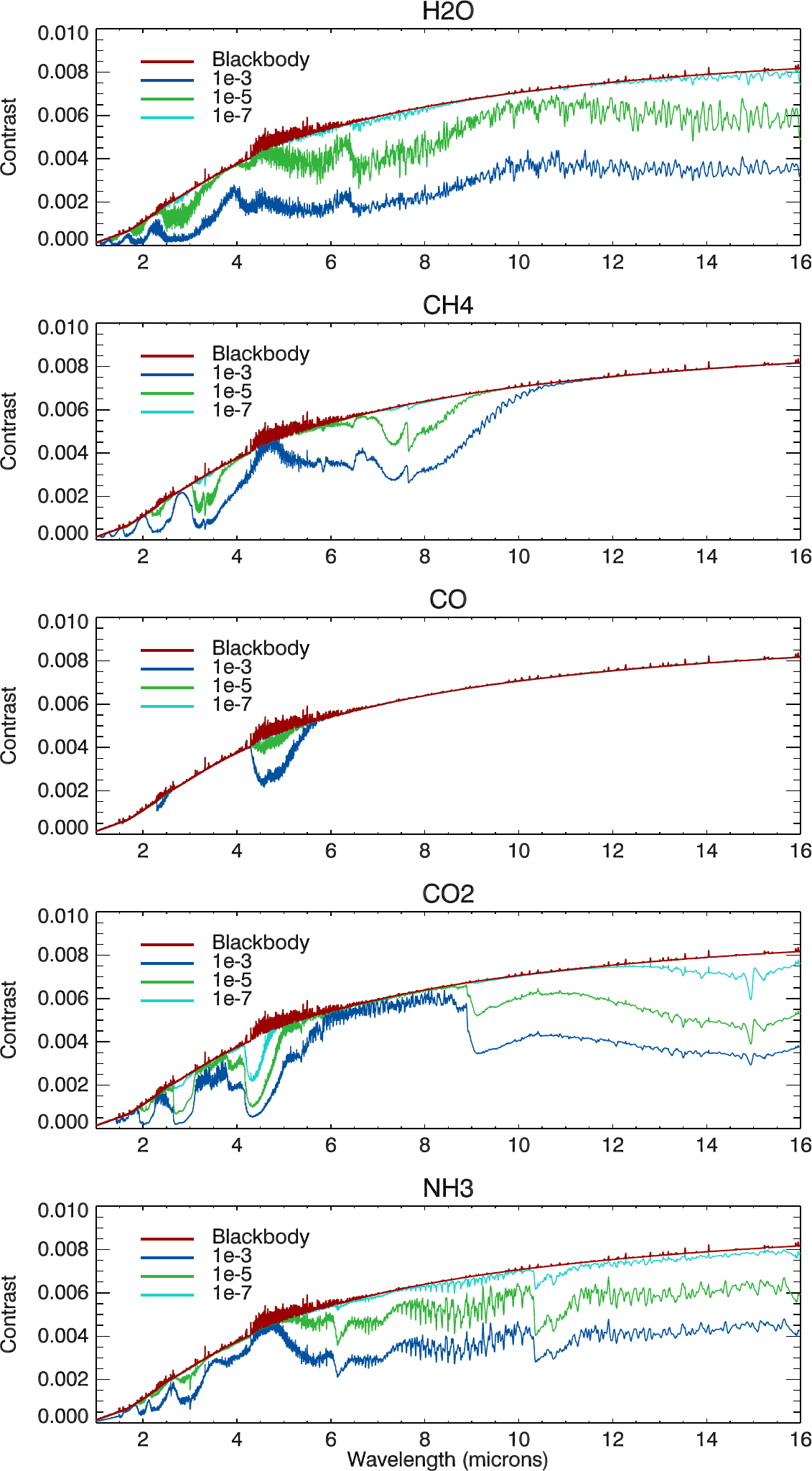}\includegraphics[width=3.5in]{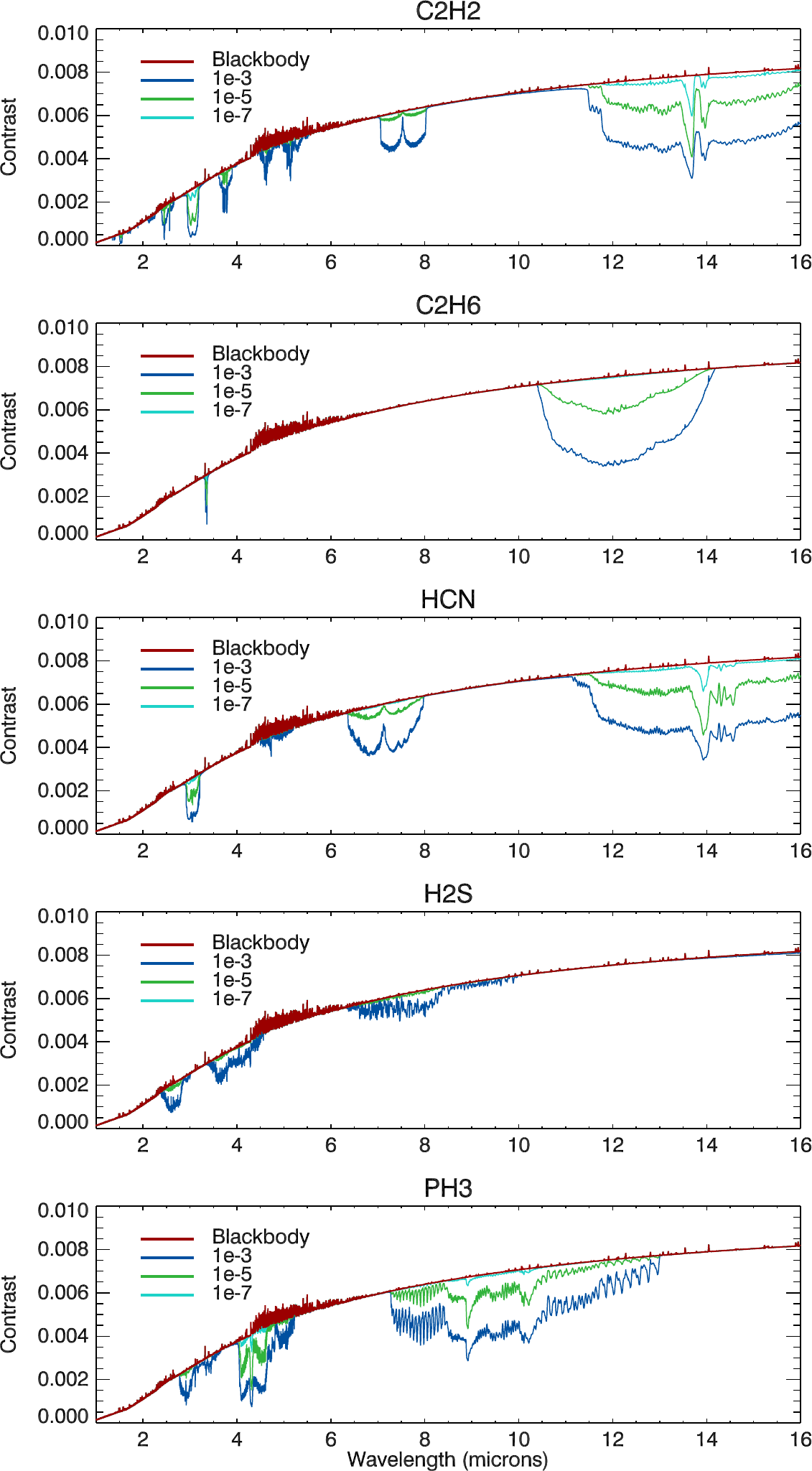}
\caption{\footnotesize Hot Jupiter: planet/star contrast spectra simulating the effect of the 10 considered molecules: $CH_4$, $CO$, $CO_2$, $NH_3$, $H_{2}O$, $C_{2}H_2$, $C_{2}H_6$, $HCN$, $H_{2}S$ and $PH_3$.
The red line shows a planetary blackbody emission with no molecules present, divided by a stellar spectrum.
The green-blue colored lines depict the molecule features at varying abundances. For clarity purposes, only three abundances are plotted of the five calculated. }
\label{fig:hj_molecules}
\end{figure}

We apply the procedure explained in section \ref{sec:molecules} to the hot Jupiter case.
Molecular spectra and minimum detectable abundances as a function of SNR are presented in Figure~\ref{fig:hj_molecules} and Table~\ref{tab:hj_fixed}.
\begin{table}[!h]
\hspace*{-0.3in}
\footnotesize
\begin{tabular}{lcclcclccclcc}
\hline
\hline
 & \multicolumn{2}{c}{$CH_4$} &  & \multicolumn{2}{c}{$CO$} & & \multicolumn{3}{c}{$CO_2$} & & \multicolumn{2}{c}{$PH_3$} \\
SNR \,\,\,\,\, & 3.3 $\mu m$ & 8 $\mu m$ &  \,\,\,\,\, & 2.3 $\mu m$ & 4.6 $\mu m$ & \,\,\,\,\, & 2.8 $\mu m$ & 4.3 $\mu m$ & 15 $\mu m$ & \,\,\,\,\, & 4.3 $\mu m$ & 10 $\mu m$ \\
\hline
20 & $10^{-6}$ & $10^{-5}$ 	& &	$10^{-3}$	& $10^{-5}$ 	& & $10^{-6}$ & $10^{-7}$ & $10^{-6}$ & & $10^{-7}$ & $10^{-5}$ \\
10 & $10^{-6}$ & $10^{-5}$ 	& &	- 		& $10^{-4}$ 	& & $10^{-6}$ & $10^{-7}$ & $10^{-6}$ & & $10^{-6}$ & $10^{-4}$ \\
5 & $10^{-6}$ & $10^{-4}$ 	& &	- 		& $10^{-3}$ 	& & $10^{-6}$ & $10^{-7}$ & $10^{-5}$ & & $10^{-6}$ & $10^{-3}$ \\
\hline
\\
\end{tabular}\\
\hspace*{-0.26in}
\begin{tabular}{lccclccclccc}
\hline
\hline
 & \multicolumn{3}{c}{$NH_3$} &  & \multicolumn{3}{c}{$HCN$} & & \multicolumn{3}{c}{$H_{2}O$} \\
SNR \,\,\,\,\, & 3 $\mu m$ & 6.1 $\mu m$ & 10.5 $\mu m$ &  \,\,\,\,\, & 3 $\mu m$ & 7 $\mu m$ & 14 $\mu m$ & \,\,\,\,\, & 2.8 $\mu m$ & 5 - 8 $\mu m$ & 11 - 16 $\mu m$ \\
\hline
20 & $10^{-5}$ & $10^{-5}$ & $10^{-6}$	& &	$10^{-6}$	& $10^{-4}$ 	& $10^{-5}$	& & $10^{-6}$ & 	$10^{-6}$	&  $10^{-5}$\\
10 & $10^{-5}$ & $10^{-5}$ &$10^{-5}$	& &	$10^{-5}$	& $10^{-3}$ 	& $10^{-4}$	& & $10^{-5}$ & 	$10^{-5}$	& $10^{-5}$\\
5 & $10^{-4}$ & $10^{-4}$ & $10^{-4}$	& &	$10^{-4}$	& -			& $10^{-3}$	& & $10^{-5}$   & 	$10^{-4}$	& $10^{-4}$\\
\hline
\\
\end{tabular}\\
\hspace*{0.05in}
\begin{tabular}{lcclccclccc}
\hline
\hline
 & \multicolumn{2}{c}{$C_{2}H_6$} &  & \multicolumn{3}{c}{$H_{2}S$} & & \multicolumn{3}{c}{$C_{2}H_2$}\\
SNR \,\,\,\,\, & 3.3 $\mu m$ & 12.2 $\mu m$ &  \,\,\,\,\, & 2.6 $\mu m$ & 4.25 $\mu m$ & 8 $\mu m$ & \,\,\,\,\, & 3 $\mu m$ & 7.5 $\mu m$ & 13.7 $\mu m$ \\
\hline
20 & $10^{-4}$ & $10^{-5}$	& &	$10^{-4}$	& $10^{-3}$ & -	& & $10^{-7}$ & $10^{-3}$ & $10^{-4}$\\
10 & $10^{-4}$ & $10^{-4}$	& &	$10^{-4}$	& $10^{-3}$ & -	& & $10^{-6}$ & $10^{-3}$ & $10^{-4}$\\
5 & $10^{-3}$ & $10^{-3}$	& &	$10^{-3}$	& 		- & -	 & & $10^{-6}$ & - 		& $10^{-4}$ \\
\hline
\\
\end{tabular}
\caption{\footnotesize Hot Jupiter: Minimum detectable abundances at fixed SNR=5, 10 and 20.}
\label{tab:hj_fixed}
\end{table}
\\

\newpage
\subsection{Hot and Temperate Super-Earth}
\label{sec:superearths}
\begin{figure}[h!]
\begin{center}
\includegraphics[width=3.8in]{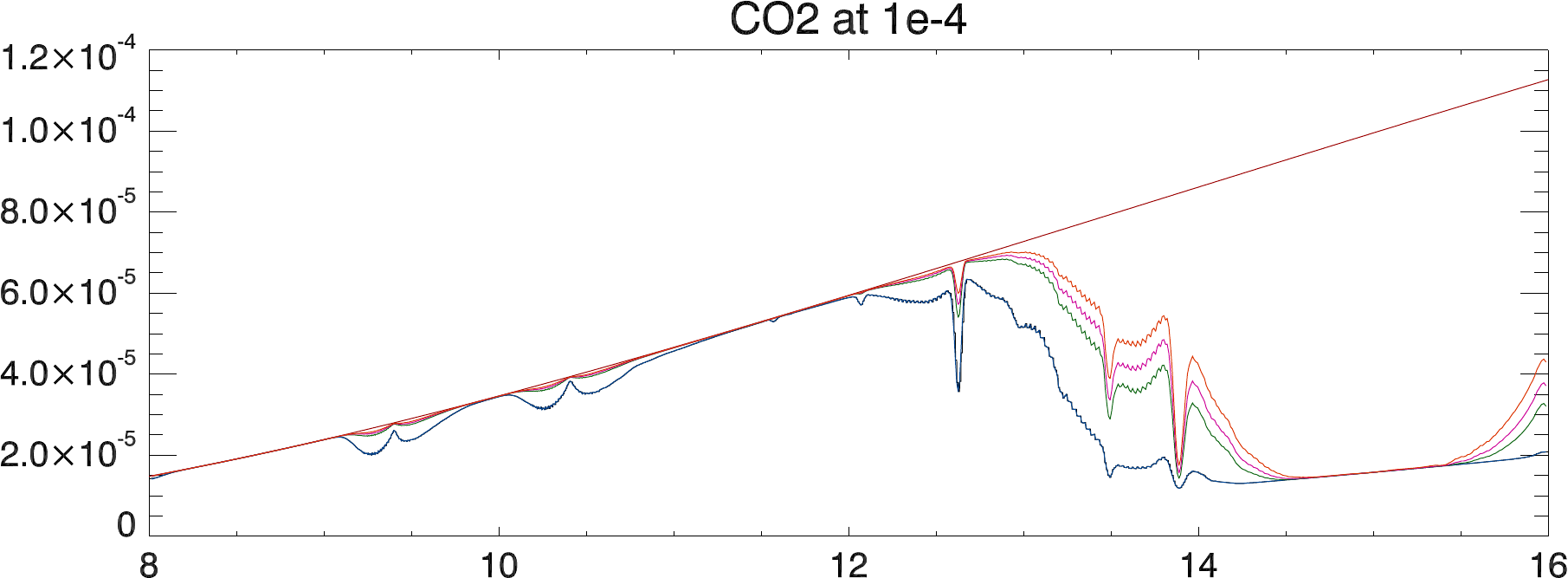}\\
\includegraphics[width=3.8in]{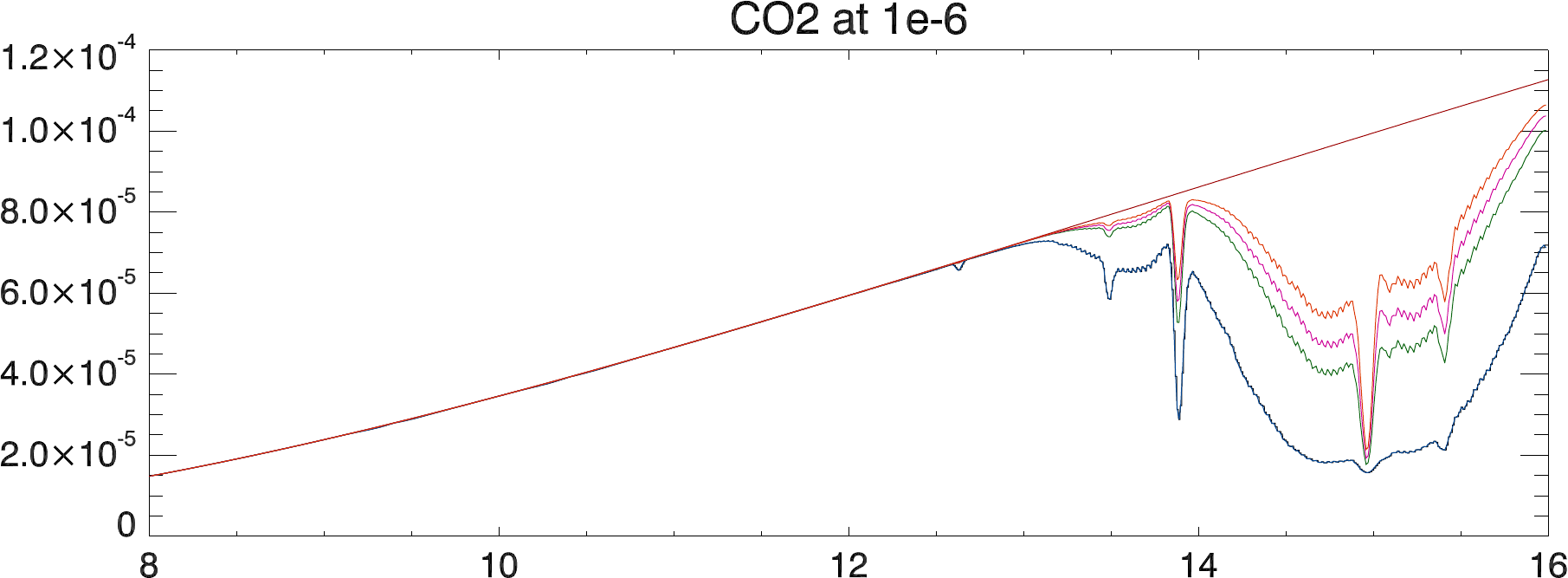}\\
\includegraphics[width=3.8in]{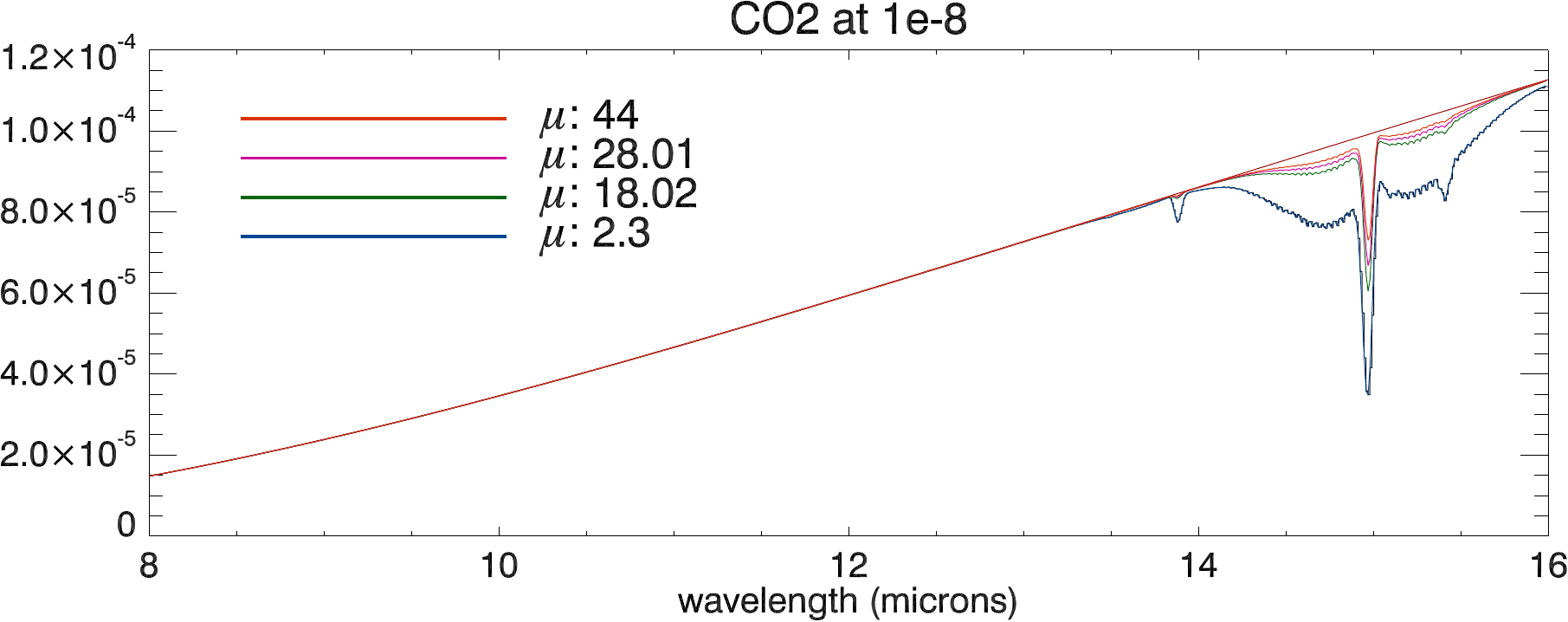}\\
\caption{\footnotesize Temperate super-Earth: planet/star contrast spectra showing the impact of the mean molecular weight of the atmosphere ($\mu$) on the detectability of $CO_2$ at abundances $10^{-4}$,$10^{-6}$,$10^{-8}$, from top to bottom. The four values for $\mu$ are: 2.3 (hydogen), 18.02 (water vapour), 28.01 (nitrogen) and 44 (carbon dioxide). The small differences between the latter three cases are hardly detectable, while a hydrogen dominated atmosphere will offer improved detectability performances. For our study we select a nitrogen dominated atmosphere.}
\label{fig:mutest}
\end{center}
\end{figure}

\begin{figure}[h!]
\hspace*{-0.7in}
\begin{minipage}[t]{3.5in}
\vspace{0pt}
\includegraphics[width=3.5in]{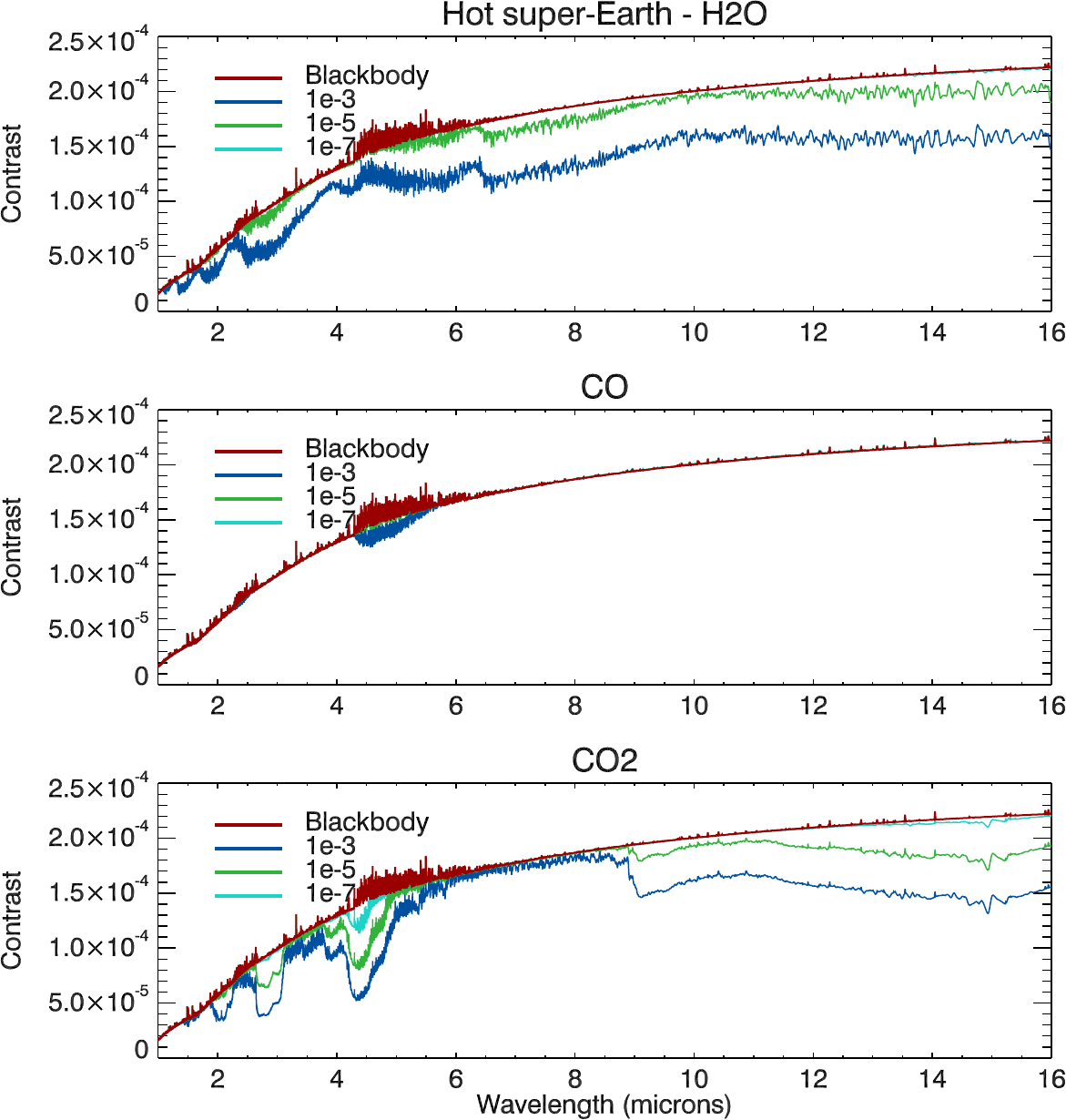}
\end{minipage}
\begin{minipage}[t]{3.5in}
\vspace{0pt}
\includegraphics[width=3.5in]{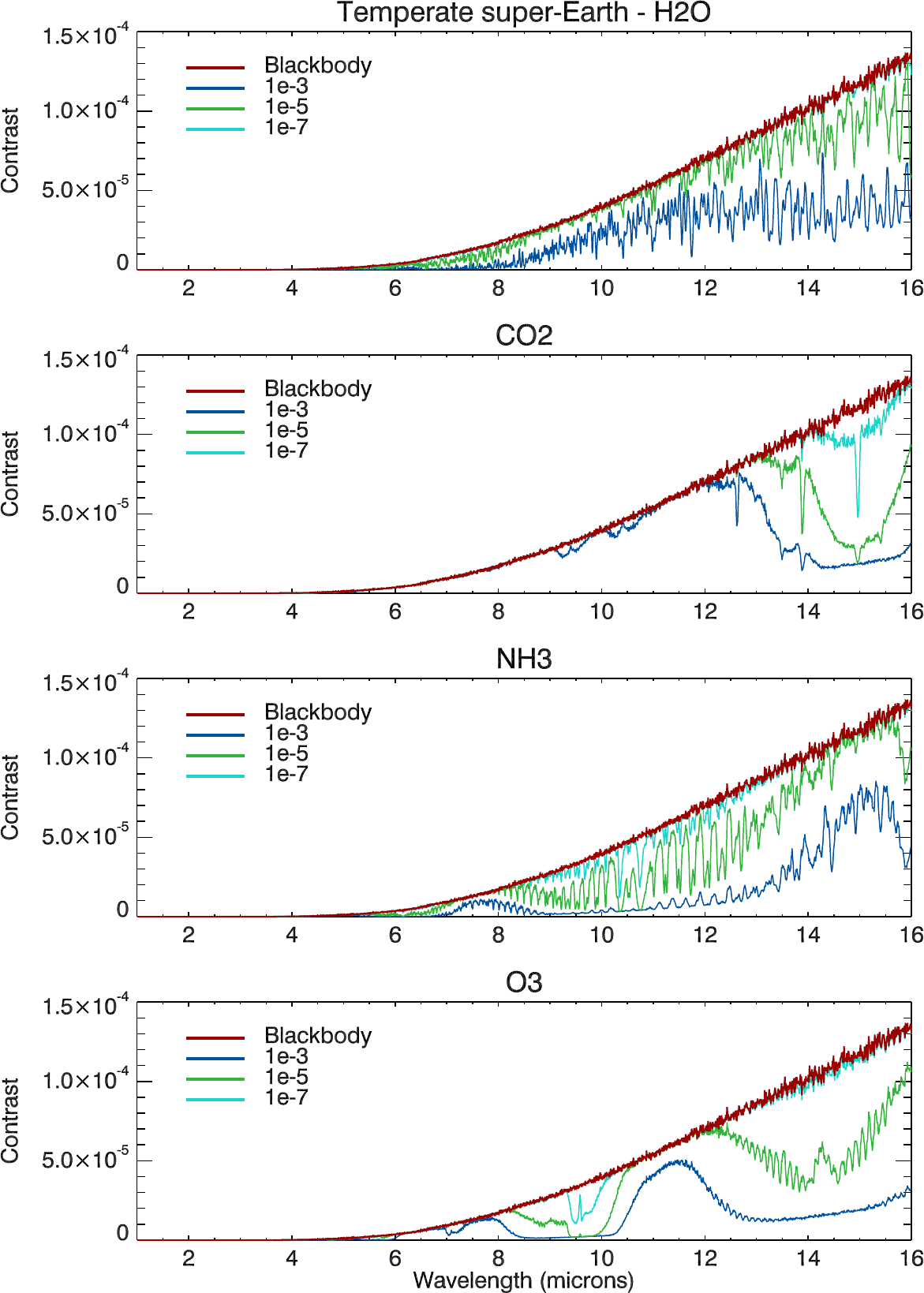}
\end{minipage}
\caption{\footnotesize Hot (\emph{left}) and temperate (\emph{right}) super-Earth: planet/star contrast spectra simulating the effect of the considered molecules: $H_{2}O$, $CO$ and $CO_2$ for the hot planet, and $H_{2}O$, $CO_2$, $NH_3$ and $O_3$ for the temperate case.
The red line shows a planetary blackbody emission with no molecules present, divided by a stellar spectrum.
The green-blue colored lines depict the molecule features at varying abundances. For clarity purposes, only three abundances are plotted out of the five calculated.}
\label{fig:se_molecules}
\end{figure}

We present two categories for the super-Earth cases: a hot super-Earth like Cancri 55 e, with a surface temperature of $\sim$2400K and orbiting a G type star, and a temperate super-Earth with a surface temperature of 320K, orbiting a late M type star. 
Given the different temperatures, we expect different components to be present in those atmospheres. In the hot case, we consider $H_{2}O$, $CO$ and $CO_2$, and in the temperate case, $H_{2}O$, $CO_2$, $NH_3$ and $O_3$.
In the case of the temperate super-Earth, we have estimated the impact for different main atmospheric components, we show in Figure \ref{fig:mutest} the detectability of $CO_2$ with three different abundances ($10^{-4}$,$10^{-6}$,$10^{-8}$).
At the SNR and resolutions considered in this paper, the small differences between the water vapour, nitrogen and carbon dioxide dominated atmospheres are hardly detectable, with the exception of the hydrogen-rich atmosphere.
For these reasons and in analogy with the Earth, we adopt a nitrogen dominated atmosphere with a wet adiabatic lapse rate for the temperate super-Earth. For the hot super-Earth, we consider a water vapour-dominated atmosphere, as can be expected in this mass/radius range \citep{fressin2013,valencia2013}.
Figure \ref{fig:se_molecules} shows the simulated spectra for the two planet categories, and Tables \ref{tab:se_fixed} and \ref{tab:se_fixed2} report the minimum abundances detectable.
We do not consider SNR=20 for the temperate super-Earth, given the challenge such a measure would present for current and short-term observatories.
Our results in the appendix show the SNR values that can be expected for such a planet at various distances.
\begin{table}[!h]
\hspace*{-0.08in}
\footnotesize
\begin{tabular}{lccclccclcc}
\hline
\hline
 & \multicolumn{3}{c}{$H_{2}O$} &  & \multicolumn{3}{c}{$CO_2$} & & \multicolumn{2}{c}{$CO$} \\
SNR \,\,\,\,\, & 2.8 $\mu m$ & 5 - 8 $\mu m$ & 11 - 16 $\mu m$ &  \,\,\,\,\, & 2.8 $\mu m$ & 4.3 $\mu m$ & 15 $\mu m$ & \,\,\,\,\, & 2.3 $\mu m$ & 4.6 $\mu m$ \\
\hline
20 & $10^{-4}$ & $10^{-4}$ & $10^{-4}$	& &	$10^{-5}$	& $10^{-7}$ & $10^{-5}$	& & - & -\\
10 & $10^{-4}$ & $10^{-3}$ &  $10^{-3}$	& &	$10^{-5}$	& $10^{-6}$ & $10^{-4}$	& & - & - \\
5    & $10^{-3}$ 	& - 		&   - 				& &	$10^{-4}$	& $10^{-5}$ & -	& & - & - \\
\hline
\\
\end{tabular}
\caption{\footnotesize Hot super-Earth, around a G type star: Minimum detectable abundances at fixed SNR=5, 10 and 20. In this specific example, CO is not detectable. The bulk composition of the planet atmosphere in this simulation is $H_{2}O$.}
\label{tab:se_fixed}
\end{table}
\begin{table}[!h]
\footnotesize
\begin{tabular}{lcclclcclcc}
\hline
\hline
 & \multicolumn{2}{c}{$H_{2}O$} &  &$CO_2$& & \multicolumn{2}{c}{$NH_3$} & & \multicolumn{2}{c}{$O_3$} \\
SNR \,\,\,\,\, & 5 - 8 $\mu m$ & 11 - 16 $\mu m$ &  \,\,\,\,\, &  15 $\mu m$ & \,\,\,\,\, & 6.1 $\mu m$ & 10.5 $\mu m$ & \,\,\,\,\, & 9.6 $\mu m$ & 14.3 $\mu m$ \\
\hline
10 &  $10^{-5}$ &  $10^{-4}$	& & $10^{-6}$	& & $10^{-6}$ & $10^{-6}$ & & $10^{-7}$ & $10^{-5}$ \\
5    & $10^{-5}$ & $10^{-4}$	& & $10^{-6}$	& & $10^{-5}$ & $10^{-6}$ & & $10^{-7}$ & $10^{-5}$ \\
\hline
\\
\end{tabular}
\caption{\footnotesize Temperate super-Earth, around a late M type star: Minimum detectable abundance at fixed SNR=5 and 10. The bulk composition of the planet atmosphere in this simulation is $N_{2}$.}
\label{tab:se_fixed2}
\end{table}
%
%
\newpage
\subsection{Temperate Jupiter}
\begin{figure}[h!]
\hspace*{-0.6in}
\begin{minipage}[t]{3.5in}
\vspace{0pt}
\includegraphics[width=3.5in]{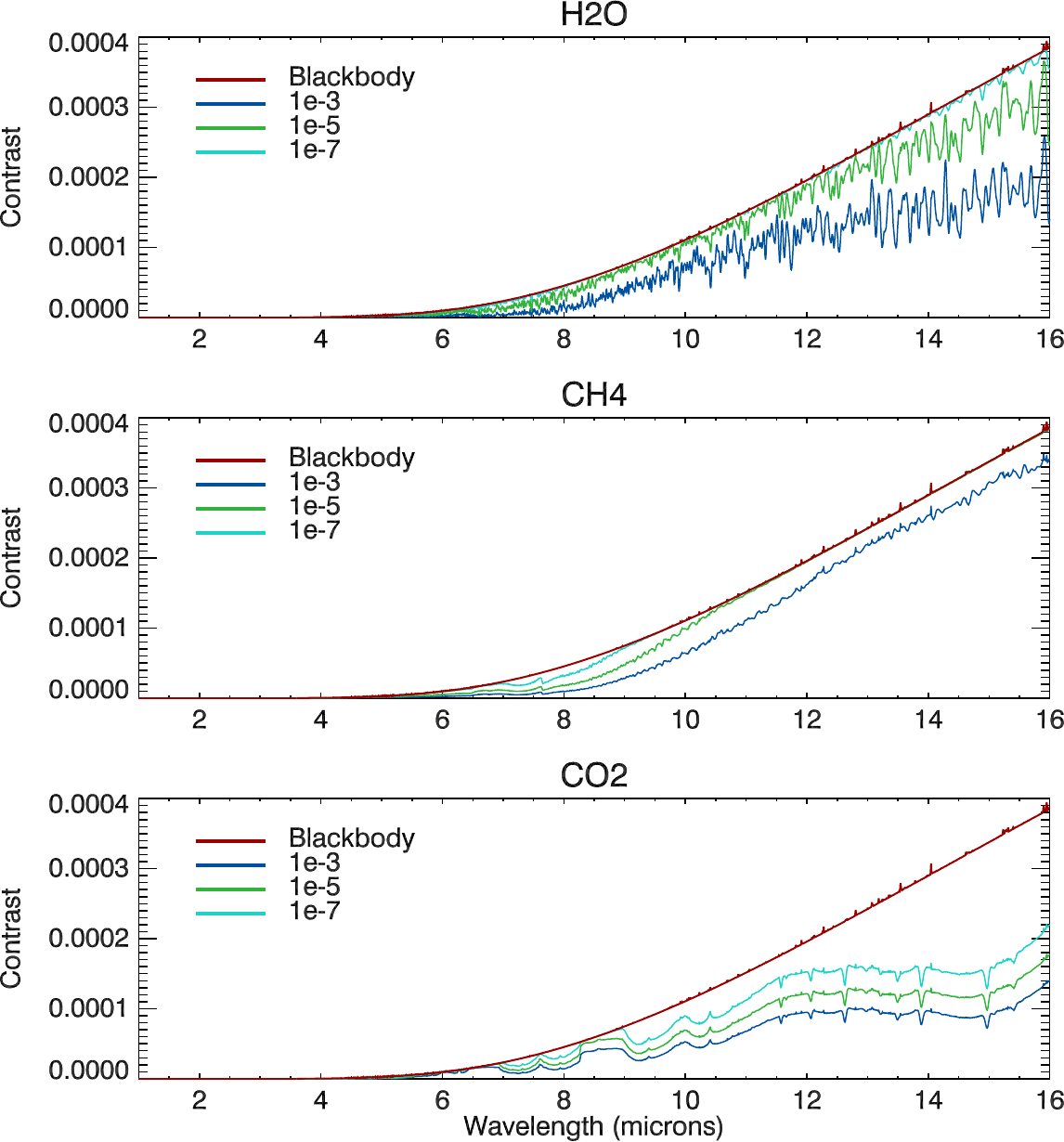}
\end{minipage}
\begin{minipage}[t]{3.5in}
\vspace{0pt}
\includegraphics[width=3.5in]{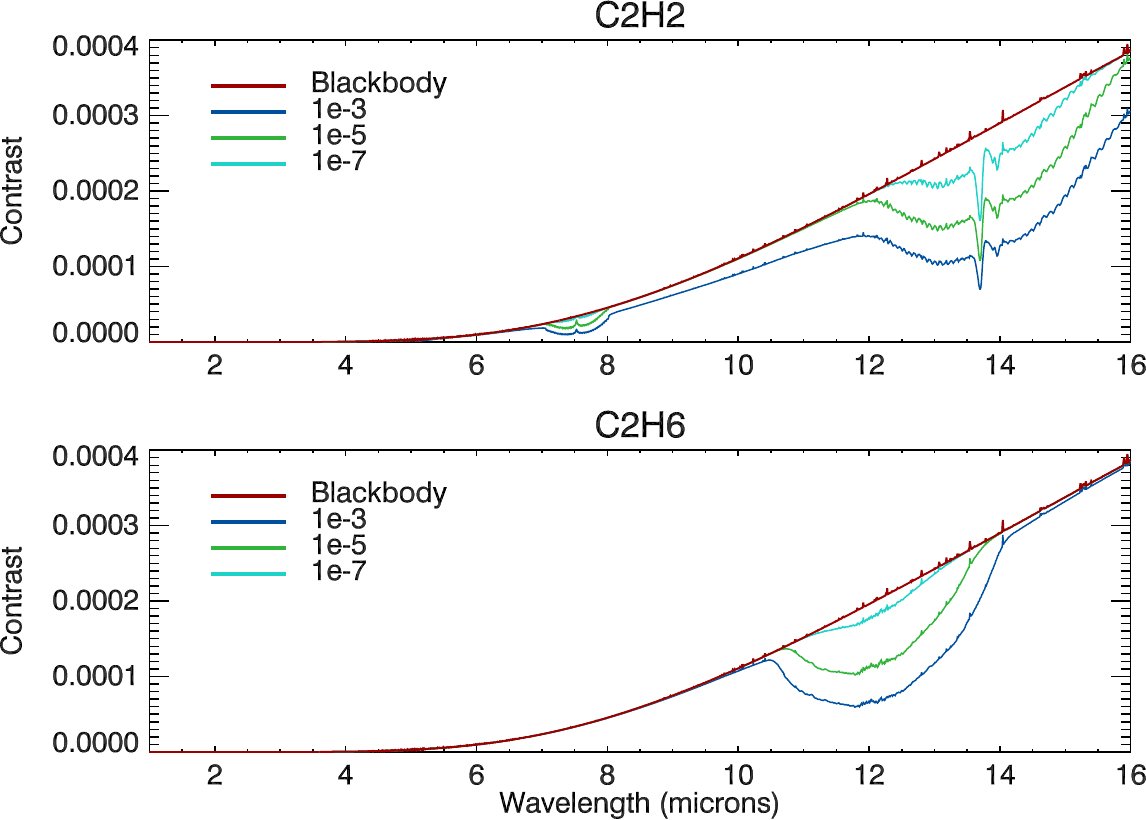}
\end{minipage}
\caption{\footnotesize Temperate Jupiter: planet/star contrast spectra simulating the effect of the 5 considered molecules: $H_{2}O$, $CH_4$, $CO_2$, $C_{2}H_2$ and $C_{2}H_6$.
The red line shows a planetary blackbody emission with no molecules present, divided by a stellar spectrum.
The green-blue colored lines depict the molecule features at varying abundances. For clarity purposes, only three abundances are plotted out of the five calculated.}
\label{fig:tjup_molecules}
\end{figure}

We consider here five molecules:  $H_{2}O$, $CH_4$, $CO_2$, $C_{2}H_{2}$ and $C_{2}H_{6}$. The spectral simulations are presented in Figure \ref{fig:tjup_molecules}, and Table \ref{tab:tjup_fixed} shows the minimum abundances detectable for this planet.
\begin{table}[!h]
\footnotesize
\begin{tabular}{lcclclclcclc}
\hline
\hline
 & \multicolumn{2}{c}{$H_{2}O$} &  & $CO_2$ & & $CH_4$ & &  \multicolumn{2}{c}{$C_{2}H_2$} &  & \multicolumn{1}{c}{$C_{2}H_6$} \\
SNR\,\,\,\,\,&  5 - 8 $\mu m$ & 11 - 16 $\mu m$ &  \,\,\,\,\, &  15 $\mu m$ &\,\,\,\,\,& 8 $\mu m$  & & 7.5 $\mu m$ & 13.7 $\mu m$ &  \,\,\,\,\, & 12.2 $\mu m$  \\
\hline
20 &  $10^{-6}$ &  $10^{-5}$	& &	 $10^{-7}$	& &$10^{-7}$ & &   $10^{-6}$ &  $10^{-7}$& &	 $10^{-6}$ \\
10 &  $10^{-6}$ &  $10^{-4}$	& &	 $10^{-7}$	& &$10^{-7}$ & &  $10^{-5}$ &  $10^{-6}$	& &	 $10^{-6}$ \\
5    &  $10^{-5}$ & $10^{-3}$	& &	 $10^{-7}$	& &$10^{-6}$ & &   $10^{-4}$ & $10^{-5}$	& &	 $10^{-5}$ \\
\hline
\\
\end{tabular}
\caption{\footnotesize Temperate Jupiter: Minimum detectable abundances at fixed SNR=5, 10 and 20.}
\label{tab:tjup_fixed}
\end{table}
\section{Results II - Comparison with Likelihood Ratio}
\label{sec:results2}	
\begin{figure}[!h]
\hspace*{-0.4in}
\includegraphics[width=3.2in]{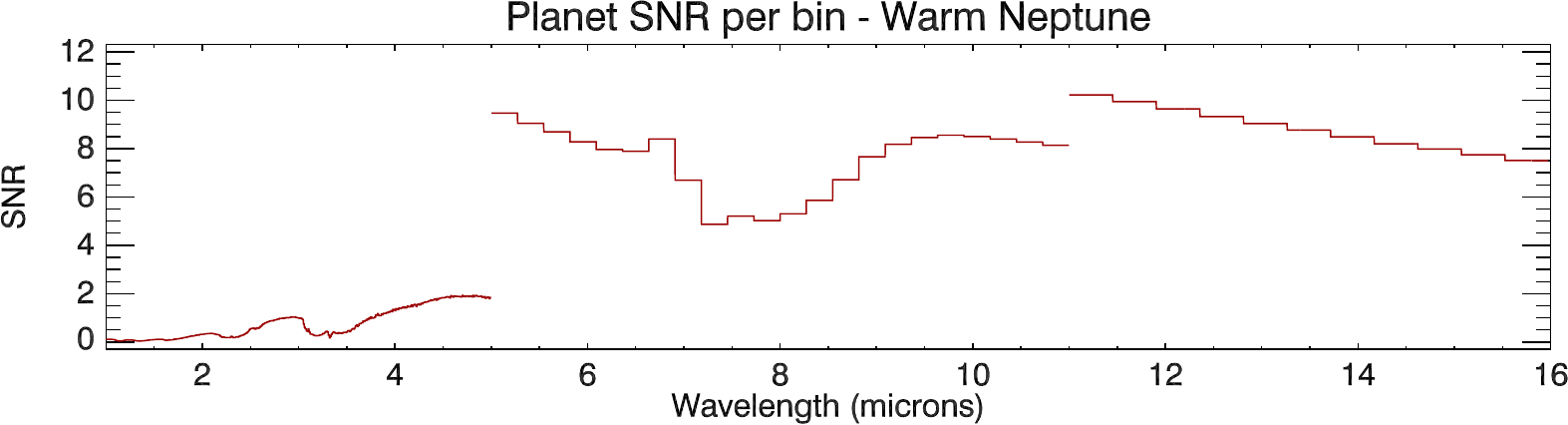}\includegraphics[width=3.2in]{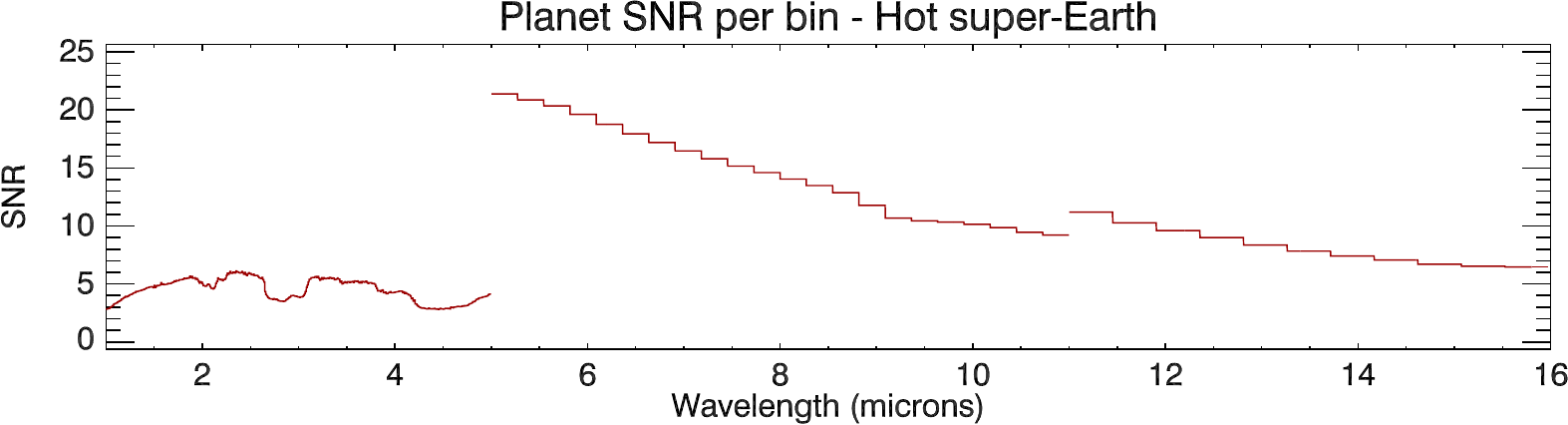}\\
\hspace*{-0.4in}
\includegraphics[width=3.2in]{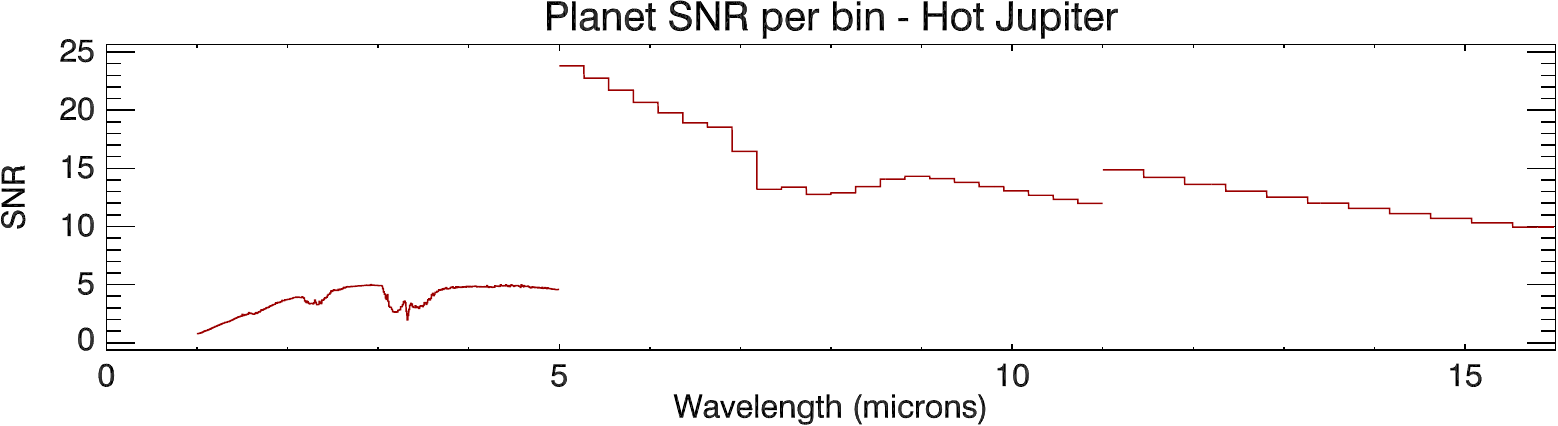}\includegraphics[width=3.2in]{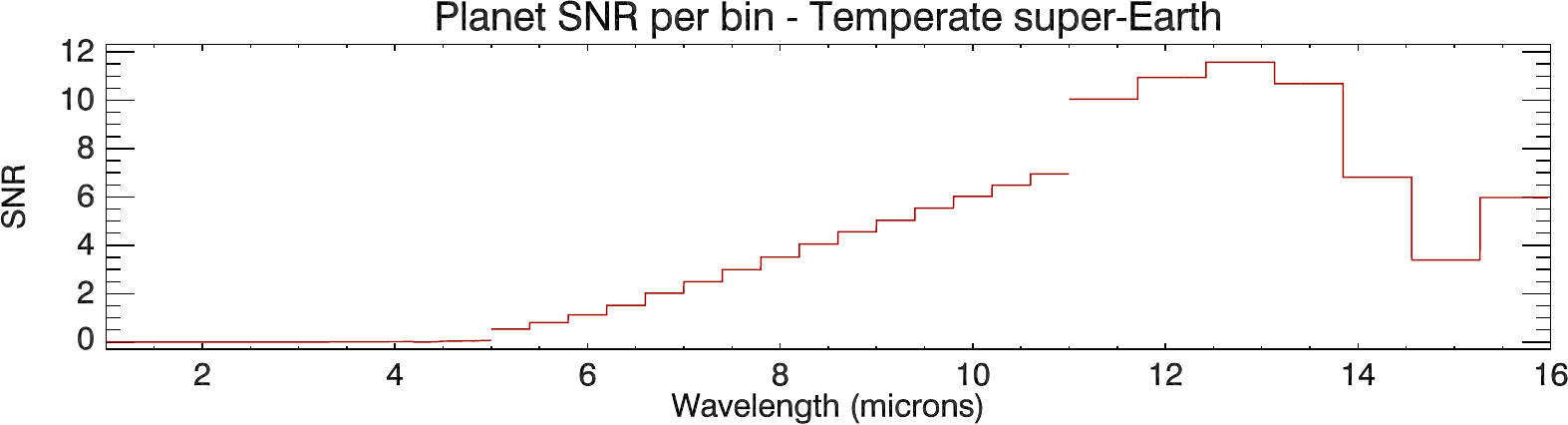}\\
\caption{\footnotesize SNR value per bin for the four planets considered. \emph{Top left:} a warm Neptune planet located at 13.5pc, observed for one transit. In this plot we show the SNR per bin for $CH_4$ in the atmosphere with an abundance of $10^{-5}$. The peak SNR value is of $\sim 10$ and the spectral feature near 7.5 microns has a SNR value of $\sim$ 5. \emph{Bottom left:} a hot Jupiter planet located at 150pc, observed for one transit, with $CH_4$ in the atmosphere with an abundance of $10^{-5}$. The peak SNR value is slightly over 20 and the spectral feature near 7.5 microns has a SNR value of $\sim$ 10. \emph{Top Right:} a hot super-Earth located at 12.34pc, observed for five transits, with $CO_2$ in the atmosphere with an abundance of $10^{-4}$. \emph{Bottom Right:} a temperate super-Earth located at 6pc and observed for 200 transits. This high number of transits and proximity are required to obtain a peak SNR of $\sim 10$, more distant planets can be observed with a lower peak SNR value. The atmosphere of this case is with $CO_2$ at an abundance of $10^{-5}$.}
\label{fig:snr13}
\end{figure}

We compare the results obtained with the likelihood ratio test and the individual bin method by applying the two methods to four examples: 
a warm Neptune, a hot Jupiter and a hot and temperate super-Earth.
These targets are placed at an optimal distance from the observer, where the SNR may reach $\sim 5$, 10 or 20 (see Appendix A) to facilitate the comparison with the results in section \ref{sec:results}.
The likelihood ratio test, in fact, can not be run with artificially fixed SNRs.\\
The SNR values per bin are shown in Figure \ref{fig:snr13}.
Table \ref{tab:comparison} shows the smallest abundances detectable for each method.
For the individual bin case, any feature providing a 3-sigma detection will be counted as a detection, while the smallest abundance which allows the rejection of the null hypothesis with 3-sigma confidence will be counted as a detection for the likelihood ratio test.
For most cases, the likelihood ratio test improves the sensitivity to the presence of molecular features and the statistical confidence of such detections.
\begin{table}[!h]
\hspace*{-0.6in}
\footnotesize
\begin{tabular}{lclclclclclclclclclc}
\multicolumn{20}{l}{Warm Neptune at 13.5pc, 1 transit}\\
\hline
\hline
Method \,\,\, & \multicolumn{1}{c}{$PH_3$} &  & \multicolumn{1}{c}{$CO$} & & \multicolumn{1}{c}{$CO_2$} & &
\multicolumn{1}{c}{$CH_4$} &  & \multicolumn{1}{c}{$NH_{3}$} & & \multicolumn{1}{c}{$HCN$} & & \multicolumn{1}{c}{$C_{2}H_2$} &  & \multicolumn{1}{c}{$C_{2}H_6$}&  & \multicolumn{1}{c}{$H_{2}S$}&  & \multicolumn{1}{c}{$H_{2}O$}\\
\hline
Individual bins		& $10^{-5}$ 	& &	$10^{-3}$	 	& & $10^{-5}$ & &
				    $10^{-5}$ 	& &	$10^{-5}$	 	& & $10^{-5}$ & &
				    $10^{-5}$ 	& &	$10^{-5}$	 	& & $10^{-3}$ & & $10^{-5}$ \\
LRT 	& $10^{-6}$  	& &	$10^{-4}$		& & $10^{-7}$& &
				    $10^{-6}$ 	& &	$10^{-7}$	 	& & $10^{-6}$ & &
				    $10^{-6}$ 	& &	$10^{-5}$	 	& & $10^{-3}$ & & $10^{-6}$ \\
\hline
\\
\multicolumn{20}{l}{Hot Jupiter at 150pc, 1 transit}\\
\hline
\hline
Method \,\,\, & \multicolumn{1}{c}{$PH_3$} &  & \multicolumn{1}{c}{$CO$} & & \multicolumn{1}{c}{$CO_2$} & &
\multicolumn{1}{c}{$CH_4$} &  & \multicolumn{1}{c}{$NH_{3}$} & & \multicolumn{1}{c}{$HCN$} & & \multicolumn{1}{c}{$C_{2}H_2$} &  & \multicolumn{1}{c}{$C_{2}H_6$}&  & \multicolumn{1}{c}{$H_{2}S$}&  & \multicolumn{1}{c}{$H_{2}O$}\\
\hline
Individual bins		& $10^{-5}$ 	& &	$10^{-4}$	 	& & $10^{-6}$ & &
				    $10^{-5}$ 	& &	$10^{-5}$	 	& & $10^{-5}$ & &
				    $10^{-5}$ 	& &	$10^{-4}$	 	& & $-$ & & $10^{-5}$ \\
LRT 				& $10^{-6}$  	& &	$10^{-5}$		& & $10^{-7}$& &
				    $10^{-6}$ 	& &	$10^{-6}$	 	& & $10^{-6}$ & &
				    $10^{-6}$ 	& &	$10^{-5}$	 	& & $10^{-4}$ & & $10^{-6}$ \\
\hline
\end{tabular}
\hspace*{-0.1in}
\begin{tabular}{lclclc}
\\
\multicolumn{6}{l}{Hot super-Earth at 12.34pc, 5 transits}\\
\hline
\hline
Method \,\,\, & \multicolumn{1}{c}{$H_{2}O$} &  & \multicolumn{1}{c}{$CO_2$} & & \multicolumn{1}{c}{$CO$} \\
\hline
Individual bins		&  $10^{-4}$ & &	$10^{-3}$	 	& & ---\\
LRT 				& $10^{-5}$  	& &	$10^{-7}$		& & $10^{-3}$ \\
\hline
\\
\end{tabular} \, \, \,
\begin{tabular}{lclclclc}
\\
\multicolumn{8}{l}{Temperate super-Earth at 6pc, 200 transits}\\
\hline
\hline
Method \,\,\, & \multicolumn{1}{c}{$H_{2}O$} &  & \multicolumn{1}{c}{$CO_2$} & & \multicolumn{1}{c}{$NH_3$} & & $O_3$*\\
\hline
Individual bins		& $10^{-4}$ 	& &	$10^{-6}$	 	& & $10^{-5}$ & & $10^{-7}$\\
LRT 	& $10^{-5}$  	& &	$10^{-6}$		& & $10^{-6}$ & & $10^{-6}$\\
\hline
\\
\end{tabular}
\caption{Comparison of minimum abundance detectable by the individual bin method and the Likelihood Ratio Test (LRT) method, for three planet cases, a warm Neptune, a hot Jupiter and a temperate super-Earth. For the three planet cases the likelihood ratio method typically improves the detectability of the limiting abundances. *: Note that in this example, for the case of ozone on a temperate super-Earth, the LRT performs less well than the individual bin method; the signal consists of a single small feature appearing in one bin only.}
\label{tab:comparison}
\end{table}

\section{Results III - Detectability Limits in a Wet Atmosphere}	
\label{sec:results3}
\begin{figure}[h!]
\hspace*{-0.45in}
\includegraphics[trim=0 124 0 0,clip,width=3.3in]{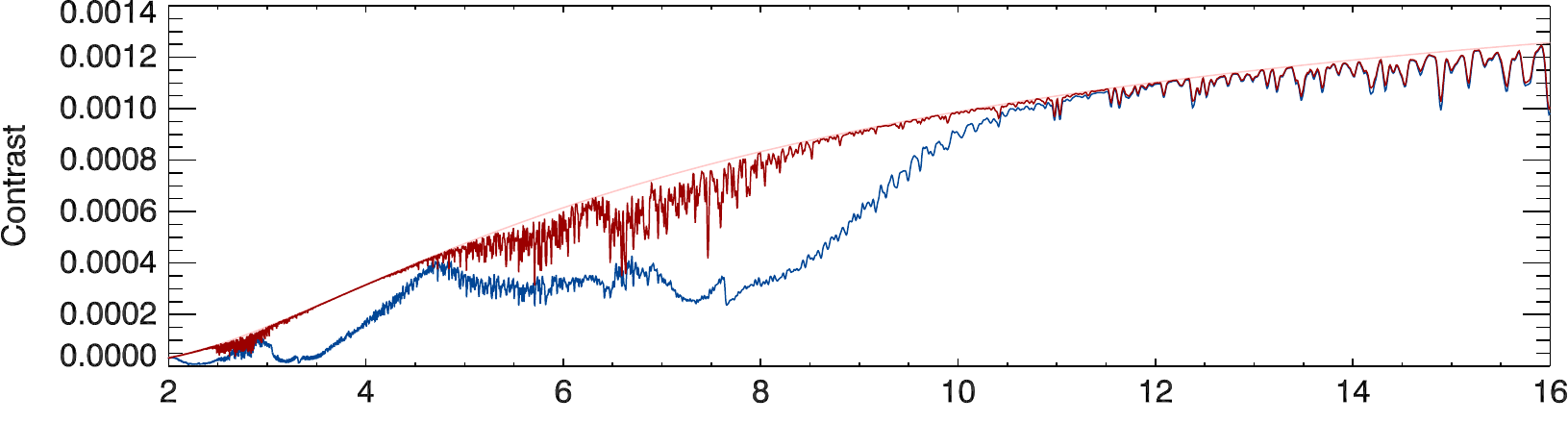}
\includegraphics[trim=0 124 0 0,clip,width=3.3in]{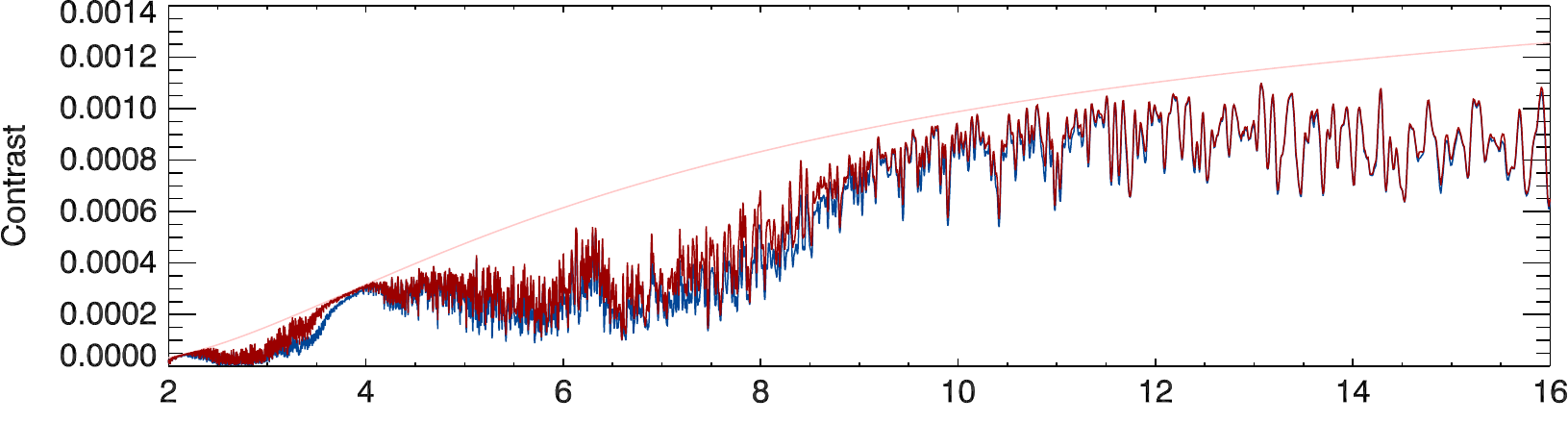}
\caption{\footnotesize Warm Neptune: Planet/star contrast spectra simulating the effect of methane with the addition of water (Left: Water at mixing ratio $10^{-6}$ and $CH_4$ at $10^{-4}$; Right: Water at mixing ratio $10^{-4}$ and $CH_4$ at $10^{-6}$). }
\label{fig:wn_water_ch4_2}
\end{figure}

\begin{figure}[h!]
\hspace*{-0.45in}
\includegraphics[trim=0 124 0 0,clip,width=3.3in]{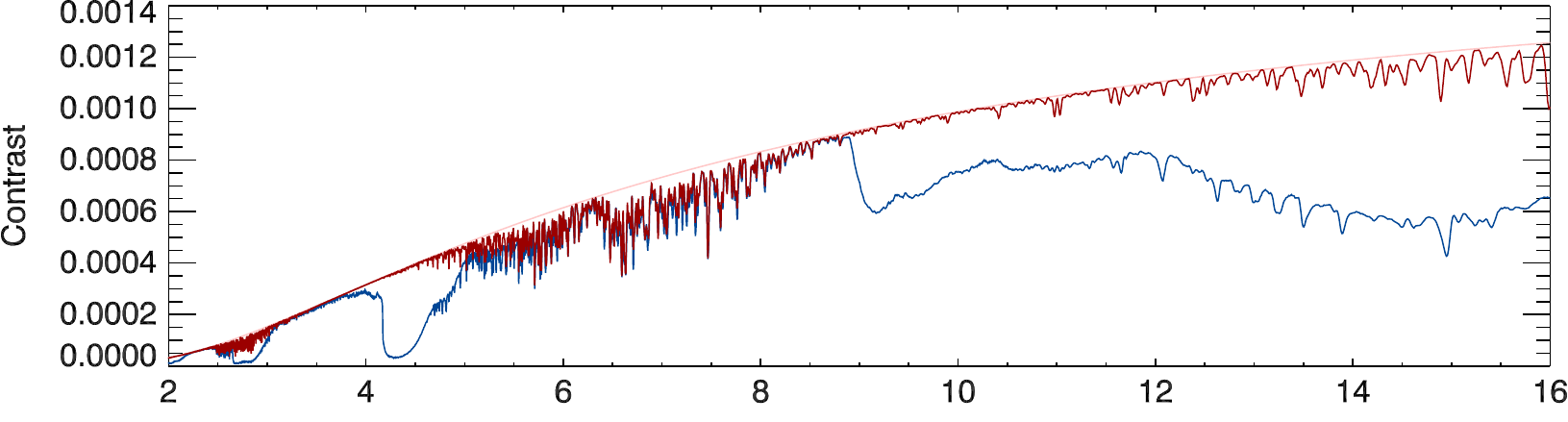}
\includegraphics[trim=0 124 0 0,clip,width=3.3in]{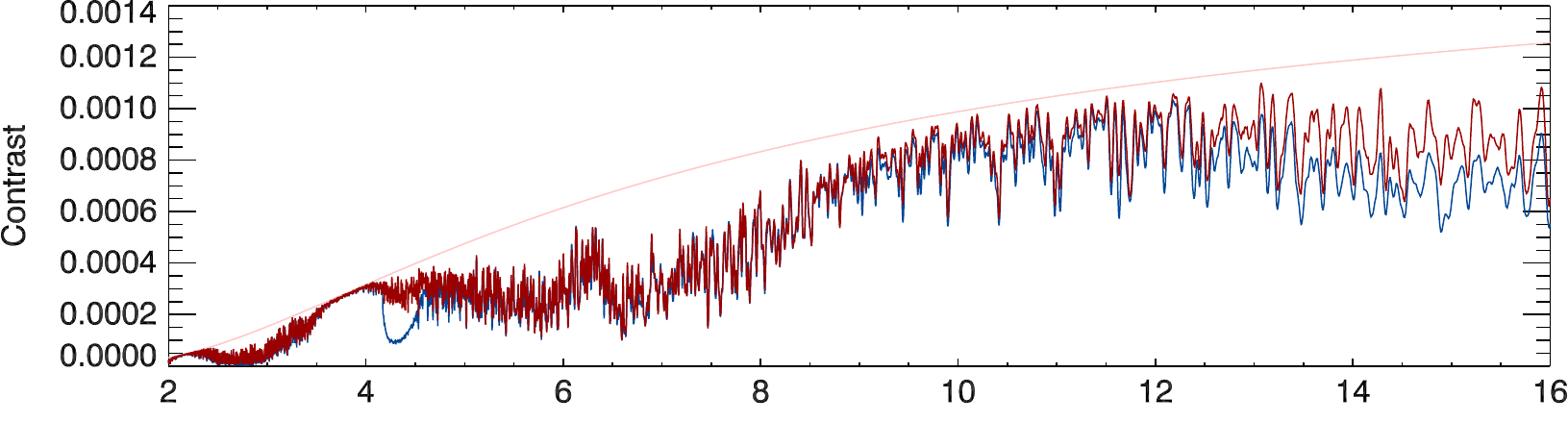}
\caption{\footnotesize Warm Neptune: Planet/star contrast spectra simulating the effect of carbon dioxide with the addition of water (Left: Water at mixing ratio $10^{-6}$ and $CO_2$ at $10^{-4}$; Right: Water at mixing ratio $10^{-4}$ and $CO_2$ at $10^{-6}$). }
\label{fig:wn_water_co2_2}
\end{figure}

As described in section \ref{sec:methods}, we show here the impact of a water vapour signal on the detectability of key molecules ($CO$, $CO_2$, $CH_4$ and $NH_3$).
We consider a warm Neptune planet case with water vapour abundances ranging from $10^{-3}$ to $10^{-7}$. The deviations of the combined ($H_{2}O$ + molecule) spectra from the water vapour only spectrum are tested for detectability (see Figures \ref{fig:wn_water_ch4_2} and \ref{fig:wn_water_co2_2}).
The minimum detectable abundances are presented in Table \ref{tab:wn_watercombo} as a function of SNR, wavelength and water vapour abundance.
For all the molecules considered, water vapour abundances of $10^{-5}$ or less do not significantly interfere with the molecular detectablity.
Larger water vapour abundances start to mask the absorption features of other molecules, with a clear impact on detectability limits.
These effects can sometimes be mitigated with an increased SNR.
\begin{table}[!h]
\hspace*{-0.5in}
\footnotesize
\begin{tabular}{llcclcclccclccc}
\multicolumn{15}{c}{SNR=5}\\
\hline
\hline
 & & \multicolumn{2}{c}{$CH_4$} & &\multicolumn{2}{c}{$CO$}&   & \multicolumn{3}{c}{$CO_2$}&   & \multicolumn{3}{c}{$NH_3$}\\
$H_{2}O$ &\,\, & 3.3 $\mu m$ & 8 $\mu m$ &\,\, & 2.3 $\mu m$ & 4.6 $\mu m$ & \,\, & 2.8 $\mu m$ & 4.3 $\mu m$ & 15 $\mu m$& \,\, & 3 $\mu m$ & 6.1 $\mu m$ & 10.5 $\mu m$\\ 
\hline
\emph{0} && $10^{-7}$ & $10^{-5}$ & &$10^{-3}$ & $10^{-4}$ & &  $10^{-6}$ & $10^{-7}$ & $10^{-5}$ & &  $10^{-5}$ & $10^{-5}$ & $10^{-5}$\\
\hline
$10^{-7}$ && $10^{-7}$ & $10^{-5}$ & &$10^{-3}$ & $10^{-4}$ & &  $10^{-6}$ & $10^{-7}$ & $10^{-5}$ & &  $10^{-5}$ & $10^{-5}$ & $10^{-5}$\\ 
$10^{-6}$ && $10^{-7}$ & $10^{-5}$ & &$10^{-3}$ & $10^{-4}$ & &  $10^{-6}$ & $10^{-7}$ & $10^{-5}$ & &  $10^{-5}$ & $10^{-5}$ & $10^{-5}$\\
$10^{-5}$ && $10^{-7}$ & $10^{-5}$  & &$10^{-3}$ & $10^{-4}$ & &  $10^{-6}$ & $10^{-7}$ & $10^{-5}$ & &  $10^{-5}$ & $10^{-5}$ & $10^{-5}$\\
$10^{-4}$ && $10^{-7}$ & $10^{-4}$  & &$10^{-3}$ & $10^{-4}$ & &  $10^{-6}$ & $10^{-7}$ & $10^{-4}$ & &  $10^{-5}$ & $10^{-4}$ & $10^{-5}$\\
$10^{-3}$ && $10^{-7}$ & $10^{-3}$  & &$10^{-3}$ & $10^{-4}$ & &  $10^{-6}$ & $10^{-7}$ & $-$ 		& &  $10^{-4}$ & $10^{-3}$ & $10^{-5}$\\
\hline
\\
\end{tabular}\\
\hspace*{-0.5in}
\begin{tabular}{llcclcclccclccc}
\multicolumn{15}{c}{SNR=10}\\
\hline
\hline
 & &\multicolumn{2}{c}{$CH_4$} & &\multicolumn{2}{c}{$CO$} &   & \multicolumn{3}{c}{$CO_2$}&   & \multicolumn{3}{c}{$NH_3$}\\
$H_{2}O$ &\,\, & 3.3 $\mu m$ & 8 $\mu m$ &\,\, & 2.3 $\mu m$ & 4.6 $\mu m$ & \,\, & 2.8 $\mu m$ & 4.3 $\mu m$ & 15 $\mu m$& \,\, & 3 $\mu m$ & 6.1 $\mu m$ & 10.5 $\mu m$\\ 
\hline
\emph{0} && $10^{-7}$ & $10^{-6}$ & &$10^{-3}$ & $10^{-5}$ & &  $10^{-6}$ & $10^{-7}$ & $10^{-6}$ & &  $10^{-6}$ & $10^{-6}$ & $10^{-6}$\\
\hline
$10^{-7}$ && $10^{-7}$ & $10^{-6}$ & & $10^{-3}$ & $10^{-5}$ & & $10^{-6}$ & $10^{-7}$ & $10^{-6}$ & &  $10^{-6}$ & $10^{-6}$ & $10^{-6}$\\
$10^{-6}$ && $10^{-7}$ & $10^{-6}$ & & $10^{-3}$ & $10^{-5}$ & & $10^{-6}$ & $10^{-7}$ & $10^{-6}$ & &  $10^{-6}$ & $10^{-6}$ & $10^{-6}$\\
$10^{-5}$ && $10^{-7}$ & $10^{-6}$  & & $10^{-3}$ & $10^{-5}$ & & $10^{-6}$ & $10^{-7}$ & $10^{-6}$ & &  $10^{-6}$ & $10^{-5}$ & $10^{-6}$\\
$10^{-4}$ && $10^{-7}$ & $10^{-6}$  & & $10^{-3}$ & $10^{-5}$ & & $10^{-6}$ & $10^{-7}$ & $10^{-5}$ & &  $10^{-6}$ & $10^{-5}$ & $10^{-6}$\\
$10^{-3}$ && $10^{-7}$ & $10^{-5}$  & & $10^{-3}$ & $10^{-5}$ & & $10^{-6}$ & $10^{-7}$ & $10^{-4}$ & &  $10^{-4}$ & $10^{-4}$ & $10^{-6}$\\
\hline
\\
\end{tabular}\\
\hspace*{-0.5in}
\begin{tabular}{llcclcclccclccc}
\multicolumn{15}{c}{SNR=20}\\
\hline
\hline
 & &\multicolumn{2}{c}{$CH_4$}& &\multicolumn{2}{c}{$CO$} &   & \multicolumn{3}{c}{$CO_2$}&   & \multicolumn{3}{c}{$NH_3$}\\
$H_{2}O$ &\,\, & 3.3 $\mu m$ & 8 $\mu m$ &\,\, & 2.3 $\mu m$ & 4.6 $\mu m$ & \,\, & 2.8 $\mu m$ & 4.3 $\mu m$ & 15 $\mu m$& \,\, & 3 $\mu m$ & 6.1 $\mu m$ & 10.5 $\mu m$\\ 
\hline
\emph{0} && $10^{-7}$ & $10^{-6}$ & & $10^{-4}$ & $10^{-6}$ & & $10^{-7}$ & $10^{-7}$ & $10^{-7}$ & &  $10^{-7}$ & $10^{-6}$ & $10^{-7}$\\
\hline
$10^{-7}$ && $10^{-7}$ & $10^{-6}$ & &$10^{-4}$ & $10^{-6}$ & &  $10^{-7}$ & $10^{-7}$ & $10^{-7}$ & &  $10^{-6}$ & $10^{-6}$ & $10^{-7}$\\
$10^{-6}$ && $10^{-7}$ & $10^{-6}$ & & $10^{-4}$ & $10^{-6}$ & & $10^{-7}$ & $10^{-7}$ & $10^{-7}$ & &  $10^{-6}$ & $10^{-6}$ & $10^{-7}$\\
$10^{-5}$ && $10^{-7}$ & $10^{-7}$  & & $10^{-4}$ & $10^{-6}$ & & $10^{-7}$ & $10^{-7}$ & $10^{-7}$ & &  $10^{-6}$ & $10^{-6}$ & $10^{-7}$\\
$10^{-4}$ && $10^{-7}$ & $10^{-7}$  & & $10^{-4}$ & $10^{-6}$ & & $10^{-7}$ & $10^{-7}$ & $10^{-7}$ & &  $10^{-6}$ & $10^{-6}$ & $10^{-7}$\\
$10^{-3}$ && $10^{-7}$ & $10^{-7}$  & & $10^{-4}$ & $10^{-6}$ & & $10^{-7}$ & $10^{-7}$ & $10^{-7}$ & &  $10^{-5}$ & $10^{-5}$ & $10^{-7}$\\
\hline
\end{tabular}
\caption{\footnotesize Warm Neptune: Minimum detectable abundances at fixed SNR=5, 10 and 20 (top, middle and bottom) with a range of quantities of water in the atmosphere. For each SNR case, the minimum detectable abundance for each molecule without the presence of water is given as comparison (values from Table 4).}
\label{tab:wn_watercombo}
\end{table}

\newpage
\section{Discussion}
In this paper we have studied the detectability of key molecules absorbing in the atmospheres of representative exoplanet cases.
Although we consider only five types of planets, most exoplanets known today have sizes and temperatures that are within the boundaries of these, so
results for intermediate cases can be extrapolated from our tables. 
Notice that the results obtained for the super-Earths are the most sensitive the type of the stellar companion \citep{tessenyi2012}.
For this reason, we have selected one hot target around a G type star, and a temperate one around a late M star.
We have adopted thermal profiles from simulations or have extrapolated them from solar system planets.
As we focus on emission spectra, the molecular absorption and thermal structure are strongly correlated.
To asses this effect, we have repeated our calculations with extreme thermal profiles in the case of the warm Neptune, and have found that our results are reliable within an order of magnitude. 

We compared two approaches to assess molecular detectability: the individual bin method (section \ref{sec:results}) and the likelihood ratio test (section \ref{sec:results2}).
We have applied the individual bin method to all the planet cases and key molecules.
We fixed the planet SNR artificially to obtain results which are independent of instrument design, observation duration and sources of noise.
The individual bin method is robust but very conservative and not optimised for most detections. In particular:\\
1) the method doesn't take advantage of spectral features that span across multiple bins.
Combining the information from multiple bins could increase the level of detection certainty, and allow smaller abundances to be detectable at limiting cases.\\
2) the confidence level of the detection does not change significantly when distinct features of the same molecule are considered.\\
By contrast, the likelihood ratio test method is able to combine effectively information from multiple bins and multiple features.
The results in section 4 show a consistent improvement on the detection sensitivity over the individual bin method for most of the cases.

We compared our results with the ones calculated by \citep{Barstow13} with an automatic retrieval method. The test case was a hot Jupiter observed for a single eclipse with an EChO-like mission (see Appendix A).
We obtained consistent results for all the molecules with the exception of $CO$ and $NH_3$, for which we predict easier detectability. For ammonia, the explanation lies in the different line lists used: HITRAN08 \citep{hitran08} for \citet{Barstow13}, and Exomol BYTe \citep{byte} at high temperatures in our case. 
In the case of $CO$, the spectral features overlap in some spectral regions with $CH_4$ or $CO_2$, so it may be harder to detect when not isolated from other species, as it is assumed in this paper. In section \ref{sec:results3} we considered the case of a wet atmosphere given that water vapour is almost ubiquitous in warm and hot atmospheres and its signal extends from the visible to the infrared. We found that our conclusions for a dry atmosphere are still valid provided the water abundance does not exceed $\sim 10^{-5}$.

By examining predictions about compositions of hot and warm gaseous planets currently available in the literature \citep{moses2011,venot2012,line2010}, the abundances retrievable with SNR$\sim$10 are sufficient to discriminate among the different scenarios proposed.
Moreover, at SNR$\sim$10, most of the molecules are detectable in multiple regions of the spectrum, indicating that good constraints on the vertical thermal profile can be obtained.

\section{Conclusions}
In this paper we have addressed the question of molecular detectability in exoplanet atmospheres, for a range of key planet types and key molecules.
The five cases considered --- hot Jupiter, hot super-Earth, warm Neptune, temperate Jupiter and temperate super-Earth --- cover most of the exoplanets characterisable today or in the near future.
For other planets, the minimum detectable abundances can be extrapolated from these results.\\
We used a conservative and straightforward method, with which we delimit the objective criteria that need to be met for claiming 3$\sigma$ detections.
By artificially fixing the signal-to-noise per wavelength bin, we showed the limits in molecular detectability independently of instrument parameters, observation duration and sources of noise. 
We assumed simulated thermal profiles for the planet atmospheres, but investigated more extreme alternative profiles to quantify their effect on our results.
We focused on key atmospheric molecules such as $CH_4$, $CO$, $CO_2$, $NH_3$, $H_{2}O$, $C_{2}H_2$, $C_{2}H_6$, $HCN$, $H_{2}S$ and $PH_3$.
We found that for all planet cases, SNR=5 is typically enough to detect the strongest feature in most molecular spectra, provided the molecular abundance is large enough (e.g. $\sim10^{-6} /10^{-7}$ for $CO_2$, $10^{-4} / 10^{-5}$ for $H_{2}O$).
In atmospheres where a molecule has abundances lower than said threshold, SNR$\sim$10 or more may be required.
For the temperate super-Earth, we also show that with SNR=5, $O_3$ can be detected with a constant abundance of $10^{-7}$ at $9.6 \mu m$, and with an abundance of $10^{-5}$ at $14.3 \mu m$ (Note that on Earth, the ozone abundance typically varies as a function of altitude in the $10^{-8}$ to $10^{-5}$ range).
Other detection methods, such as the likelihood ratio test, combine information from multiple spectral bins and distinctive features.
We often find an improved performance in detection sensitivity of $\sim$10 when using this method.\\
Finally, we tested the robustness of our results by exploring sensitivity to the mean molecular weight of the atmosphere and relative water abundances, and found that our main results remain valid except for the most extreme cases.\\
To conclude, our analysis shows that detectability of key molecules in the atmospheres of a variety of exoplanet cases is within realistic reach, even with low SNR and spectral resolution values.
With new instruments specifically designed for exoplanet spectroscopic observation planned or under construction, the coming decade is set to be a golden age for the understanding of these newly-found worlds.

\section{Acknowledgements}
The author would like to thank the anonymous referees for helpful and constructive comments, and Paul Ginzberg for very helpful discussions.

\newpage
\bibliographystyle{apalike}       
\bibliography{tessenyi13}
\newpage
\appendix	
\section{Appendix}
\label{sec:dist_snr}
	The results in section \ref{sec:results} are obtained using a fixed SNR=5, 10 and 20.
We show here what observational requirements are needed to obtain these SNR values with a dedicated space instrument similar to EChO \citep{Tinetti2012b}.\\
For the five planet cases, we show a planet with and a planet without molecular absorptions, located at 3 distances from the observer.
Photon noise and an overall optical efficiency of 0.25 (to account for possible loss of signal through the instrument) are considered.
Further sources of noise may affect the observed signal, we refer the reader to \citet{tessenyisf2a,varley2013} for a more complete analysis on instrument performances.
The resolution is set to R=300 and 30 for the 1-5 and 5-16 $\mu m$ ranges, respectively.
Because of a weaker and colder signal, we only consider the 5-16 $\mu m$ spectral interval for the temperate super-Earth, and lower the resolution to 20.
\subsection{Warm Neptune}
\begin{figure}[h!]
\hspace*{-0.6in}
\includegraphics[width=3.4in]{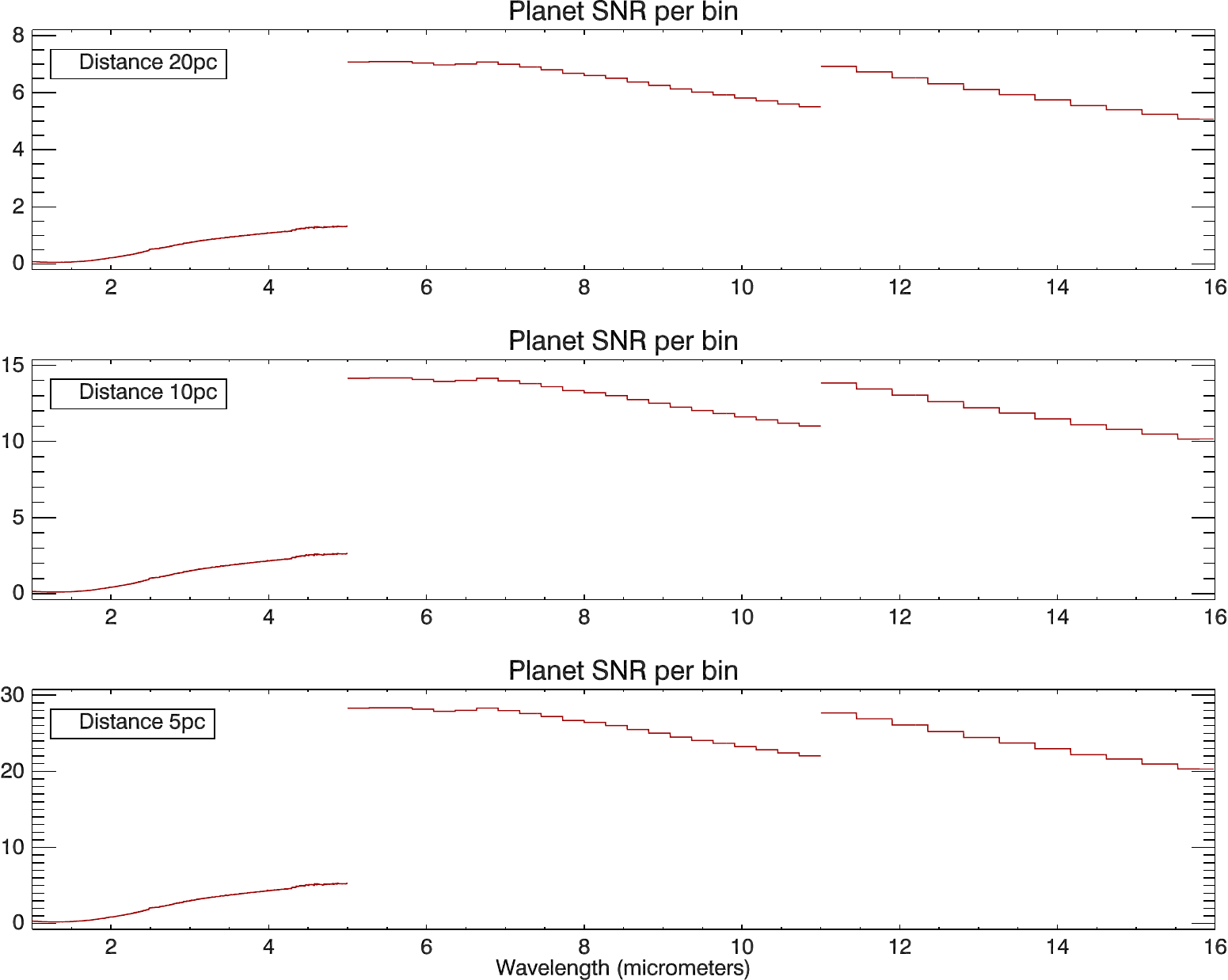}\includegraphics[width=3.5in]{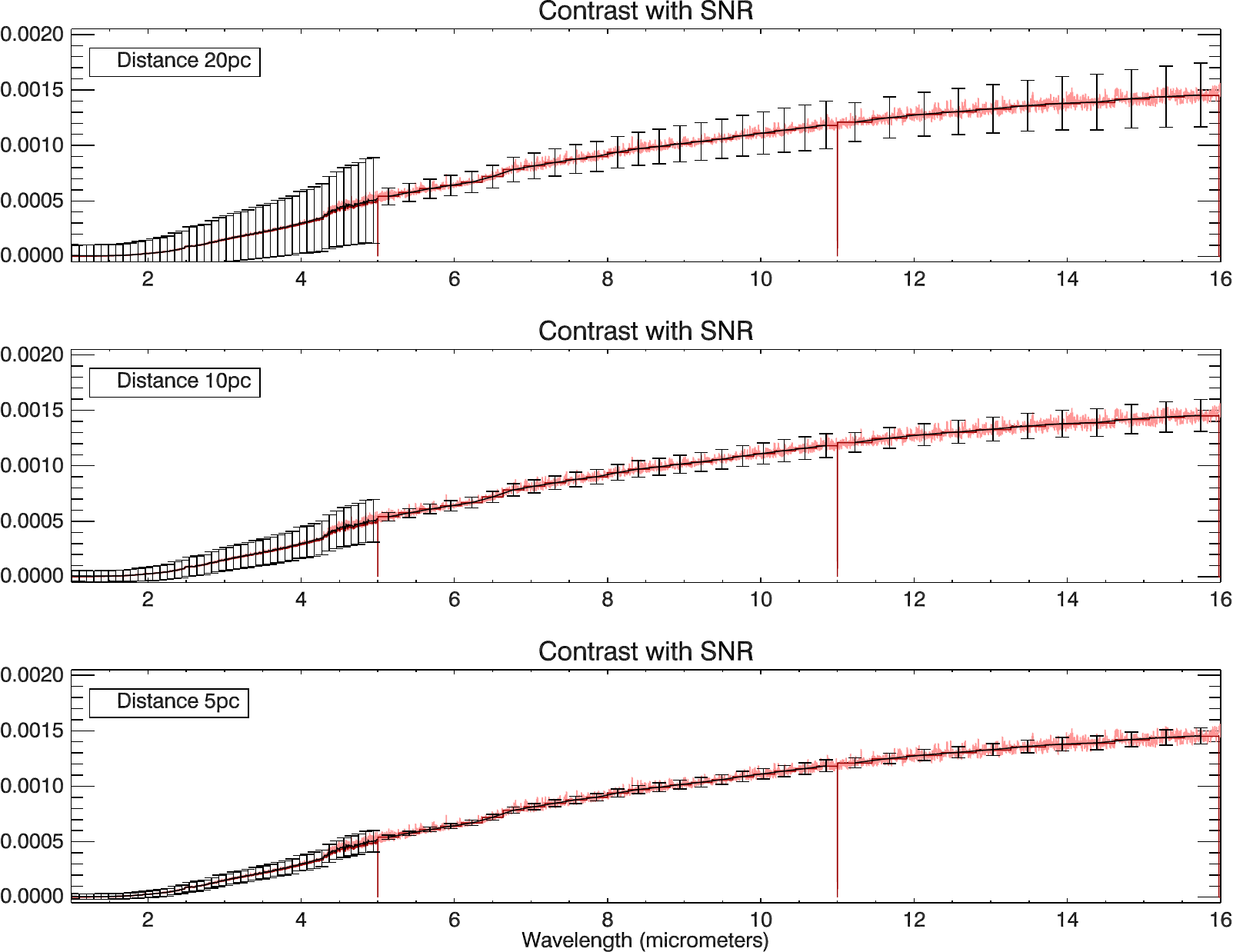}
\caption{\footnotesize A single transit of a warm Neptune with no molecules absorbing. \emph{Left:} SNR per resolution bin for a target located at 20, 10 and 5pc from the observer.
\emph{Right:} Planet/star contrast spectra with 1-sigma error bars. }
\label{fig:wn_SNR_dist}
\end{figure}

\begin{figure}[h!]
\hspace*{-0.6in}
\includegraphics[width=3.4in]{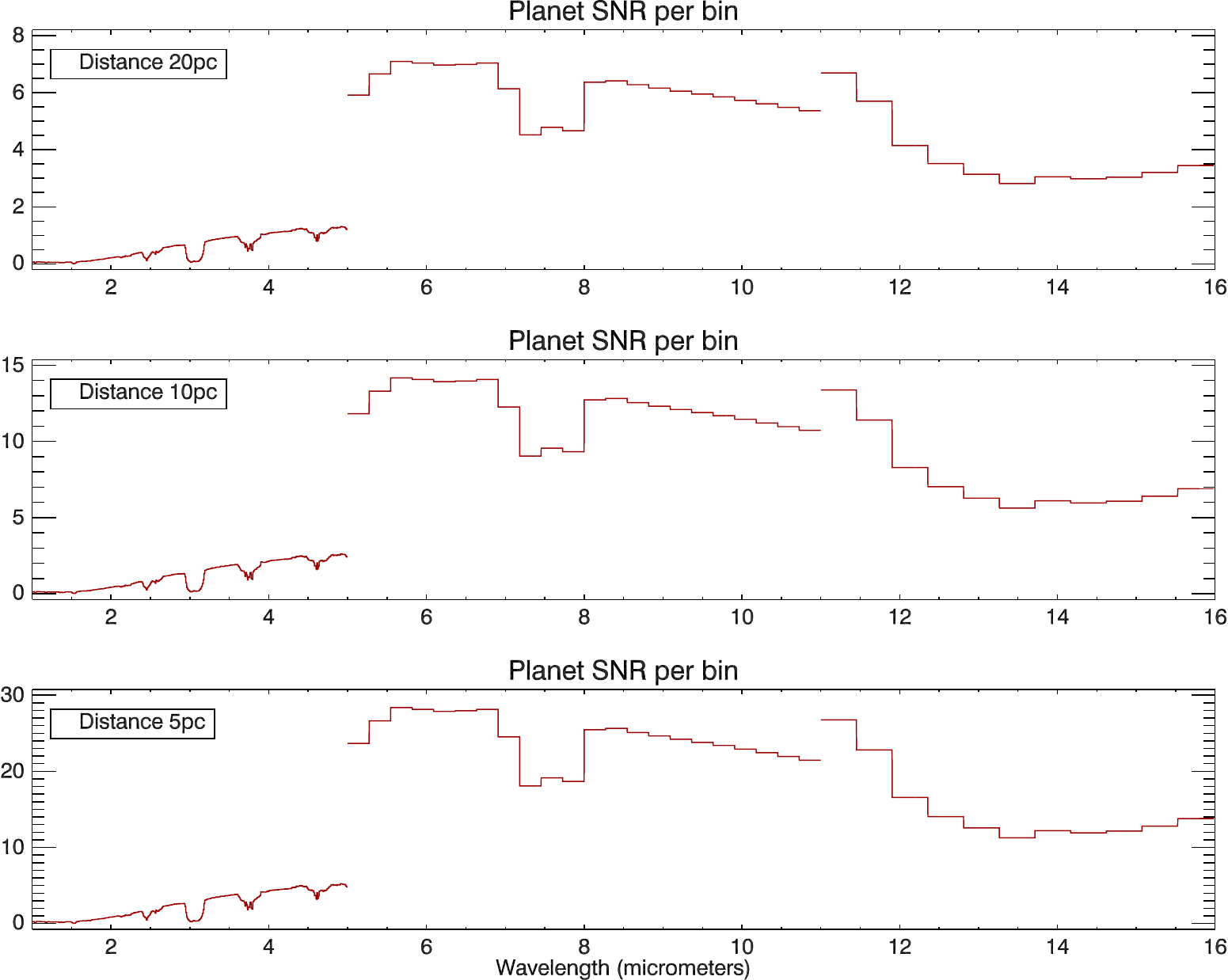}\includegraphics[width=3.5in]{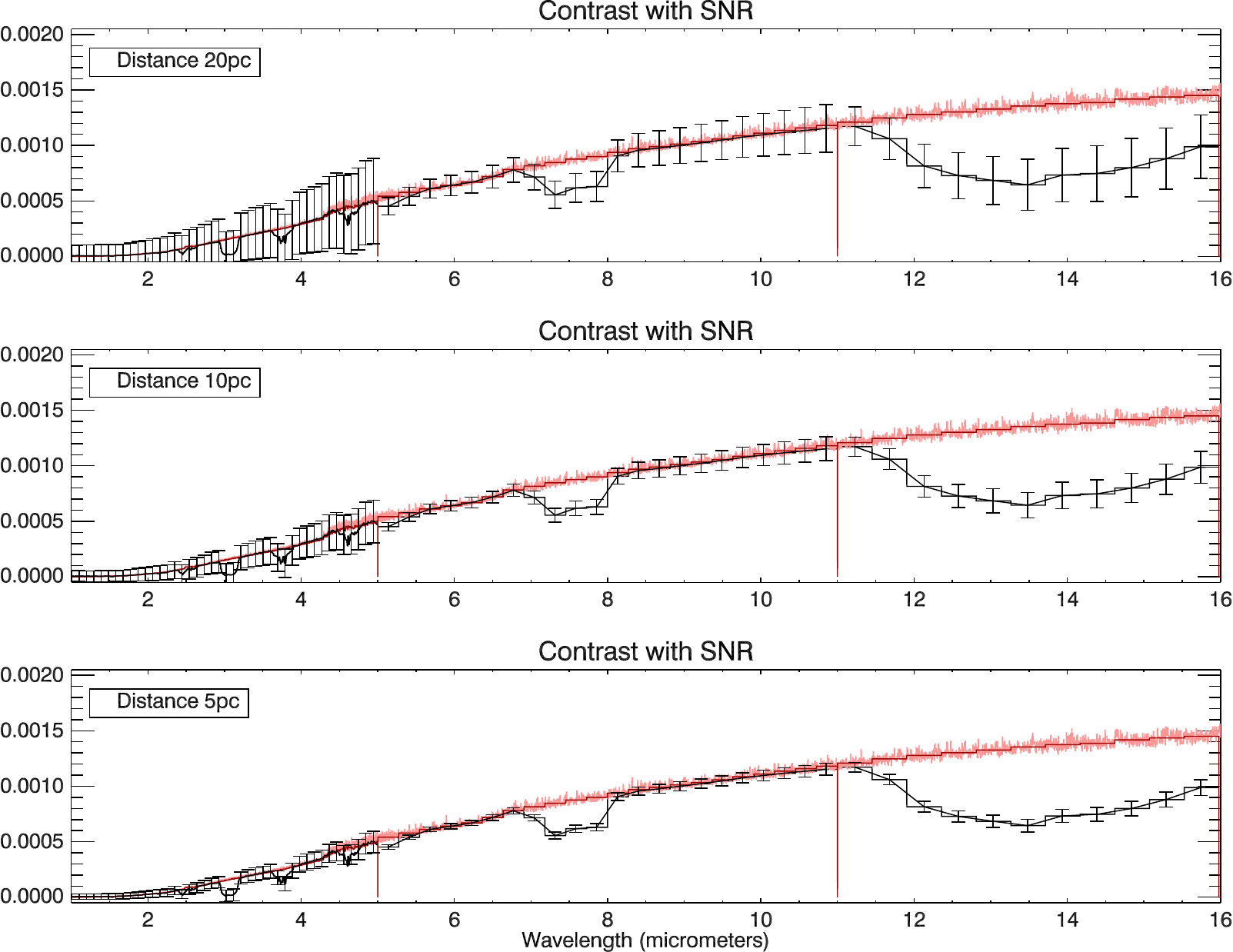}
\caption{\footnotesize A single transit of a warm Neptune with $C_{2}H_{2}$ in the atmosphere (abundance $10^{-4}$). \emph{Left:} SNR per resolution bin for a target located at 20, 10 and 5pc from the observer.
\emph{Right:} Planet/star contrast spectra with 1-sigma error bars.}
\label{fig:wn_SNR_dist_c2h2}
\end{figure}

Figures \ref{fig:wn_SNR_dist} and \ref{fig:wn_SNR_dist_c2h2} show the SNR per bin and the planet/star contrast spectra of a warm Neptune, without molecular absorption and with the presence of $C_{2}H_{2}$ at abundance $10^{-4}$. The planet is placed at three distances (5, 10 and 20pc) from the observer.
The maximum SNR value with no absorptions is $\sim$30 for the 5pc target, while the 20pc target has a maximum SNR value of $\sim$7.
With the presence of an absorbing feature at $\sim$7.5$\mu m$, the SNR drops to $\sim$5 for the 20pc target. A stronger absorbing feature will lower the SNR below 5.
With a distant Warm Neptune, the SNR may be too low for a single transit observation, and the co-adding up of multiple transits will be required. In addition, the shorter wavelength range (1 to 5 $\mu m$) will require co-adding of transits, as a single transit is not sufficient to obtain SNR of 5 or more, even for the closest target.
\subsection{Hot Jupiter}
In comparison with the warm Neptune, the signal of a hot Jupiter is stronger due to the combination of a larger and hotter planet+star, leading to higher SNR values per bin. 
\begin{figure}[h!]
\hspace*{-0.6in}
\includegraphics[width=3.4in]{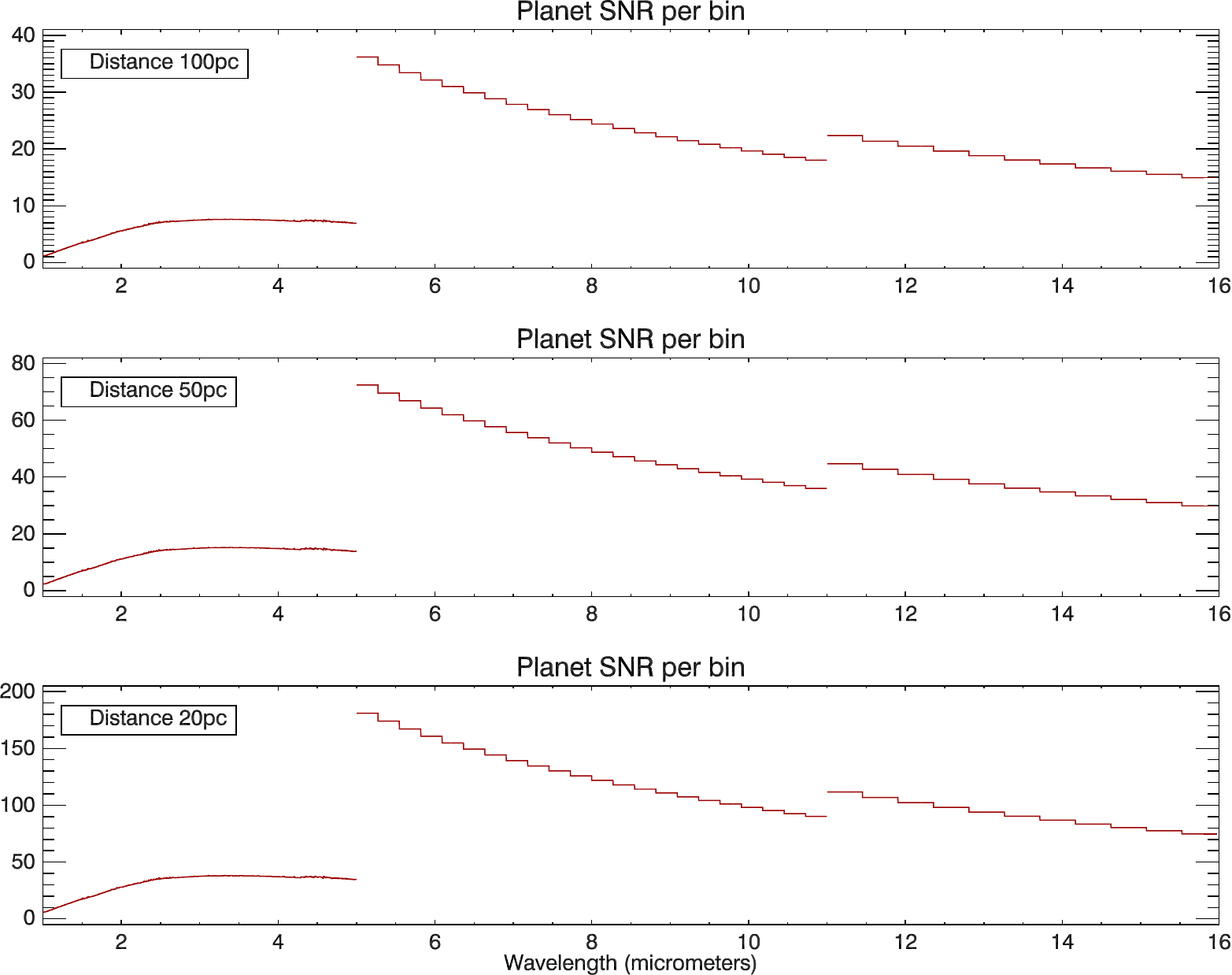}\includegraphics[width=3.5in]{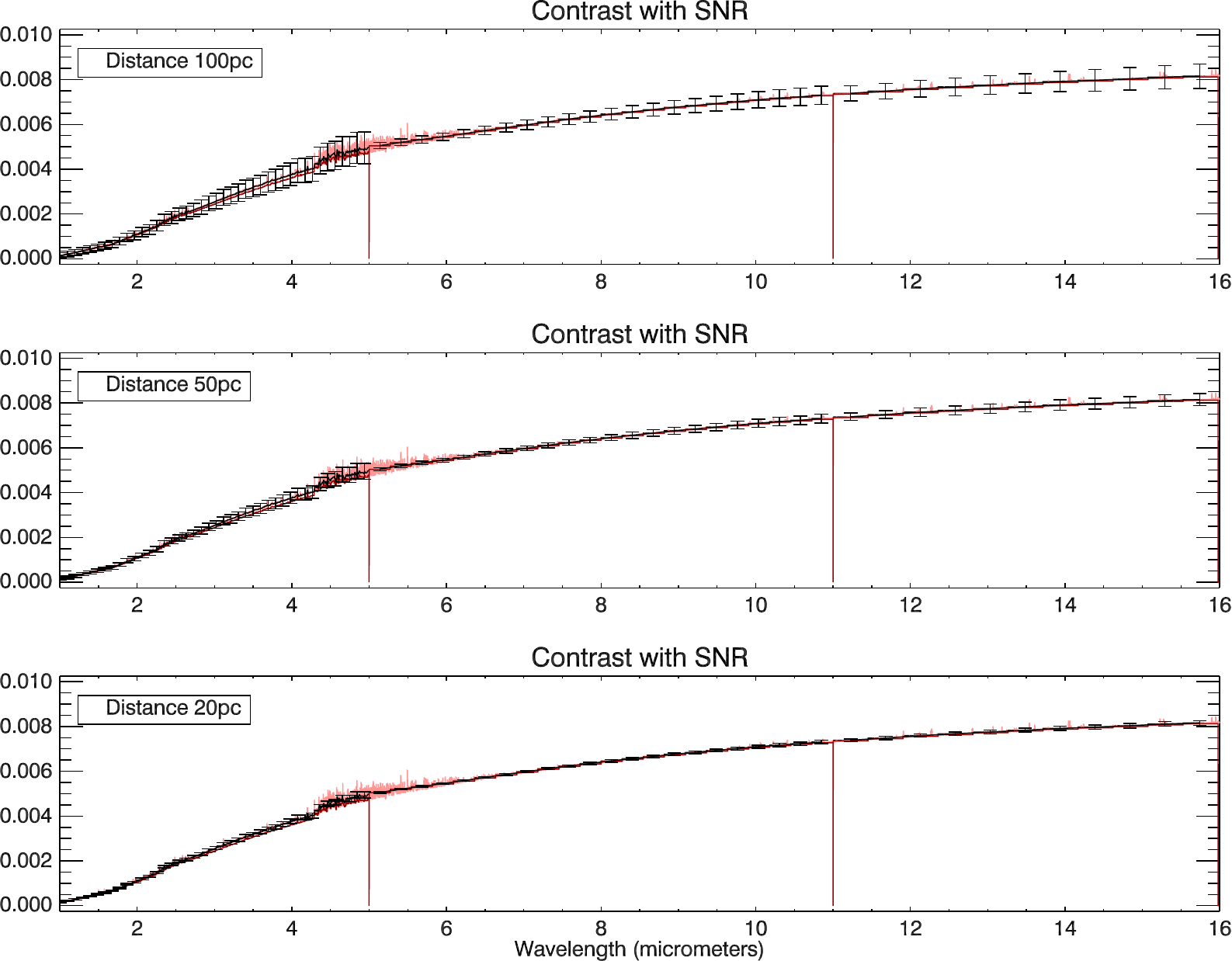}
\caption{\footnotesize A single transit of a hot Jupiter with no molecules absorbing. \emph{Left:} SNR per resolution bin for a target located at 100, 50 and 20pc from the observer.
\emph{Right:} Planet/star contrast spectra with 1-sigma error bars.
 }
\label{fig:hj_SNR_dist}
\end{figure}
\begin{figure}[h!]
\hspace*{-0.6in}
\includegraphics[width=3.4in]{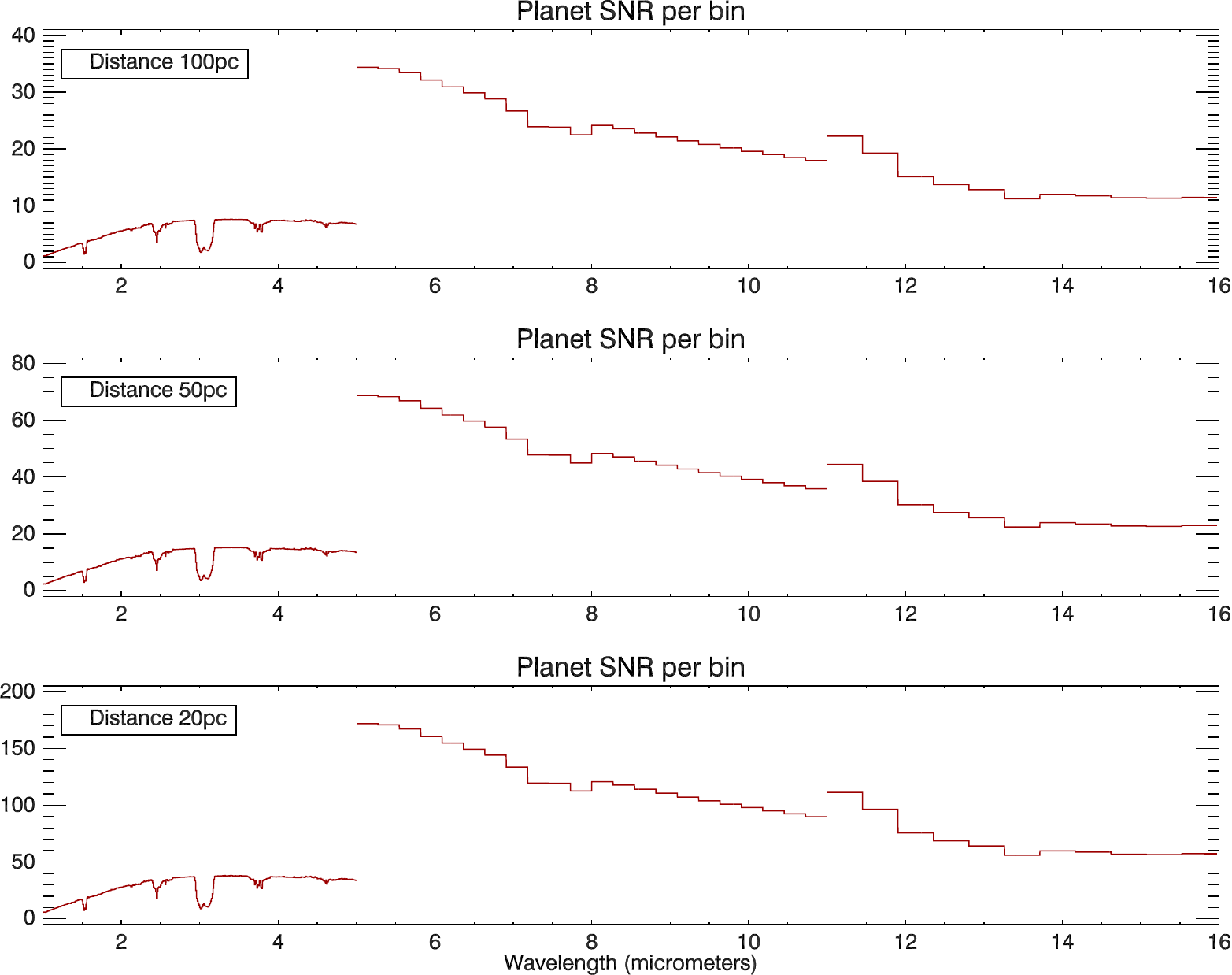}\includegraphics[width=3.5in]{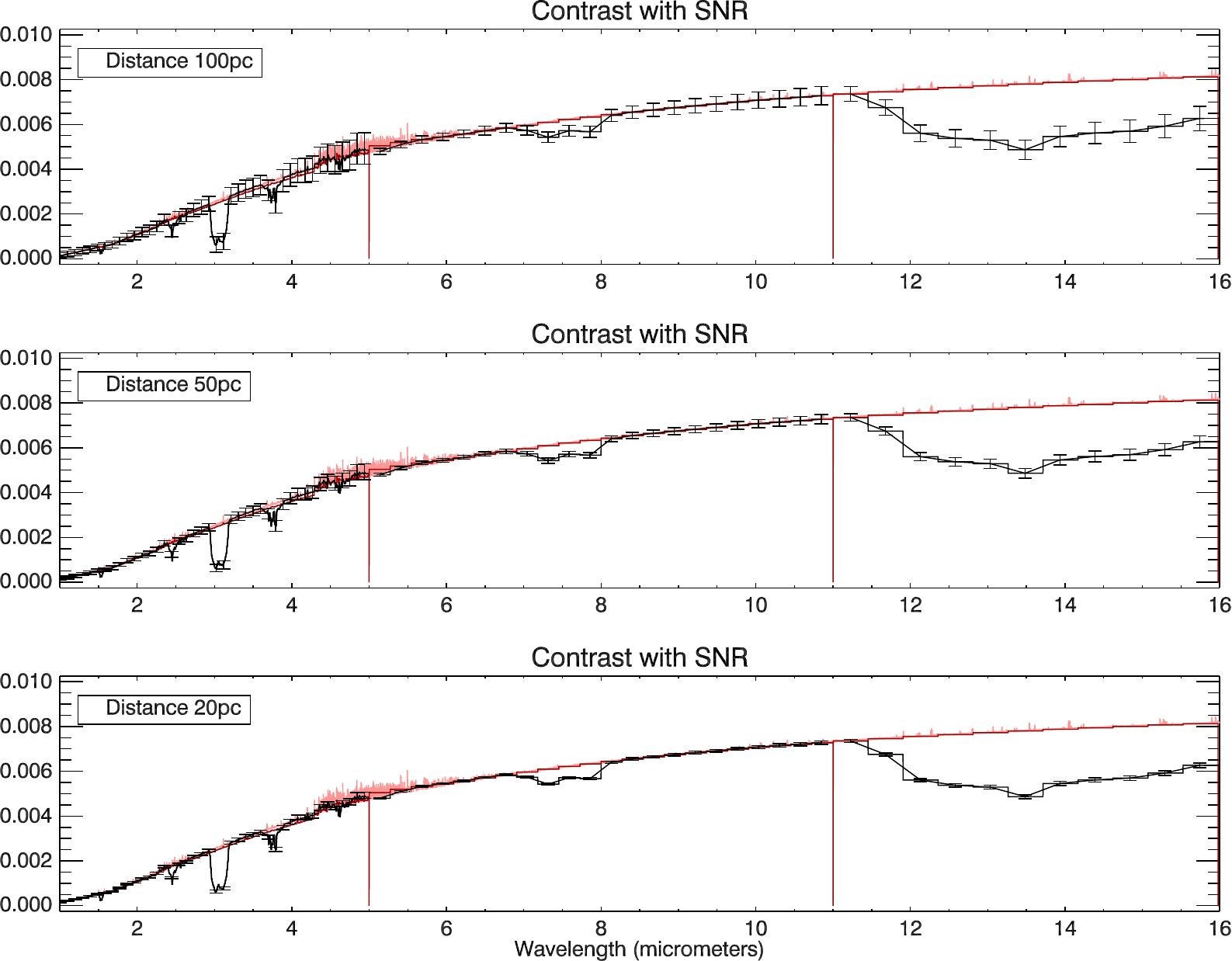}
\caption{\footnotesize 
A single transit of a hot Jupiter with $C_{2}H_{2}$ in the atmosphere (abundance $10^{-4}$). \emph{Left:} SNR per resolution bin for a target located at 100, 50 and 20pc from the observer.
\emph{Right:} Planet/star contrast spectra with 1-sigma error bars.
}
\label{fig:hj_SNR_dist_c2h2}
\end{figure}
Given the high SNR values from this planet, and to place the results from section \ref{sec:results} into context, the distances for this planet are changed to 100, 50 and 20 pc (HD189733b, our template hot Jupiter, is located at 19.3pc). 
Figure \ref{fig:hj_SNR_dist} shows the SNR per resolution bin and corresponding planet/star contrast spectra for a blackbody case, and Figure \ref{fig:hj_SNR_dist_c2h2} shows the change in SNR due to the presence of $C_{2}H_{2}$ in the atmosphere with abundance  $10^{-4}$.
\subsection{Hot super-Earth}
\begin{figure}[h!]
\hspace*{-0.6in}
\includegraphics[width=3.4in]{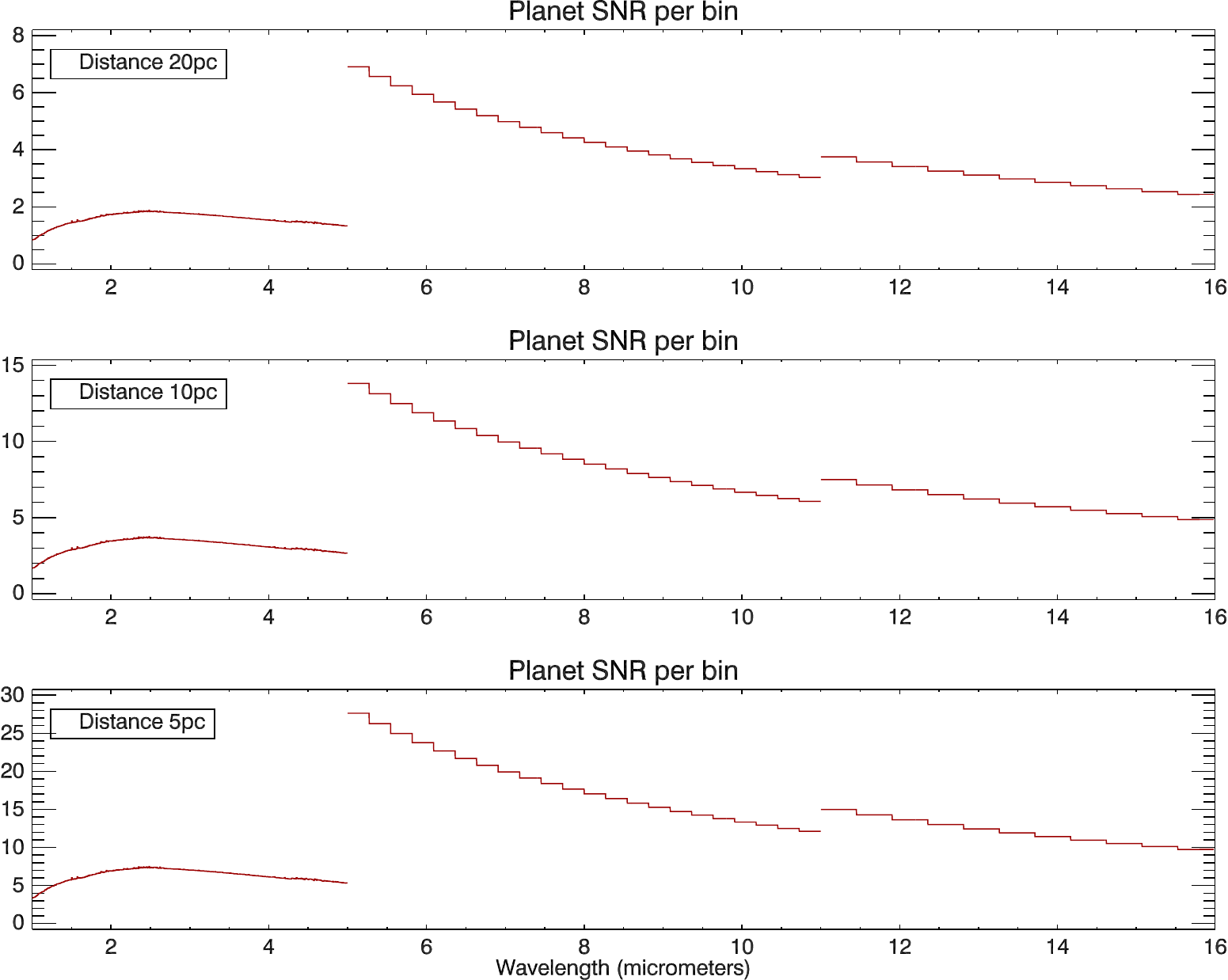}\includegraphics[width=3.5in]{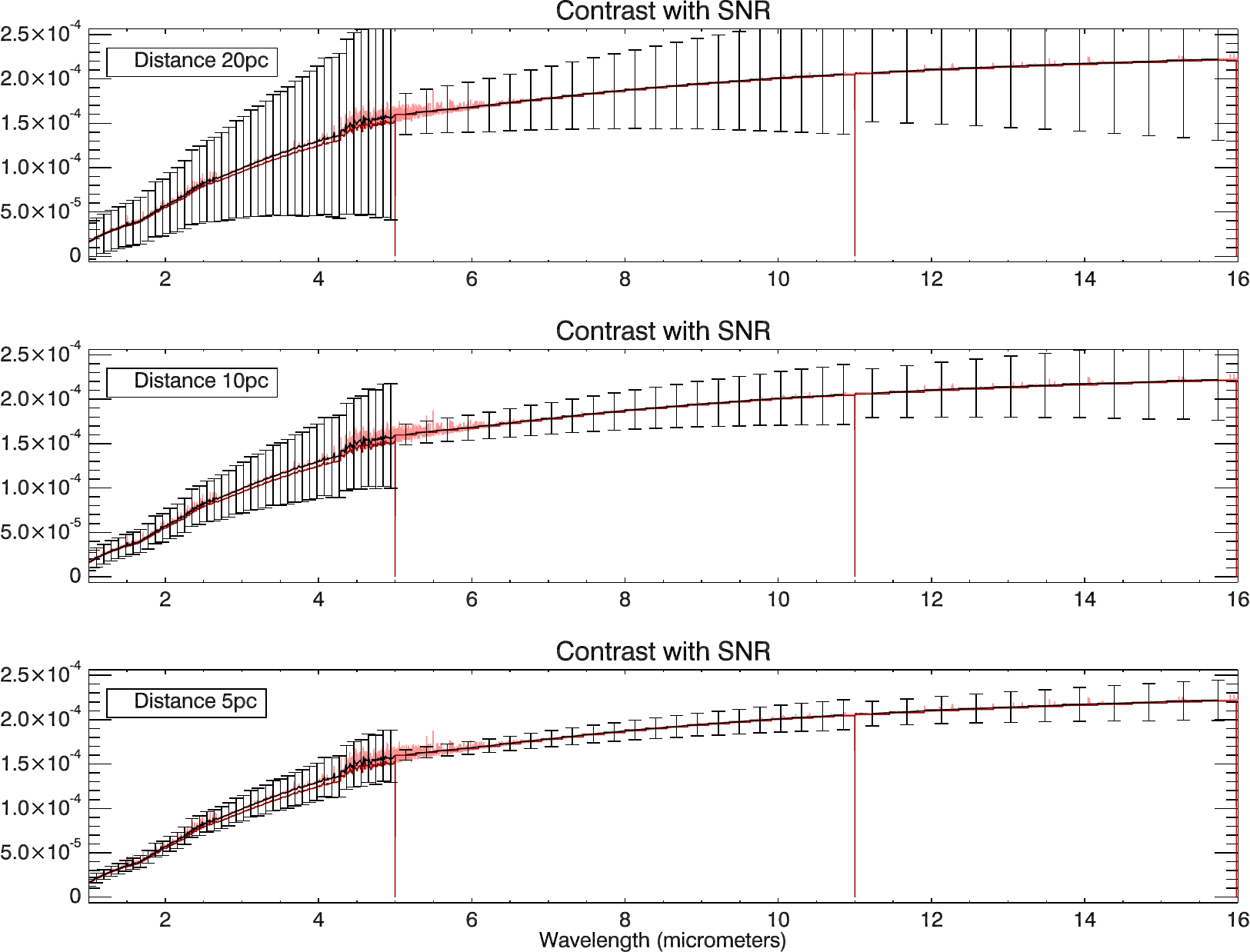}
\caption{\footnotesize A single transit of a hot super-Earth with no molecules absorbing. \emph{Left:} SNR per resolution bin for a target at 20, 10 and 5pc. \emph{Right:} Planet/star contrast spectra with 1-sigma error bars. }
\label{fig:hotse_SNR_dist}
\end{figure}

\begin{figure}[h!]
\hspace*{-0.6in}
\includegraphics[width=3.4in]{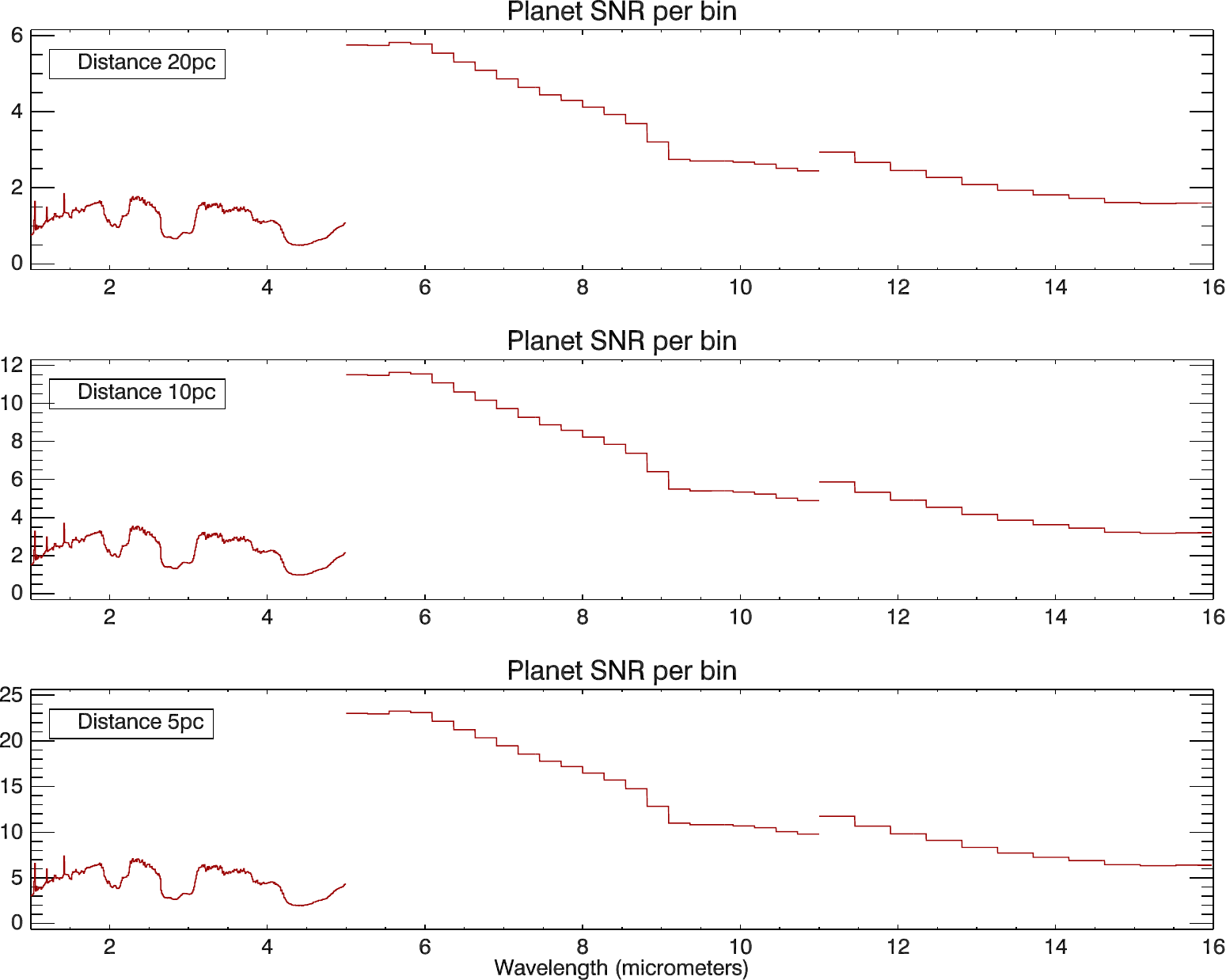}\includegraphics[width=3.5in]{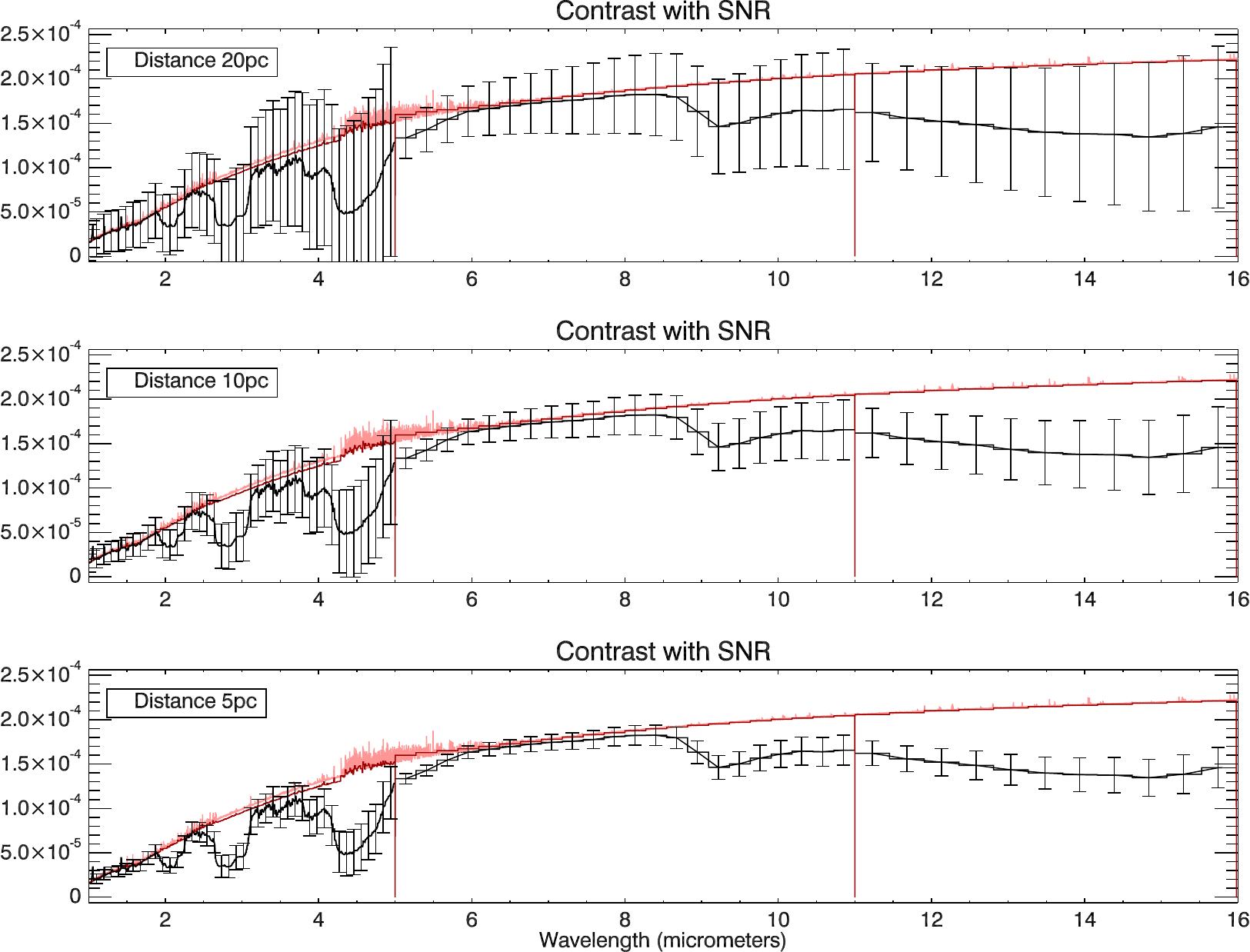}
\caption{\footnotesize A single transit of a hot super-Earth planet with only $CO_{2}$ in the atmosphere (abundance $10^{-4}$). The bulk composition of the planet atmosphere in this simulation is $H_{2}O$, see section \ref{sec:superearths} for a comparison of main atmosphere constituents. \emph{Left:} SNR per resolution bin for a target at 20, 10 and 5pc. \emph{Right:} Planet/star contrast spectra with 1-sigma error bars. }
\label{fig:hotse_SNR_dist_co2}
\end{figure}

The planet/star surface ratio is less favorable here than the warm Neptune and hot Jupiter cases, however the temperature on this planet is assumed to be 2390 K, presenting a strong emission signal. The distances thus considered are 5, 10 and 20 pc (\emph{55 Cnc} is located at 12.34 pc). The SNR per bin for a blackbody case is shown in Figure \ref{fig:hotse_SNR_dist} alongside the planet/star contrast spectra. The same planet is also shown with the presence of $CO_2$ in the atmosphere (abundance $10^{-4}$), in Figure \ref{fig:hotse_SNR_dist_co2}. 
At a distance of 20pc, co-adding of transits will be necessary to obtain higher SNR values in the longer wavelength range: in Figure \ref{fig:hotse_SNR_dist_co2}, the signature of $CO_2$ at 10 $\mu m$ gives a SNR per bin that is below 3. The 1 to 5 $\mu m$ range will need to have multiple transits added to obtain higher SNR values, even for a close-by target.
\subsection{Temperate Jupiter}
\begin{figure}[h!]
\hspace*{-0.6in}
\includegraphics[width=3.4in]{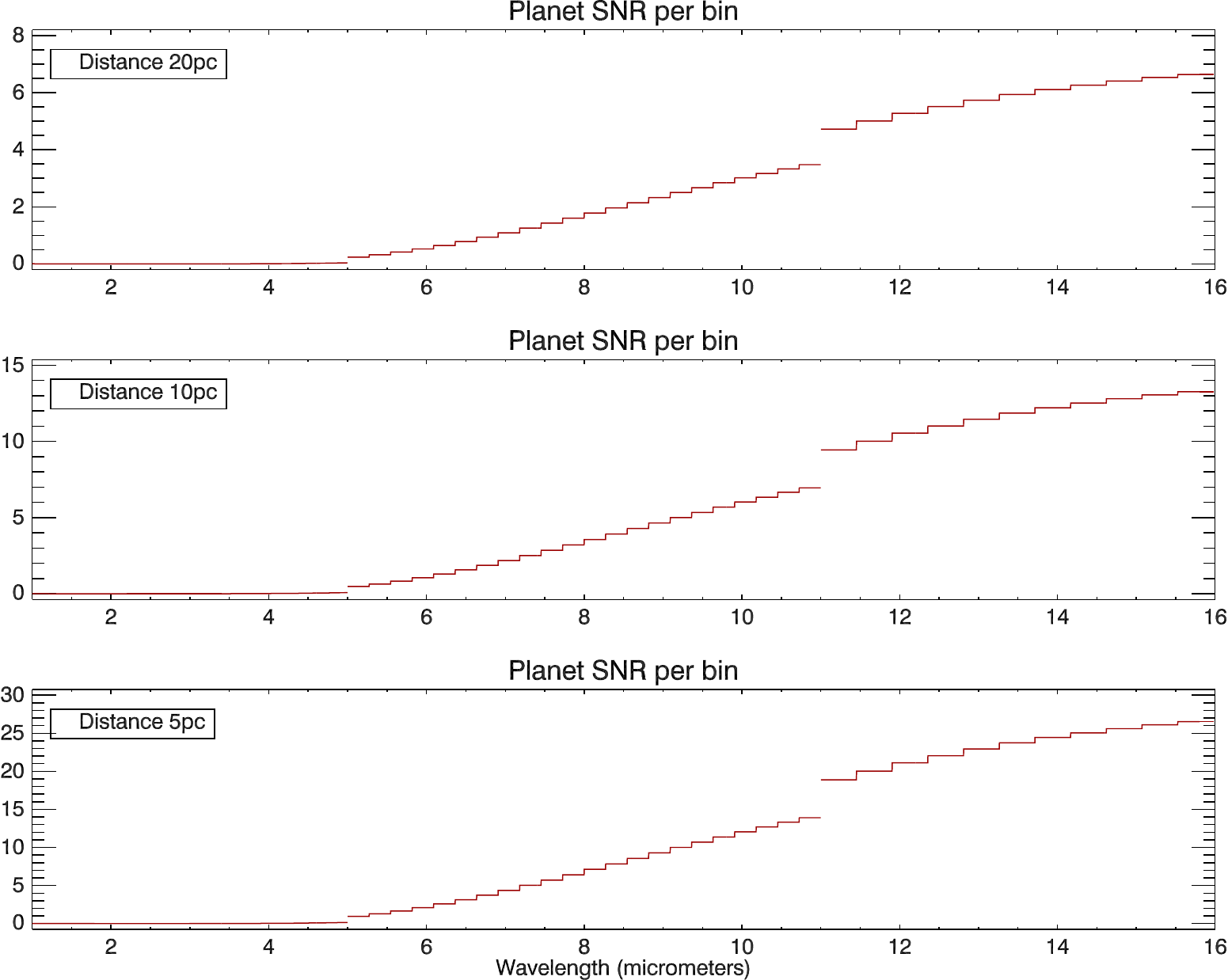}\includegraphics[width=3.5in]{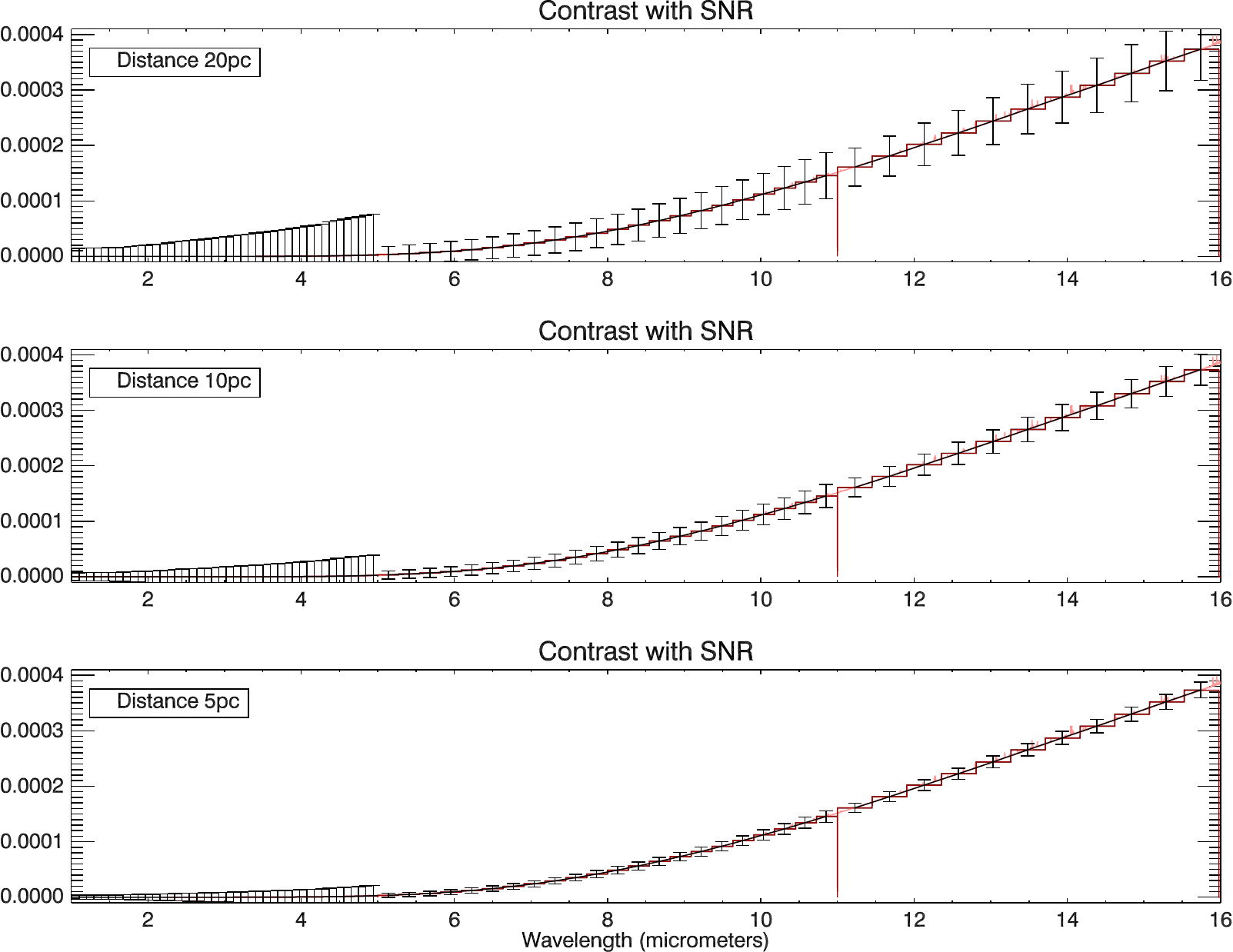}
\caption{\footnotesize A single transit of a Temperate Jupiter with no molecules absorbing.  \emph{Left:} SNR per resolution bin for a target at 20, 10 and 5pc. \emph{Right:} Planet/star contrast spectra with 1-sigma error bars. }
\label{fig:hzjup_SNR_dist}
\end{figure}

\begin{figure}[h!]
\hspace*{-0.6in}
\includegraphics[width=3.4in]{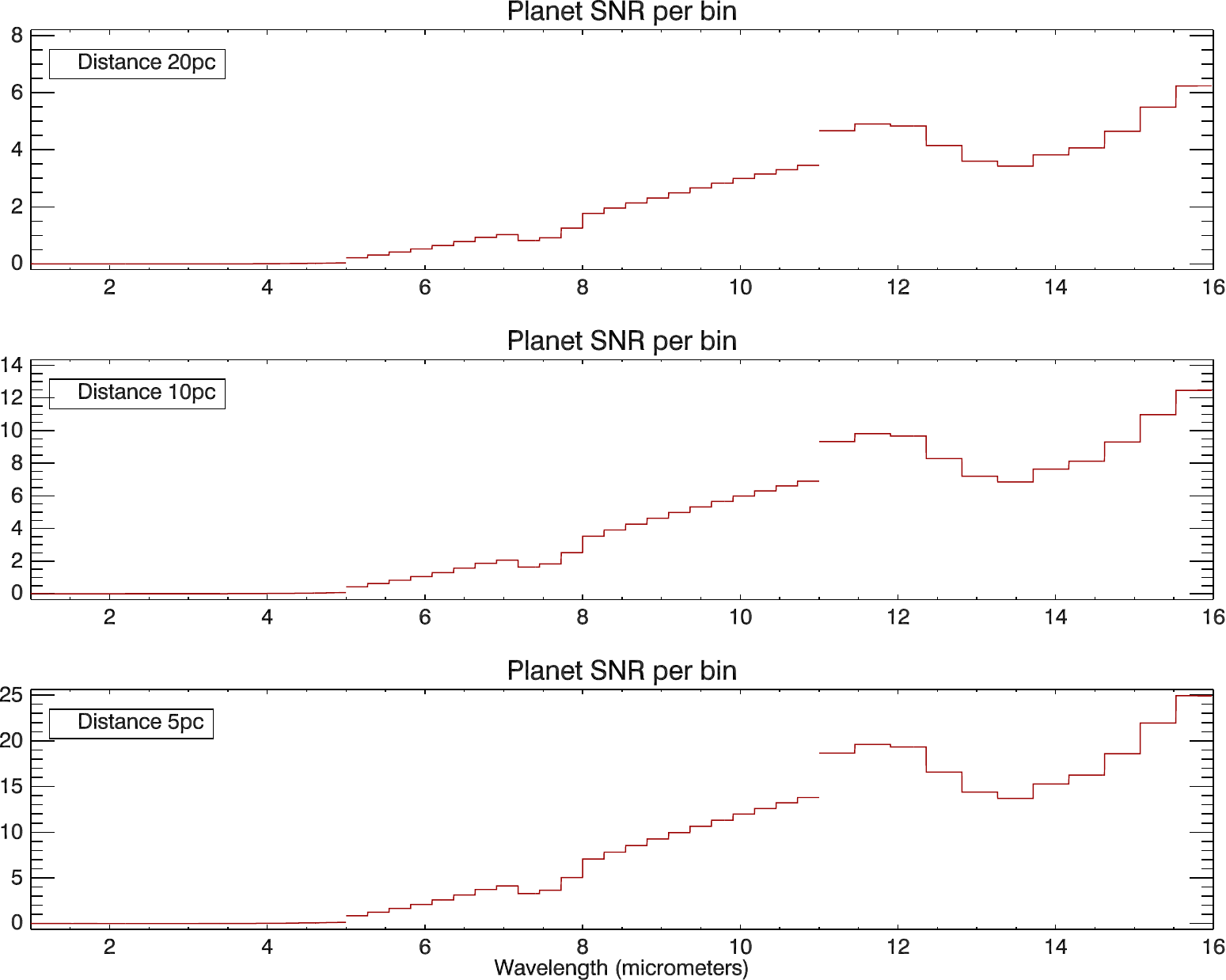}\includegraphics[width=3.5in]{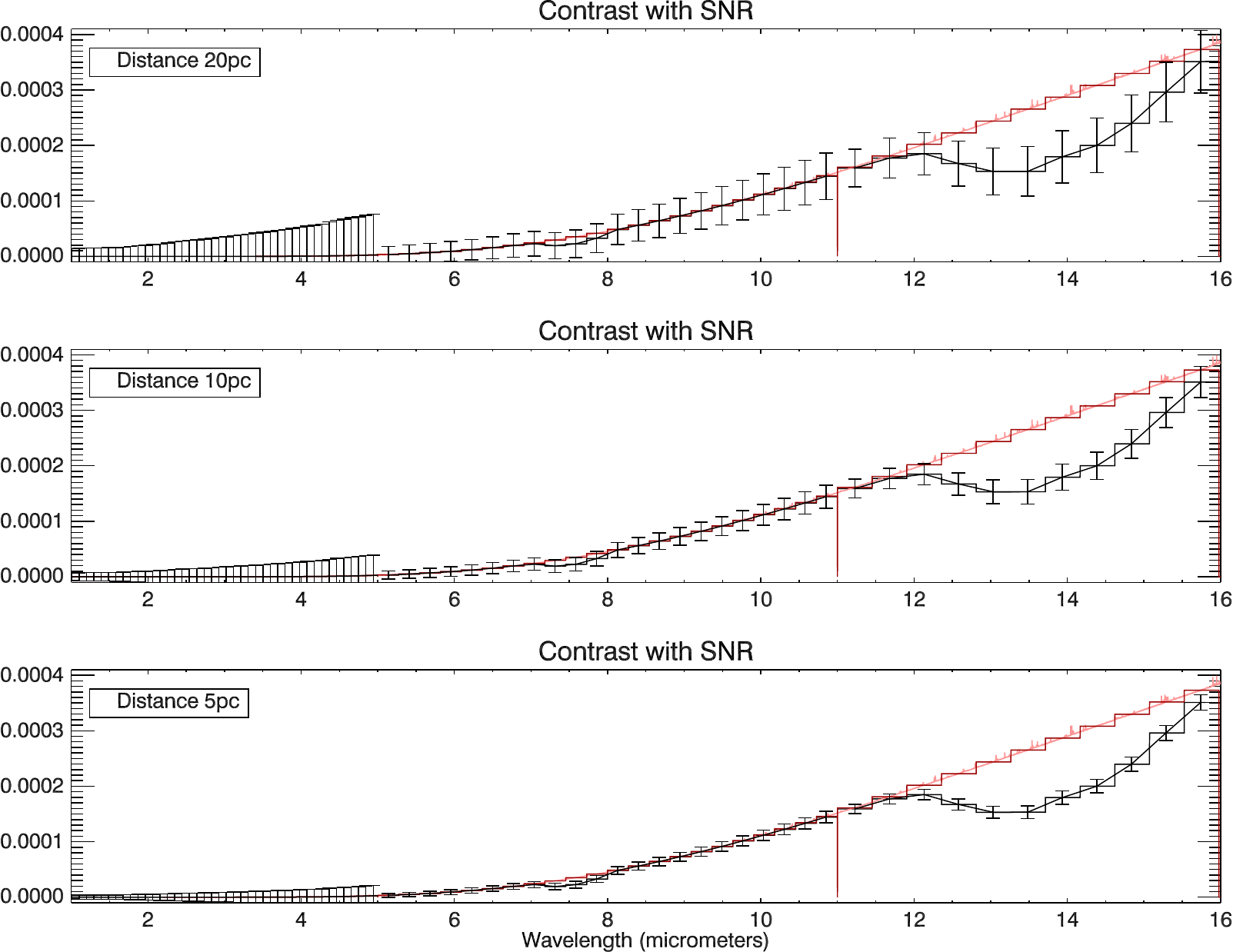}
\caption{\footnotesize A single transit of a Temperate Jupiter planet with only $C_{2}H_2$ in the atmosphere (abundance $10^{-5}$).  \emph{Left:} SNR per resolution bin for a target at 20, 10 and 5pc. \emph{Right:} Planet/star contrast spectra with 1-sigma error bars.  }
\label{fig:hzjup_SNR_dist_c2h2}
\end{figure}

Of the five planet cases, the Temperate Jupiter has the strongest planet/star surface ratio. In addition, a single transit of this planet lasts 7.9 hours. This allows us to consider distances of 5, 10 and 20pc, for both a blackbody continuum planet (Figure \ref{fig:hzjup_SNR_dist}) and a planet with C$_2$H$_2$ at abundance $10^{-5}$ in the atmosphere (Figure \ref{fig:hzjup_SNR_dist_c2h2}). The temperature of the planet at 320K will emit mostly around 10$\mu m$, and no signal will be visible at wavelengths below 5 $\mu m$. The more distant planets will require co-adding of transit observations to reach SNR values of 5 to 10 in the 5 to 11 $\mu m$ wavelength range. 
\subsection{Temperate super-Earth}
\begin{figure}[h!]
\centering
\includegraphics[width=3.4in]{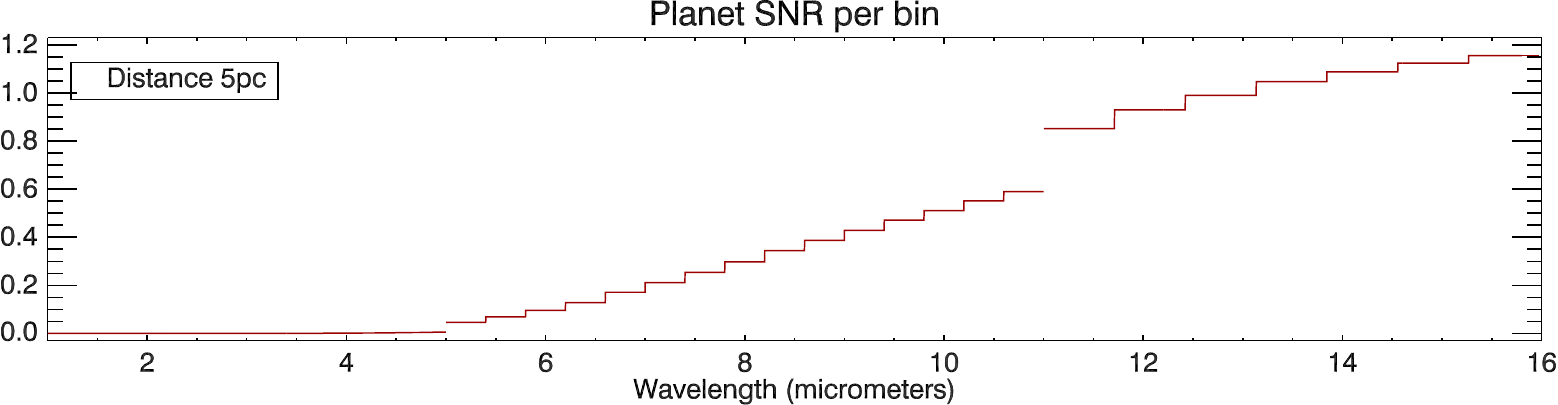}
\caption{\footnotesize A single transit of a temperate super-Earth planet with no atmosphere at 5pc. The SNR per bin is very low, and multiple transits will be needed for this type of target.}
\label{fig:hzse_SNR_dist}
\end{figure}

\begin{figure}[h!]
\hspace*{-0.6in}
\includegraphics[width=3.4in]{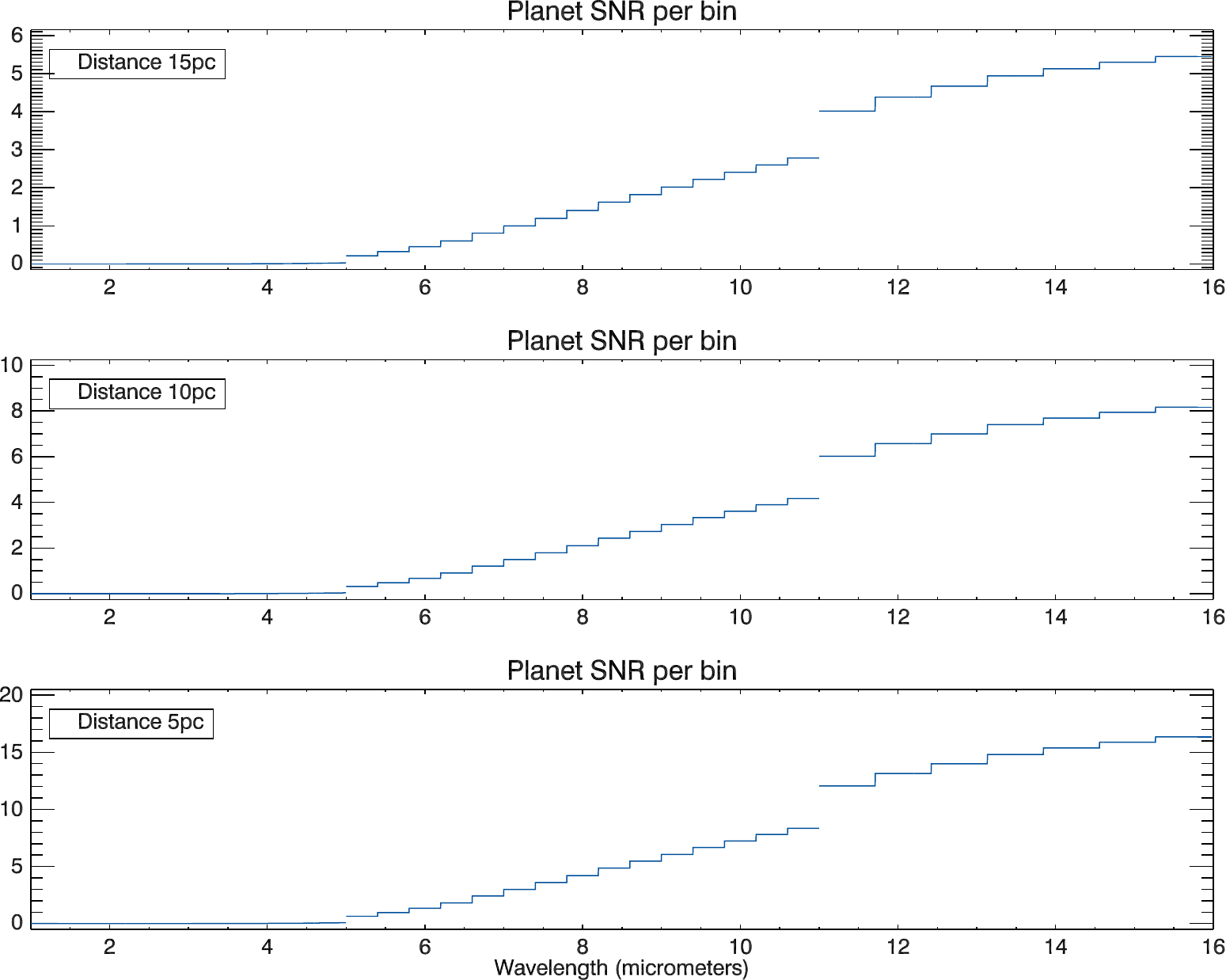}\includegraphics[width=3.5in]{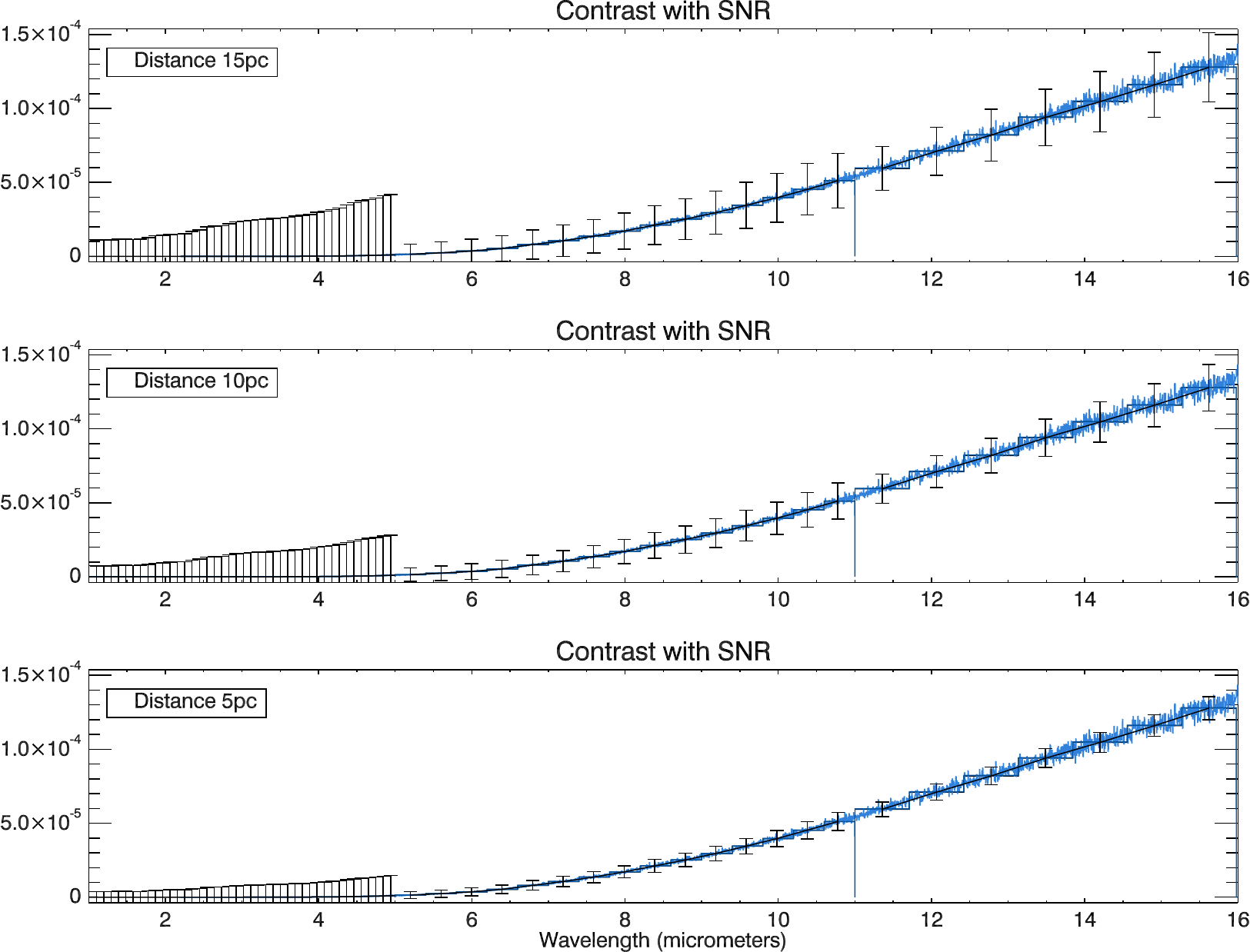}
\caption{\footnotesize 200 transits of a temperate super-Earth with no molecules absorbing. \emph{Left:} SNR per resolution bin for a target at 15, 10 and 5pc. \emph{Right:} Planet/star contrast spectra with 1-sigma error bars.  }
\label{fig:hzse200_SNR_dist}
\end{figure}

\begin{figure}[h!]
\hspace*{-0.6in}
\includegraphics[width=3.4in]{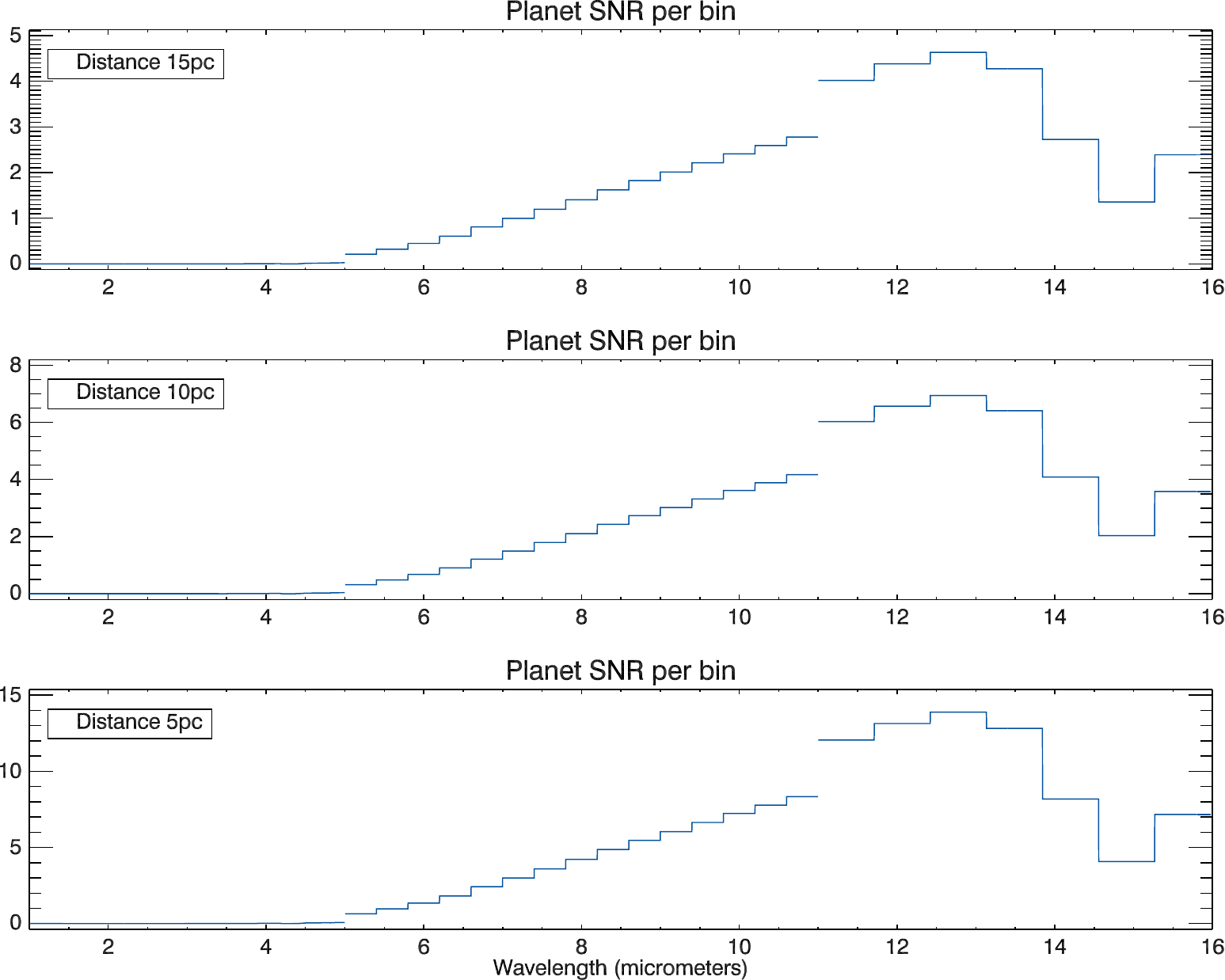}\includegraphics[width=3.5in]{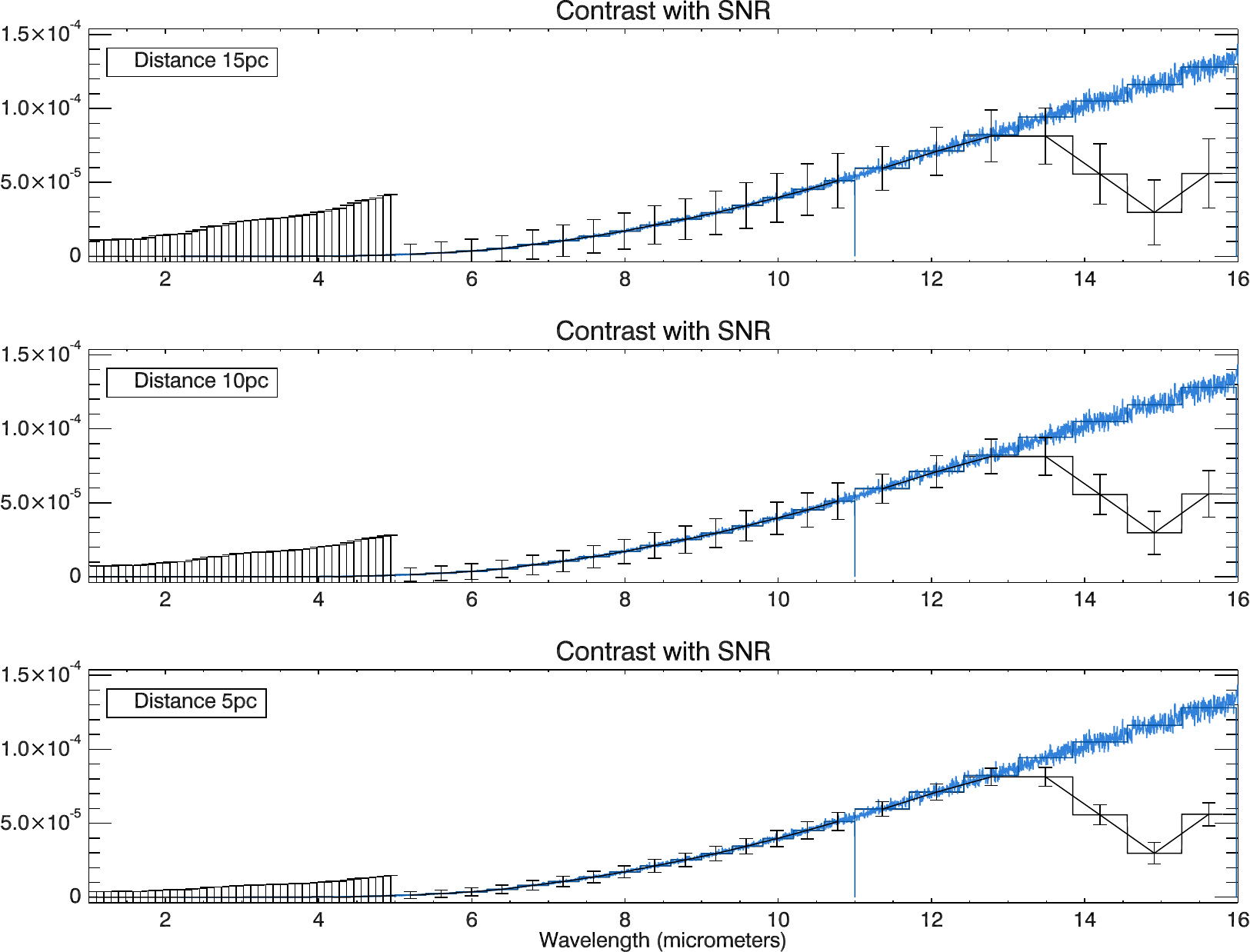}
\caption{\footnotesize 200 transits of a temperate super-Earth planet with only $CO_{2}$ in the atmosphere (abundance $10^{-4}$). The bulk composition of the planet atmosphere in this simulation is $N_{2}$, see section \ref{sec:superearths} for a comparison of main atmosphere constituents. \emph{Left:} SNR per resolution bin for a target at 15, 10 and 5pc. \emph{Right:} Planet/star contrast spectra with 1-sigma error bars. }
\label{fig:hzse200_SNR_dist_co2}
\end{figure}

We consider this planet to be a 1.8 Earth radii telluric planet orbiting a M4.5V star, with a surface ratio similar to the Warm Neptune case. However the smaller and dimmer star combined with a colder planet provide a weaker emission signal. In this case, a single transit can not be used, as the SNR values will be of the order of $10^0$, illustrated in Figure \ref{fig:hzse_SNR_dist}, with a nearby (5pc) target. 
We present here the results of co-added transits (200) to obtain SNR values that are similar to the other target cases, for a target located at 5, 10 and 15 pc (Figure \ref{fig:hzse200_SNR_dist}).
This is the most challenging case, even a small variation in stellar of planetary parameters might impact the observability of this target, see \citet{tessenyi2012}.
We show the SNR per resolution bin and the planet+star contrast spectra for a blackbody continuum planet and a planet with CO$_2$ (abundance $10^{-4}$) the atmosphere (Figure \ref{fig:hzse200_SNR_dist_co2}).
As in the Temperate Jupiter case, this planet has a temperature of 320K, with peak emission near 10 $\mu m$, and no emission signal will be visible below 5 $\mu$m. The resolution in the 5 - 16 $\mu m$ range is lowered to 20, to maximise the number of photons.

\end{document}